\documentclass[fleqn,usenatbib]{mnras}

\usepackage{newtxtext,newtxmath}
\usepackage[T1]{fontenc}

\DeclareRobustCommand{\VAN}[3]{#2}
\let\VANthebibliography\thebibliography
\def\thebibliography{\DeclareRobustCommand{\VAN}[3]{##3}\VANthebibliography}

\usepackage{graphicx}	% Including figure files
\usepackage{amsmath}	% Advanced maths commands
\usepackage{amssymb}	% Extra maths symbols

\title[Galactic outflows in high-$z$ massive galaxies]{The inefficiency of stellar feedback in driving galactic outflows in massive galaxies at high redshift}

\author[L. Bassini et al.]{
\parbox{\textwidth}{
Luigi Bassini,$^{1}$\thanks{E-mail: luigi.bassini@uzh.ch}
Robert Feldmann,$^{1}$
Jindra Gensior,$^{1}$
Christopher C. Hayward, $^{2}$
Claude-André Faucher-Giguère,$^{3}$
Elia Cenci, $^{1}$\
Lichen Liang, $^{4}$\
Mauro Bernardini $^{1}$
}\\
$^{1}$Institute for Computational Science, University of Zurich, Zurich CH-8057, Switzerland\\
$^{2}$Center for Computational Astrophysics, Flatiron Institute, 162 Fifth Avenue, New York, NY 10010, USA\\
$^{3}$Department of Physics and Astronomy and CIERA, Northwestern University, 2145 Sheridan Road, Evanston, IL 60208, USA\\
$^{4}$Canadian Institute for Theoretical Astrophysics, University of Toronto, 60 St. George Street, Toronto, ON M5S 3H8, Canada\\
}

\date{Accepted XXX. Received YYY; in original form ZZZ}

\pubyear{2015}

\begin{document}
\label{firstpage}
\pagerange{\pageref{firstpage}--\pageref{lastpage}}
\maketitle

% Abstract of the paper
\begin{abstract}
 
Recent observations indicate that galactic outflows are ubiquitous in high redshift galaxies, including normal star forming galaxies, quasar hosts, and dusty star forming galaxies (DSFGs). However, the impact of outflows on the evolution of their hosts is still an open question. Here, we analyse the star formation histories (SFH) and galactic outflow properties of galaxies in massive haloes ($10^{12}M_{\odot}<M_{\rm vir} <5\times 10^{12}M_{\odot}$) at $z\gtrsim5.5$ in three zoom-in cosmological simulations from the MassiveFIRE suite, as part of the Feedback In Realistic Environments (FIRE) project. The simulations were run with the FIRE-2 model, which does not include feedback from active galactic nuclei (AGN). The simulated galaxies resemble $z>4$ DSFGs, with SFRs of $\sim 1000\ M_{\odot}\rm yr^{-1}$ and molecular gas masses of $M_{\rm mol}\sim 10^{10}\ M_{\odot}$. However, the simulated galaxies are characterised by higher circular velocities than those observed in high-z DSFGs. The mass loading factors from stellar feedback are of the order of $\sim 0.1$, implying that stellar feedback is inefficient in driving galactic outflows and gas is consumed by star formation on much shorter time-scales than it is expelled from the interstellar medium (ISM). We also find that stellar feedback is highly inefficient in self-regulating star formation in this regime, with an average integrated star formation efficiency (SFE) per dynamical time of $30\%$. Finally, compared to FIRE-2 galaxies hosted in similarly massive haloes at lower redshift, we find lower mass loading factors and higher SFEs in the high redshift sample. We argue that both effects originate from the higher total and gas surface densities that characterise high$-z$ massive systems.
\end{abstract}

% Select between one and six entries from the list of approved keywords.
% Don't make up new ones.

%%%%%%%%%%%%%%%%%%%%%%%%%%%%%%%%%%%%%%%%%%%%%%%%%%

%%%%%%%%%%%%%%%%% BODY OF PAPER %%%%%%%%%%%%%%%%%%

\section{Introduction}

Dusty star forming galaxies (DSFGs) represent an important stage in the evolution of massive galaxies. They are typically massive ($M_{\star}\gtrsim 10^{10} M_{\odot}$), gas rich ($M_{\rm gas}\gtrsim 10^{10} M_{\odot}$), and bright at far-infrared (FIR) wavelengths ($L_{\rm FIR}\sim 10^{12}-10^{13}L_{\odot}$, e.g. \citealt{2013BOTHWELL}, \citealt{2013RIECHERS}, \citealt{2016ARAVENA}, \citealt{2018ZAVALA}, \citealt{2018MARRONE}, \citealt{2020SPILKER1}, \citealt{2020RIECHERS}). Their number density peaks at $z\sim 2.5$ (e.g., \citealt{2005CHAPMAN}), and given their high star formation rates (SFRs) and stellar masses they significantly contribute to the cosmic star formation rate density and stellar mass density at cosmic noon (about $\sim 20$ per cent and  $\sim30-50$ per cent respectively, e.g., \citealt{2010MICHALOWSKI}, \citealt{2017SMITH}). Although more rare, DSFGs have been successfully detected up to redshift $z\sim7$ (e.g., \citealt{2018MARRONE}), and they are thought to represent the natural progenitor of the massive quiescent galaxies that are found at redshifts as high as $z>3$ (e.g., \citealt{2017GLAZEBROOK}, \citealt{2018SCHREIBER}, \citealt{2020FORRESTb}, \citealt{2020VALENTINO}, \citealt{2021DEUGENIO}). The derived evolutionary path of these galaxies requires them to form the majority of their stellar content on short time-scales ($\ll 1$ Gyr) at a rate of hundreds to thousands $M_{\odot}\ \rm yr^{-1}$ (e.g., \citealt{2020VALENTINO}, \citealt{2020FORRESTb}). While the physical processes responsible for the quenching of massive galaxies are still debated (see, e.g.,  \citealt{2018MAN}, \citealt{2021HAYWARD}), the short time-scale involved at high redshift makes the quenching problem even more puzzling.

Galactic outflows, generated by feedback processes, either in the form of energy released by supernovae (SNe) or active galactic nuclei (AGN), likely play an important role. Their strength is usually parameterised through the mass loading factor
\begin{equation}
    \eta = \frac{\dot{M}_{\rm out}}{\rm SFR},
\end{equation}
where $\dot{M}_{\rm out}$ is the mass outflows rate. 

From a theoretical perspective, it is widely accepted that stellar feedback is the main driver of galactic outflows in low mass galaxies, while AGN feedback can power energetic winds in massive haloes ($M_{\rm vir}\gtrsim 10^{12}M_{\odot}$, e.g., \citealt{1974LARSON}, \citealt{1998SILK}, \citealt{2019NELSON}, \citealt{2020MITCHELL}). However, the physical conditions characterising high redshift DSFGs are very peculiar: they are very compact (effective radii of $\sim 2$ kpc, e.g., \citealt{2021PANTONI}), and have extremely high SFRs. Analytical works have shown that in starburst galaxies the SFRs can be high enough to expel all the gas from the galaxy (e.g., \citealt{2005MURRAY}, \citealt{2017HAYWARD}). Moreover, there is observational evidence of strong galactic outflows in compact massive starburst ($M_{\star}\sim 10^{11}M_{\odot}$), whose origin is likely related to stellar feedback rather than AGN (e.g., \citealt{2021PERROTTA}). A similar conclusion has been reached in the study of massive post-starburst galaxies ($\log M_{\star}\sim10.3-10.7$), where the time delay between the starburst phase and subsequent AGN activity suggests a secondary role of the latter process in the initial quenching of the galaxy (\citealt{2014YESUF}).

In the last two decades important efforts to characterise galactic outflows from the observational point of view have been made (see \citealt{2005VEILLEUX}, \citealt{2015ERB}, \citealt{2020VEILLEUX} for reviews). The methods commonly used to detect cold and warm outflowing gas rely on the detection of blue-shifted absorption features in the rest-frame ultraviolet (UV) and optical bands, with respect to the redshift of the host galaxy. This technique has been successfully used to measure the properties of galactic outflows both in the local Universe (e.g., \citealt{2014ARRIBAS}, \citealt{2015CHISHOLM}, \citealt{2016CICONE}) and at intermediate redshift, $z\gtrsim3-4$ (e.g., \citealt{2010STEIDEL}, \citealt{2014RUBIN}, \citealt{2015HECKMAN}, \citealt{2017TALIA}). 

At higher redshifts, methods based on absorption features become infeasible, as both the emission and the following absorption become weaker as the distance from the source increases. Therefore, observations rely on the detection of faint wings in the emission lines at far infrared (FIR) wavelengths. Instruments like the Atacama Large Millimeter/submillimeter Array (ALMA) and the NOrthern Extended Millimeter Array (NOEMA)
have enabled the detection of emission lines such as $[\rm CII]$ $158\mu m$, $[\rm OIII]$ $88\mu m$, and $\rm OH$ $119\mu m$ (e.g., \citealt{2018HASHIMOTO}, \citealt{2018CARNIANI}, \citealt{2019MATTHEE}, \citealt{2019FUJIMOTO}). The analysis of the profiles associated with these lines has then been used to probe the presence, mass, and velocities of cold and warm outflows in normal star forming galaxies (e.g., \citealt{2020GINOLFI}, \citealt{2018GALLERANI}, \citealt{2021HERRERA}), quasar hosts (e.g., \citealt{2012MAIOLINO}, \citealt{2015CICONE}, \citealt{2019BISCHETTI}), and also in DSFGs (\citealt{2020SPILKER2}) in the redshift range $4<z<7$, hinting that feedback mechanisms were already operating and expelling gas even at these early times. 

Despite this large effort, a complete characterisation of galactic outflows as a function of stellar mass and redshift is still lacking, especially at high redshift where studies of statistically relevant samples of galaxies are not feasible. A tool to complement observations and to theoretically study galactic outflows are hydrodynamical cosmological simulations. However, in order to self-consistently simulate the effect of feedback, a spatial numerical resolution of $\ll 1\ \rm kpc$ is needed, while state-of-the-art large cosmological box simulations can achieve a resolution of $\sim 100\ \rm pc$ at best (\citealt{2020VOGELSBERGER}). Therefore, to simulate galactic outflows, cosmological box simulations rely on sub-resolution models. These typically lean on free parameters that are tuned to reproduce a few observational constraints, such as the stellar mass function in the local Universe (see \citealt{2015SOMERVILLE} and \citealt{2017NAAB} for recent reviews). In the case of stellar feedback, the free parameters include the values of $\eta$ and the velocity of the wind, $v_{w}$. $v_w$ is commonly set to a characteristic velocity of the system, such as the galaxy circular velocity (e.g., \citealt{2016DAVE}, \citealt{2019DAVE}), or the one dimensional local dark matter (DM) velocity dispersion (e.g., \citealt{2018PILLEPICH}, \citealt{2018HENDEN}). The values of $\eta$ are either constant (e.g., \citealt{2003SPRINGEL}), or a decreasing function of either galaxy stellar mass (e.g., \citealt{2016DAVE}, \citealt{2019DAVE}) or metallicity (e.g., \citealt{2018PILLEPICH}). In the case of AGN feedback, the parameter space is even larger (e.g., \citealt{2022WELLONS}). While this enables the correct reproduction of average galactic properties, galactic outflows properties resulting from these models do not represent a prediction, as they are strongly dependent on the choice of the free parameters.

For the study of galactic outflows as driven by stellar feedback, a possible work-around is the use of zoom-in cosmological simulations or small cosmological volumes (e.g., \citealt{2021DUBOIS}, \citealt{2022FELDMANN}) which rely on more sophisticated feedback models. Modern zoom-in simulations can achieve a mass resolution of $\sim 1000\ M_{\odot}$, and a spatial resolution of $\sim 10 \rm \ pc$ for Milky-Way like galaxies evolved to $z=0$ (e.g., \citealt{2011GUEDES}, \citealt{2013STINSON}, \citealt{2014HOPKINS}, \citealt{2015AGERTZ}, \citealt{2016WETZEL}, \citealt{2018HOPKINS}, \citealt{2021APPLEBAUM}). These simulations can resolve the multi-phase structure of the ISM, and the formation of GMCs. However, the resolution is still too low to resolve the formation of single stars, and sub-resolution models still need to be implemented. For the feedback from SNe, which is thought to be the main source of energy and momentum for stellar feedback driven outflows (e.g., \citealt{1977MCKEE}, \citealt{1986DEKEL}, \citealt{1997SILK}), modern zoom-in simulations (e.g., \citealt{2014HOPKINS}, \citealt{2018HOPKINS}) rely on sub-resolution models that are based on high-resolution simulations of isolated SNe remnants (e.g., \citealt{2015MARTIZZI}). These simulations provide fitting formulas for the energy injected (both thermal and kinetic) as a function of the ISM density and metallicty at the scales resolved by the simulation (e.g., \citealt{2015MARTIZZI}, \citealt{2015KIM}). Therefore, even though the explosion of single SNe remains unresolved, the energy is injected locally and the resulting scaling of galactic outflows with galaxy properties are a prediction of the sub-resolution model. 

An example of such zoom-in simulations are the ones developed within the Feedback In Realistic
Environments project\footnote{https://fire.northwestern.edu/} (FIRE; \citealt{2014HOPKINS}, \citealt{2018HOPKINS}). FIRE is a well tested model that aims to improve the predictive power of galaxy formation simulations by reducing the reliance that the implemented sub-resolution models have on adjustable parameters. In different works, the FIRE model was shown to naturally reproduce observational constraints such as the integrated and resolved Kennicutt-Schmidt relation (\citealt{2014HOPKINS}, \citealt{2018ORR}), the Elmegreen-Silk relation (\citealt{2018ORR}), the SFR-$M_{\star}$ relation (\citealt{2017SPARRE}), and the mass-metallicity relation (\citealt{2016MA}) over a wide halo mass range ($10^9M_{\odot}\lesssim M_{\rm vir} \lesssim 2 \times 10^{12} M_{\odot}$). Finally, FIRE simulations reproduce galactic outflows over a wide range of scales, and their resulting scaling with halo/galactic properties are implemented as sub-resolution models in state-of-the-art cosmological simulations (e.g., \citealt{2019DAVE}).

In this paper we aim to constrain the capability of stellar feedback driven galactic outflows to (at least temporarily) quench high redshift starbursts. To this end, we extend previous works regarding the role of stellar feedback in driving galactic outflows from the FIRE collaboration (\citealt{2015MURATOV}, \citealt{2017ALCAZAR}, \citealt{2021PANDYA}) to massive systems (haloes masses $M_{\rm vir}\gtrsim 10^{12}\ M_{\odot}$) at high redshift ($z\gtrsim 6$). This choice ensures the selection of haloes that are likely to host DSFGs at high redshift, which observations suggest are galaxies at $z\gtrsim 5$ hosted in haloes with masses of few times $10^{11}M_{\odot}$ to few times $10^{12}M_{\odot}$ (e.g., \citealt{2018MARRONE}). A similar conclusion is also reached by exploiting abundance matching techniques, which find that at $z>4$ galaxies with SFR $\gtrsim 100\ M_{\odot}\ \rm yr^{-1}$ are hosted in haloes with $M_{\rm vir}\gtrsim10^{12}$ (e.g., \citealt{2015AVERSA}, \citealt{2016MANCUSO}). 

In this work, we focus on stellar feedback only. We exclude AGN feedback for two main reasons. Firstly, theoretical and observational evidence implies that stellar feedback could be sufficient to drive galactic outflows in starburst galaxies (e.g., \citealt{2005MURRAY}, \citealt{2014YESUF}, \citealt{2021PERROTTA}). Secondly, gas accretion onto SMBHs and AGN feedback are still poorly understood and sub-grid models for these processes rely on a large number of free parameters. These include the seeding of SMBHs particles, their pinning to the centre of the galaxy, SMBHs merger, gas accretion into SMBHs, and different energy injection channels into the ISM together with their numerical implementation. However, different models and models parameters are now under investigation in the framework of the FIRE-2 model (e.g., \citealt{2017ALCAZAR_BH}, \citealt{2022CATMABACAK}, \citealt{2021SU}, \citealt{2022WELLONS}), and therefore the role of AGN feedback in the evolution of high redshift massive galaxies will be investigated in future work.

This paper is structured as follows: in Sect.~\ref{sec:sim_set_up} we describe how the initial conditions for our simulations are created, and we summarise the main aspect of the sub-resolution models used to evolve them. Moreover, we also describe the methods used to compute quantities relevant for this work. In Sect.~\ref{sec:simulation_properties} we describe the main properties of our simulations, and we compare them with available observational data. In Sect.~\ref{sec:outflow_section} we quantify the mass loading factors for our simulations, while in Sect.~\ref{sec:SFE} we quantify the role of stellar feedback in self-regulating the SF in simulated galaxies. Finally, we summarise our main conclusions in Sect.~\ref{summary}.

\section{Methods}\label{sec:sim_set_up}
\subsection{Initial conditions}
In this work, we make use of three high-resolution cosmological zoom-in simulations which are part of the MassiveFIRE suite (\citealt{2016FELDMANN}, \citealt{2017FELDMANN}), and have already been presented in \cite{2022CATMABACAK}, from which we keep the same nomenclature (i.e., D9, D7, and D3). All three simulations were run with the FIRE-2 model, whose main features are described in Sect.~\ref{sec:subgrid_model}. 

The three galaxies analysed in this work were extracted from a dark-matter (DM) only cosmological box of $400\ h^{-1}$ comoving Mpc (cMpc) a side. From this box, three of the most massive haloes at $z=6$ ($M_{\rm vir}>10^{12}\ M_{\odot}$, see Table~\ref{tab:example_table}) have been selected. For these selected regions, the initial conditions have been created with MUlti-Scale Initial Conditions (MUSIC, \citealt{2011HAHN}), using a Lagrangian region of three times the virial radius of the haloes at $z=6$. Outside this high resolution region, the matter field is sampled with DM particles at lower resolution. While two of our simulations (D3 and D9) were evolved from $z=120$ to $z=5.5$\footnote{Since the final redshift is lower than the redshift at which the ICs have been created, we made sure that low resolution DM particles contributed less then $1\%$ of the total halo mass at $z=5.5$.}, D7 became too compact. This means that the particles with the highest densities were evolved at extremely small time-steps, such that it became infeasible to run it down to our target redshift. Therefore, the lowest redshift reached by this simulation is $z=7.2$. 

\begin{figure*}
	\includegraphics[width=1.0\linewidth]{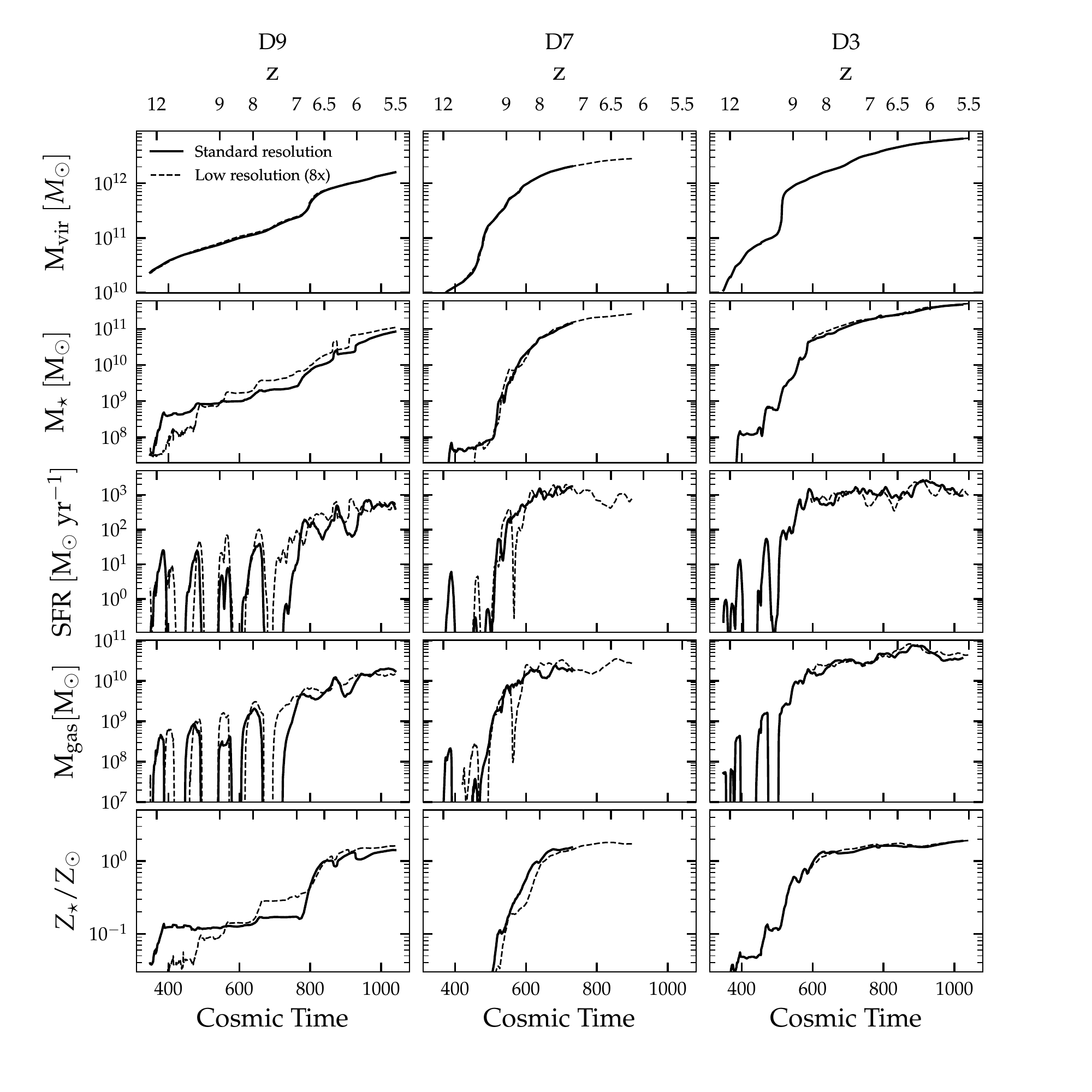}
    \caption{Halo mass, stellar mass, SFR$_{10\rm Myr}$, total gas mass, and stellar metallicity as a function of cosmic time for our three simulations. Each column refers to a different simulation (D9, D7, D3). Simulations are ordered by increasing halo mass. All quantities but the halo mass are computed within $0.1\ R_{\rm vir}$. While D3 and D9 are run down to $z=5.5$, D7 is run down to $z=7.2$ (see Sect.~\ref{sec:sim_set_up} for more details). Apart from our standard resolution, we also plot results obtained at a mass resolution 8 times lower (low resolution) as dashed lines. From this comparison we conclude that galaxy integrated properties are converged at the mass resolution used in this work. }
    \label{fig:general_properties}
\end{figure*}

At the resolution adopted for this work, the mass of the particles is $1.9\times 10^5 \ M_{\odot}$ and $3.6\times 10^4\ M_{\odot}$ for DM and gas respectively.  Stellar particles, which are spawned from gas particles, preserve their progenitor's mass. The gravitational softenings are kept fixed in comoving (physical) coordinates at $z\gtrsim 9$ ($z\lesssim 9$) and are equal to $65$ pc for DM particles, $6.3$ pc for star particles, while the minimum softening length for gas particles is $0.8$ pc\footnote{In GIZMO the gravitational softening for gas particles is fully adaptive, meaning that the gravitational softening is set equal to the smoothing length.}. To show that results are numerically converged, in Fig.~\ref{fig:general_properties} we also show results from the same set of initial conditions run at 8 times lower mass resolution (2 times lower spatial resolution). We will refer to this set of simulations as low resolution (LR).

The cosmological parameters chosen to run the simulations are: $\Omega_{\rm M}=0.3089$, $\Omega_{\Lambda}=0.6911$, $\Omega_{\rm b}=0.0486$, $H_0=67.74 \rm \  km\ s^{-1}\ Mpc^{-1}$, $\sigma_8= 0.8159$, $n_s=0.9667$ (consistent with \citealt{2020PLANCK}). 

\subsection{The FIRE-2 physics model}\label{sec:subgrid_model}

The simulations are run with the cosmological code GIZMO\footnote{\url{http://www.tapir.caltech.edu/~phopkins/Site/GIZMO.html}}, using the meshless finite-mass (MFM, \citealt{2015HOPKINS}) scheme for hydrodynamics. This method conserves the natural adaptivity of Lagrangian schemes like Smoothed Particle Hydrodynamics (SPH), essential in cosmological simulations given the large dynamical range covered, while avoiding the introduction of artificial dissipative terms thanks to the implementation of a Riemann solver. 

The physics related to gas cooling, star formation, chemical evolution, and stellar feedback is treated using the FIRE-2 model for galaxy evolution, which is described in \cite{2018HOPKINS}. Here we only highlight the main features, while we refer to the original paper for a more detailed description. 

Radiative heating and cooling combine different physical processes which include free–free,
photo-ionization/recombination, Compton, photoelectric, metal-line, molecular and fine-structure processes calculated in the temperature range $10-10^{10}\,$K and for all the 11 tracked chemical species (H, He, C, N, O, Ne, Mg, Si, S, Ca, Fe). The model also accounts for photo-heating both by a UV background \citep{2009FAUCHER} and local sources, as well as self-shielding.

Star formation occurs only for gas particles that exceed a density threshold of $n_{\rm min}=1000\,{\rm cm^{-3}}$, are self-gravitating, molecular and self-shielding (following \citealt{2011KRUMHOLZ}), and are Jeans unstable. If all these criteria are fulfilled, star formation occurs at a rate $\dot{\rho}_{\star}=f_{\rm mol}\rho / t_{\rm ff}$, where $f_{\rm mol}$ is the molecular gas fraction, $\rho$ is the gas density, and $t_{\rm ff}$ is the local free-fall time. Therefore, in this model star formation happens locally with a $100$ per cent efficiency per free-fall time. As has been shown several times before, the average few per cent efficiency observed on galactic scales arises as a consequence of stellar feedback (e.g., \citealt{2014HOPKINS}, \citealt{2018ORR}). 

At the formation stage, each star particle is considered as a single stellar population with a given mass, age, and metallicity. Given these properties, all feedback event rates, luminosities and energies, mass-loss rates, and all other quantities are tabulated directly from stellar evolution models ({\sc starburst99}; \citealt{1999LEITHERE}), assuming a \citet{2001KROUPA} initial mass function. In particular, we explicitly model the following stellar feedback mechanisms:
(1) local and long-range momentum flux from radiation pressure (in the initial UV/optical single-scattering, and re-radiated light in the IR), (2) energy, momentum, mass, and metal injection from supernovae (SNe, Types Ia and II) and stellar winds (both from OB and AGB), and (3) photo-ionization and photo-electric heating. 

\subsection{Derived physical properties}\label{sec:methods}

In this subsection, we describe how all the quantities that will be used throughout this work are computed.

We identify dark matter haloes with the publicly available Amiga Halo Finder\footnote{\url{http://popia.ft.uam.es/AHF/Download.html}} (AHF, \citealt{2004GILL}, \citealt{2009AKNOLLMANN}). AHF provides the position, velocity, virial mass ($M_{\rm vir}$), and virial radius ($R_{\rm vir}$, following the virial overdensity definition of \citealt{1998BRYAN}) of each indentified halo with $\gtrsim 100$ particles. The merger tree implemented in AHF allow us to link the haloes identified at the lowest redshift reached by each simulation to their main progenitors at higher redshift.

SFRs are calculated as:
\begin{equation}\label{eq:SFR}
    \text{SFR}_{\Delta T} = \frac{\sum_{i} m_{\star,i}}{\Delta T},
\end{equation}
where $m_{\star,i}$ is the mass of star particles at the time of their formation, $\Delta T$ is the time interval over which the SFR is averaged, and the sum is extended to all star particles younger than $\Delta T$. This approach is useful since results from simulations will be compared to observations in the FIR wavelengths, which are sensitive to stars with ages up to $100\ \rm Myrs$ (e.g., \citealt{2012KENNICUTT}) in actively star forming galaxies (\citealt{2014HAYWARD}). If not stated differently, the star formation rate is computed within $0.1\ \rm R_{\rm vir}$.

The mass loading factor is:
\begin{equation}\label{eq:eta}
    \eta = \frac{\dot{M}_{\rm out}}{\text{SFR}_{\Delta T}},
\end{equation}
where $\dot{M}_{\rm out}$ is the mass outflow rate, and the adopted value of $\Delta T$ will be explicitly defined in the section of interest. In order to quantify the mass outflow rate, we follow the definition of \cite{2015MURATOV}. Specifically, we define outflowing (inflowing) gas particles as particles with a positive (negative) radial velocity. With this definition, the mass outflow rate through a spherical shell of thickness $\rm \Delta L$ will be:
\begin{equation}
    \dot{M}_{\rm out} = \frac{\sum_i m_i v_{r,i}}{\rm \Delta L},
    \label{eq:out_def}
\end{equation}
where $m_i$ is the gas particle mass, $v_{r,i}$ is the radial velocity of the particle, and the sum is extended to all gas particles with positive radial velocity within the shell. Similarly, we define the total inflow rate by selecting only gas particles with a negative radial velocity. Following previous works (\citealt{2015MURATOV}, \citealt{2021PANDYA}), we consider spherical shells of thickness $\rm \Delta L = 0.1 \ \rm R_{\rm vir}$. While the innermost shell, $[0-0.1]\times \rm R_{\rm vir}$, traces the ISM, the outer shells are considered to be CGM halo gas. If not stated differently, we will adopt the spherical shell $[0.2-0.3]\times \rm R_{\rm vir}$ for the CGM, which enables a direct comparison with previous theoretical works. In this paper, we will refer to a particular spherical shell by its radial mid-point. For example, we will refer to the shell $[0.2-0.3]\times \rm R_{\rm vir}$ as $0.25\ \rm R_{\rm vir}$.

Finally, in Sect.~\ref{sec:observations} we compare simulation results with available observations. Since most of the information about gas content in high redshift galaxies comes from observations of CO emission lines, which is a tracer of molecular gas, we need to compute the molecular gas mass in our simulations. In FIRE-2 this information is computed at run-time, but is not stored as an output. Therefore, we compute the molecular gas fraction $f_{\rm mol}$ of each gas particle in post-processing following \cite{2011KRUMHOLZ} (see Appendix C of \citealt{2018HOPKINS} for more details). The total molecular gas mass within a sphere of radius R is
\begin{equation}
    M_{H_2} = \sum_i f_{\rm mol, i} \times m_i,
\end{equation}
the sum being extended to all gas particles within R. 

\section{Simulation properties}\label{sec:simulation_properties}

\subsection{Main properties of the simulated galaxies}

\begin{table*}
	\centering
	\caption{Properties of the simulations at different redshifts. Specifically: virial mass ($M_{\rm vir}$), virial radius ($R_{\rm vir}$), stellar mass within $0.1 R_{\rm vir}$ ($M_{\star}$), and final redshift reached by the simulation ($z_f$).}
	\label{tab:example_table}
	\begin{tabular}{lcccccccccc} % four columns, alignment for each
		\hline
		Name & $M_{\rm vir}\ [M_{\odot}]$ & $M_{\rm vir}\ [M_{\odot}]$ & $M_{\rm vir}\ [M_{\odot}]$ & $R_{\rm vir}\ [\rm kpc]$ & $R_{\rm vir}\ [\rm kpc]$ & $R_{\rm vir}\ [\rm kpc]$ & $M_{\star}\ [M_{\odot}]$ & $M_{\star}\ [M_{\odot}]$ & $M_{\star}\ [M_{\odot}]$ & $z_f$ \\
		& $(z=8)$ & $(z=7)$ & $(z=6)$ & $(z=8)$ & $(z=7)$ & $(z=6)$ & $(z=8)$ & $(z=7)$ & $(z=6)$\\
		\hline
		D9 & 1.1e11 & 2.4e11 & 1.1e12 & 17.5 & 25.1 & 47.2 & 1.4e9 & 2.6e9 & 3.7e10 & 5.5\\
		D7 & 1.3e12 & 2.1e12 & - & 39.6 & 49.9 & - & 5.7e10 & 1.4e11 & - & 7.2\\
		D3 & 1.7e12 & 3.4e12 & 5.8e12 & 42.5 & 60.7 & 82.7 & 7.1e10 & 1.8e11 & 3.8e11 & 5.5\\
		\hline
	\end{tabular}
\end{table*}

\begin{table}
	\centering
	\caption{Stellar half-mass radii, half-SFR radii, and molecular gas half-mass radii. Each value represents the radius that contains half of that quantity within $0.1\ R_{\rm vir}$. The values reported refer to the last snapshot analysed, whose corresponding redshift is reported in the table.}
	\label{tab:radii_table}
	\begin{tabular}{lccccccccc} % four columns, alignment for each
		\hline
		Name & $R_{\star, 1/2}$ & $R_{\rm SFR, 1/2}$ & $R_{\rm gas, 1/2}$ & $R_{\rm H_2, 1/2}$ & $z$\\
		  & [pkpc] & [pkpc] & [pkpc] & [pkpc] \\
		\hline
		D9 & 0.32 & 1.82 & 3.45 & 2.37 & 5.5\\
		D7 & 0.15 & 0.02 & 3.08 & 1.53 & 7.2\\
	    D3 & 0.95 & 0.78 & 4.96 &  1.13 & 5.5\\
		\hline
	\end{tabular}
\end{table}

In Fig.~\ref{fig:general_properties} we show the main properties of our three galaxies as a function of cosmic time, with each simulation named following the nomenclature of \cite{2022CATMABACAK}. All galaxies are identified at the lowest redshift available (see Table~\ref{tab:example_table}) and are then followed back in time using the halo merger tree. As explained in Sect.~\ref{sec:sim_set_up}, D7 run at our standard resolution only reached $z=7.2$. Because the main halo in this simulation becomes extremely dense (see for example the stellar half-mass radius in Table~\ref{tab:radii_table}), it makes it too numerically costly to run it further.

The first row in Fig.~\ref{fig:general_properties} shows the evolution of the halo mass with time. This highlights both the high mass of these haloes, which reach values $\gtrsim 10^{12}M_{\odot}$ within the first Gyr of cosmic history, and the different evolutionary paths of the three selected systems. While D7 experiences a relatively smooth, but very rapid, accretion history, both D9 and D3 show the evidence of a merger at $t\sim 800\ \rm Myrs$ ($z\sim7$) and $t\sim 500\ \rm Myrs$ ($z\sim 9$) respectively. In the case of D3, this interaction is a simultaneous merger between 4 different systems (three of them having similar stellar masses, while the fourth being half as massive), whose presence is highlighted by the rapid increase of $M_{\rm vir}$ at $t\sim 500\ \rm Myrs$.

After the mergers, the haloes grow in mass by a factor of a few (D9) or even by an order of magnitude (D3), impacting all the other quantities that are shown. Of particular interest is the transition in the SFH of D9 before and after the merger, going from a low ($\sim 10^9\ M_{\odot}$) to a high ($\gtrsim 10^{10}\ M_{\odot}$) stellar mass. At $M_{\star} \lesssim 10^{10}\ M_{\odot}$ the SFH of this galaxy is extremely bursty, mirroring results of simulations at similar stellar masses but at lower redshift (e.g, \citealt{2015MURATOV}). At these stellar masses, stellar feedback is efficient in removing gas from the central region, halting both the SF and the gas inflows (the latter effect will be discussed more in depth in Sect.~\ref{sec:outflows}). On the other hand, at $M_{\star} \gtrsim 10^{10}\ M_{\odot}$, star formation is much more steady, as stellar feedback becomes quickly inefficient both in driving galactic outflows (see Sect.~\ref{sec:outflow_section}) and in regulating star formation (see Sect.~\ref{sec:SFE}), enabling all simulated galaxies to reach stellar masses of $ M_{\star} \gtrsim 10^{11}\ M_{\odot}$ in a few hundred Myrs. The transition from bursty to more steady star formation in FIRE simulations, including the connection to galactic outflows, has been studied in a number of recent papers (e.g., \citealt{2021STERN}, \citealt{2022GURVICH}). In the next two sections, we will quantitatively study the impact of stellar feedback in these two regimes, relating its efficiency to the physical properties of simulated galaxies. 

Interestingly, in the high stellar mass regime all three systems show very high SFRs, roughly in agreement with the values that are measured for high redshift ($z\gtrsim 4$) DSFGs (e.g., \citealt{2016ARAVENA}, \citealt{2018MARRONE}, \citealt{2020RIECHERS}). The similarity with high redshift DSFGs will be further explored in Sect.~\ref{sec:observations}. 

Finally, the last row of Fig.~\ref{fig:general_properties} shows the evolution of the average metallicity of star particles within the galaxy, computed by summing up all the mass in metals locked in stars within $0.1\times R_{\rm vir}$ and normalising by the total stellar mass in the same region. We see from the plot that even at this very high redshift ($z \gtrsim 6$), the stellar population is predicted to be at solar (or even super-solar) metallicity\footnote{We assume a solar metallicity $Z_{\odot}=0.0153$ (\citealt{2011CAFFAU})}. Because of the inefficient outflows in the high mass regime, the metals produced during stellar evolution are not launched at large distances but instead trapped in the star-forming region and rapidly recycled in newly born stars. This prediction regarding the metal enrinchment in high$-z$ massive galaxies can be tested directly through \textit{James Webb Space Telescope} (JWST\footnote{\url{https://jwst.nasa.gov/content/webbLaunch/index.html}}) observations.

\subsection{Comparison with observational data}\label{sec:observations}

As described in Sect.~\ref{sec:sim_set_up}, the three simulations analysed in this work have been selected from the most massive haloes in a cosmological volume of $400h^{-1}\ \rm cMpc$ at $z=6$. Given their high masses, these haloes are likely hosts of high redshift DSFGs. In this Subsection we study how closely the properties of simulated galaxies resemble the properties of high$-z$ DSFGs.

Since In DSFGs most of the emission comes from dust reprocessed UV light re-emitted at FIR and sub-millimeter wavelengths, the information that can be indirectly retrieved from scaling relations include the SFR, molecular gas mass, and dynamical mass (through the  full
width at half maximum, FWHM, of CO emission lines and galaxy sizes, see below), while information like galaxy stellar mass or metallicity, which are generally obtained from the SED at UV-NIR wavelengths, are not available for a significant fraction of the observed DSFGs. Therefore, in this section we compare simulation results with observational data in the physical space currently accessible. 

\begin{figure}
	\includegraphics[width=\columnwidth]{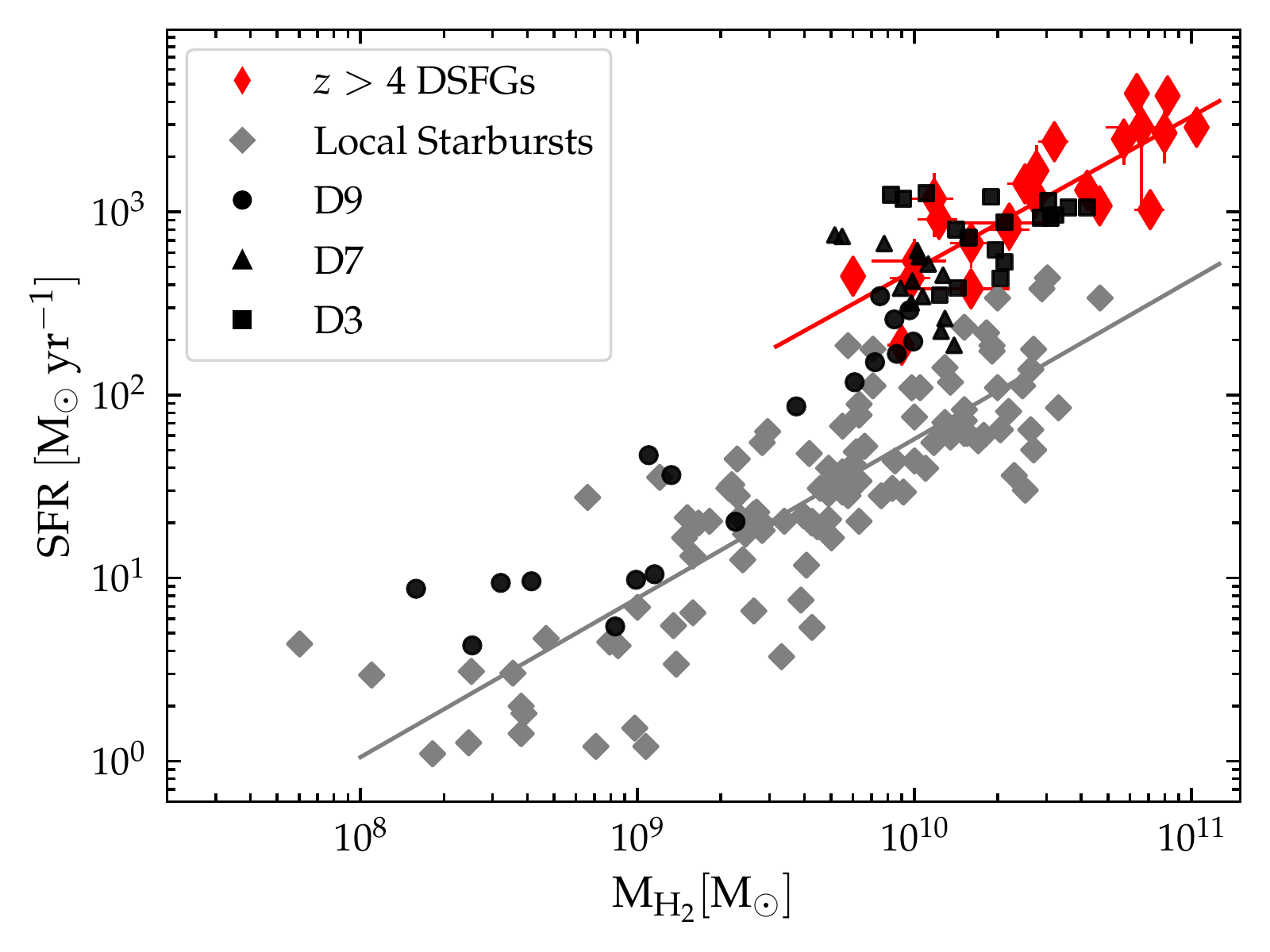}
    \caption{SFR as a function of molecular gas mass for observed and simulated galaxies. Red diamonds show a sample of DSFGs from the literature (\citealt{2016ARAVENA}, \citealt{2013RIECHERS}, \citealt{2018ZAVALA}, \citealt{2018MARRONE}, \citealt{2020SPILKER1}, \citealt{2020RIECHERS}). Grey symbols show a compilation of local starburst galaxies (\citealt{2021KENNICUTT}). Black points show simulations results: D9 (circle), D7 (triangles), and D3 (squares). For clarity, we only show simulations results at time intervals of $25\rm \ Myrs$ from a cosmic time of $600$ Myrs ($z \sim 8$) to the end of the simulation. Results of simulations are in agreement with high redshift DSFGs, and have significantly lower depletion times ($t_{\rm dep}=\rm M_{\rm H_2}/\rm SFR$) than local starbursts. }
    \label{fig:depletion_time}
\end{figure}

In Fig.~\ref{fig:depletion_time} we show the correlation between the molecular gas mass and the SFR in our simulations comparing to a sample of DSFG in the redshift range $4\lesssim z \lesssim 7$ from the literature (\citealt{2013RIECHERS}, \citealt{2016ARAVENA}, \citealt{2018ZAVALA}, \citealt{2018MARRONE}, \citealt{2020SPILKER1}, \citealt{2020RIECHERS}), and to a sample of local starbursts (\citealt{2021KENNICUTT}). In observations, the SFR is derived from the FIR luminosity in the wavelength range $8-1000\mu m$, while the molecular gas mass is estimated from the total CO luminosity assuming a conversion factor of $\alpha_{\rm CO} \sim 1 / \rm\ K\ Km\ s^{-1} pc^2$ (i.e. the value found for local compact starbursts, see \citealt{1998DOWNES}). 

For the simulations, we compute the molecular gas masses following the procedure described in Sect.~\ref{sec:methods} and within a 3D spherical volume with $R=20$ physical kpc (pkpc). This aperture roughly matches the beam size of CO observations in both \cite{2016ARAVENA} and \cite{2020SPILKER1} obtained through the Australia Telescope
Compact Array (ATCA), and comprise 13 of the 22 DSFGs included in the comparison. The SFR is computed as the average SFR over $100\ \rm Myr$. This approximately matches the time scales probed by IR emission (e.g, \citealt{2012KENNICUTT}). Indeed, in Liang et al. (in Prep). we show that our simulations follow the \cite{1998KENNICUTT} SFR$-L_{\rm IR}$ relation, the same relation which is also used to derive the SFR in observations. For consistency, we derive the SFR within the same aperture as used for the molecular gas mass, although we emphasise that the actual value of the SFR only weakly depends on the particular aperture chosen, since most of the star formation takes place in a much smaller region ($\lesssim 1$ kpc, see Table~\ref{tab:radii_table}). For clarity, we show galaxy properties at intervals of $25\rm \ Myrs$ starting from $z<8$. This redshift range cuts out most of the period of time during which D9 has a stellar mass lower than $M_{\star}\sim 10^{10}M_{\odot}$, and therefore wouldn't be suitable for a comparison with the sample of high redshift DSFGs.

Fig.~\ref{fig:depletion_time} shows that simulated galaxies have molecular gas depletion times, defined as $t_{\rm dep} = \rm M_{\rm H_2} / \rm SFR$, consistent with those measured for high redshift DSFGs. The only exception is D9, which is characterised by depletion times that are mostly in between what is measured for $z>4$ DSFGs and local starburst. This is because simulated galaxies have short depletion times ($t_{\rm dep}\sim20$ Myr) only when they reach stellar masses $M_{\star} \sim 10^{11}\ \rm M_{\odot}$, a condition that for D9 is met only in the latest stages of its evolution. A fit to the observational data yields $\rm SFR\propto M_{\rm H_2}^{0.84\pm 0.14}$, corresponding to a depletion time of $20\rm \ Myr$ for $\rm M_{H_2}=10^{10}\ M_{\odot}$. Given these very short depletion time scales, without replenishment from the large-scale environment, star formation in these galaxies should cease very rapidly. Instead, in Fig.~\ref{fig:general_properties} we see that simulated galaxies sustain star formation at a rate of $\sim 1000M_{\odot}\rm yr^{-1}$ for several hundred Myrs. This suggests that the SFR is not regulated by internal processes, i.e. by the efficiency at which the available reservoir of gas is transformed into stars, but by the amount of gas available for star formation set by the cosmological inflow rate, which is expected to be very high at such high redshift (e.g, \citealt{2009DEKEL}, \citealt{2011CLAUDEANDRE}). 

\begin{figure}
    \centering
    \includegraphics[width=\columnwidth]{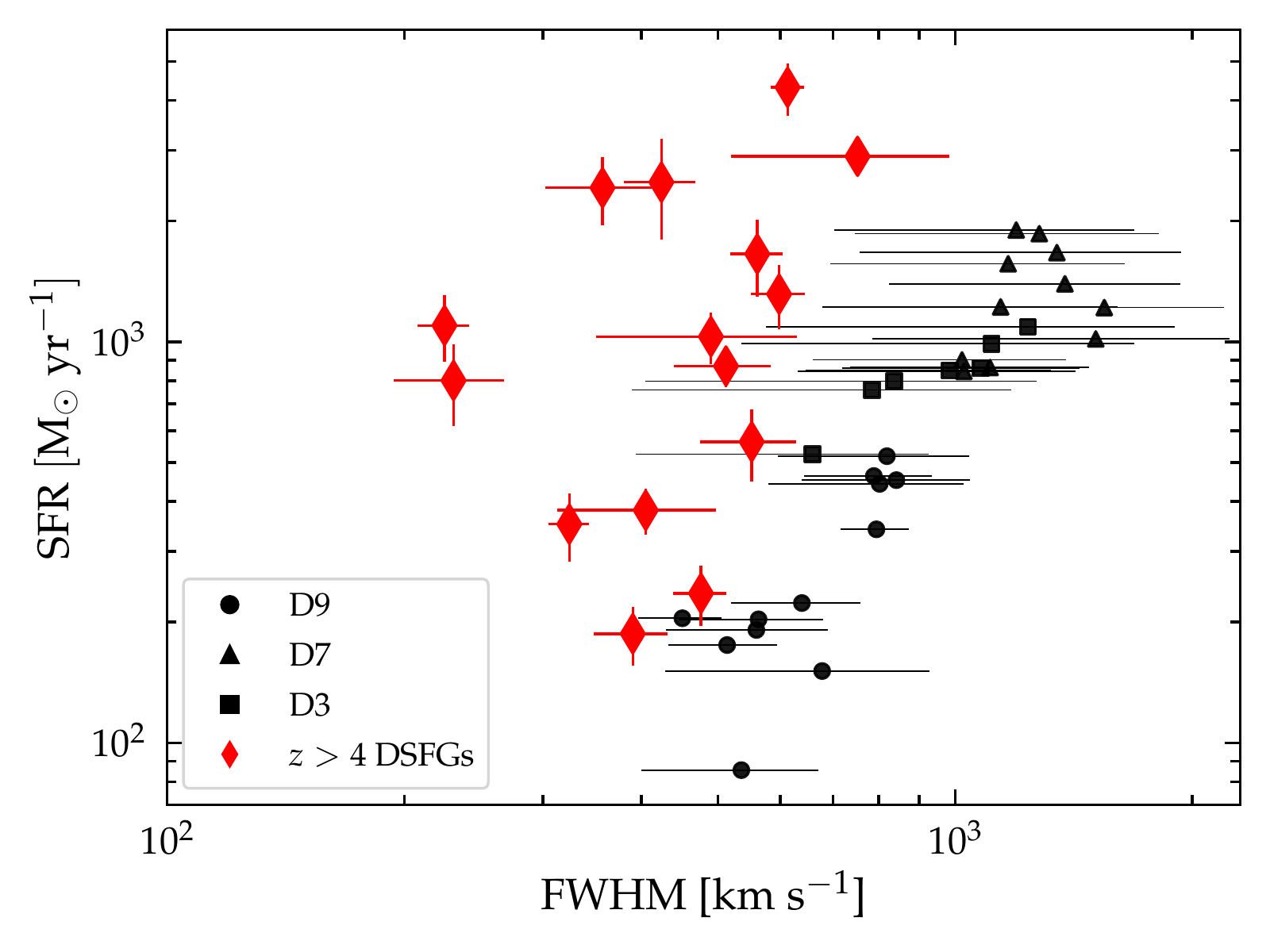}
    \caption{SFR as a function of CO emission lines full width at half maximum (FWHM; proxy for circular velocity) in observations and simulations. For the simulations we compute the FWHM from the 1D velocity distribution of molecular gas (eq.~\ref{eq:FWHM}). Error bars show the $2\sigma$ interval derived considering 1000 random lines of sights. Observations, shown as black diamonds, are the sub-sample of the systems presented in Fig.~\ref{fig:depletion_time} for which FWHM measurements are available. Simulations tend to have larger values of the FWHM, suggesting that they are either more compact, more massive, or both with respect to high redshift DSFGs.}
    \label{fig:fwhm}
\end{figure}

Under the assumption that the galaxy is virialized, the circular velocity, $v_c \equiv \sqrt{GM/R}$, can be derived from the CO emission line, as $M_{\rm dyn} \propto R \sigma^2$, where $R$ is the size of the galaxy and $\sigma$ is the 1D velocity dispersion traced by CO emission (e.g., \citealt{2013BOTHWELL}). Therefore, in Fig.~\ref{fig:fwhm} we show the correlation between the CO line FWHM\footnote{Assuming a Gaussian profile, the relation between the FWHM and $\sigma$ is FWHM = $2.35 \sigma$.} and the SFR. In observations, the FWHM can be directly computed by fitting a Gaussian profile to the CO emission lines. In simulations, this would require to model emission from CO molecules, which would introduce large uncertainties in the analysis. Therefore, instead of modelling CO emission lines we use the molecular gas mass, assuming a linear relationship between CO luminosity and $\rm M_{\rm H_2}$. In order to derive the FWHM, we select all gas particles within a cylindrical volume centred on the galactic centre with a radius of $20\rm \ pkpc$ and height equal to $2\times R_{\rm vir}$. The radius is chosen to roughly match the beam size of ATCA observations, which are used to detect CO emission lines at low rotational transitions. Finally, following \cite{2013BOTHWELL}, from the line of sight velocity distribution we compute the FWHM as:
\begin{equation}\label{eq:FWHM}
    \text{FWHM}=2.35\sigma=2.35\ \frac{\int M_{H_2,v}(v-\bar{v})^2dv}{\int (v-\bar{v})^2dv},
\end{equation}
where $M_{H_2},v$ is the molecular gas mass in each velocity bin, $v$ is the line of sight velocity, and $\bar{v}$ is the average mass-weighted line of sight velocity. This formula is exact in the case of a Gaussian profile and gives more accurate results when the profile is double-peaked (as is the case for a rotating disk), where a Gaussian fit would overestimate the real FWHM of the distribution. Finally, to take the variance related to the choice of the particular line of sight into account, we consider $1000$ random line of sights for each snapshot, and show the average value with the $2\sigma$ interval in black. 

Simulated galaxies are characterised by SFRs similar to those measured in high$-z$ DSFGs. However, the line-of-sight CO emission $\rm FWHM$ of the simulated galaxies are typically much larger than the observed galaxies, with values as large as $1500\ \rm km\  s^{-1}$ (although we also note that simulated galaxies agree with some of the observed galaxies within uncertainties). This difference suggests that, compared to observations, simulated galaxies are either too massive or too compact, or both. Indeed, \cite{2021PARSOTAN} already showed that FIRE-2 galaxies hosted in haloes of $10^{12.5}M_{\odot}$ at $z=2$ are too compact with respect to observations, suggesting that stellar feedback is not efficient enough and another source of energy like AGN feedback might be needed in this halo mass range (see also \citealt{2020WELLONS}). At $z>4$, \citealt{2016ARAVENA} measure CO and dust sizes for mm-bright ($S_{1.4}>15$ mJy) galaxies selected with the South Pole Telescope (SPT, \citealt{2011CARLSTROM}) in the range $0.5-2.5$ kpc. These values are similar to the molecular gas half-mass and half-SFR radii of D9 and D3, while D7 forms most of the stars in a much more compact region (see Tab.~\ref{tab:radii_table}). Stellar masses are known only for three galaxies in the $z>4$ DSFGs shown in Fig.~\ref{fig:fwhm}, and are in the range $[3.5-11.9]\times 10^{10}M_{\odot}$ (\citealt{2013RIECHERS}, \citealt{2018MARRONE}, \citealt{2020RIECHERS}). Since simulated galaxies form stars at a rate $\gtrsim 10^3\rm \ M_{\odot}\ yr^{-1}$ only at $M_{\star}>10^{11}M_{\odot}$, this suggests that simulated systems grow over-massive already at $z>6$. However, due to the lack of a complete sample of observed massive galaxies at such high redshifts, it is not possible to confirm that our FIRE-2 simulated galaxies are too compact and / or too massive with respect to real galaxies. Future observations with JWST will help to understand whether another source of energy (e.g., AGN feedback) is already needed at $z>6$ in this mass regime.

\section{Outflows}\label{sec:outflow_section}

In the previous section we have shown that our simulated galaxies have gas-related properties consistent with those observed for high-redshift DSFGs. In this section, we investigate whether stellar feedback driven outflows could result in quiescence following a DSFG phase. Specifically, we link the efficiency of stellar feedback in driving galactic outflows to the physical properties of simulated galaxies. Moreover, we compare the properties of our systems to previous results from the FIRE collaboration, enabling us to quantify the differences (if any) between our simulations and results obtained for systems with similar halo masses but at much lower redshift. 

\begin{figure*}
	\includegraphics[width=\linewidth]{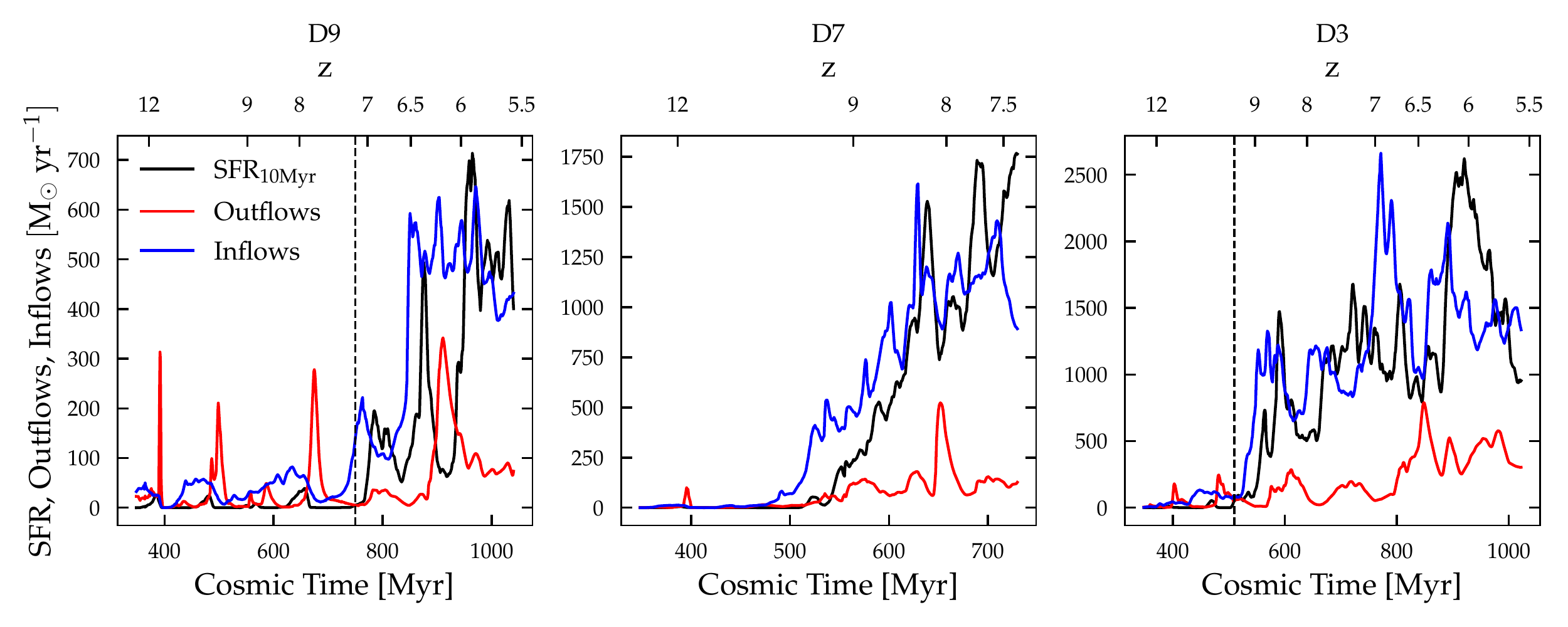}
    \caption{SFR$_{10\rm Myr}$, outflows, and inflows (black, red, and blue respectively) as a function of time for our three simulations. From left to right: D9, D7, and D3. SFR$_{10\rm Myr}$ is computed considering all star particles within $0.1\ \rm R_{\rm vir}$, while outflows and inflows are computed at $0.25\ \rm R_{\rm vir}$. The black dashed vertical lines for D9 and D3 show the time at which major merger events happened and differentiate between two regimes: at early times, the stellar mass is small and stellar feedback is efficient in driving galactic outflows; at later times, stellar mass grows and stellar feedback becomes rapidly inefficient in driving galactic outflows. }
    \label{fig:outflow_profiles}
\end{figure*}

\subsection{Outflow profiles }\label{sec:outflows}

In Fig.~\ref{fig:outflow_profiles} we show the time evolution of the SFR and of the inflow and outflow rates for our three simulations. In this analysis we compute the star formation rate by considering star particles within $0.1\times R_{\rm vir}$ and averaging over $10\ \rm Myrs$. Outflows and inflows are computed at $0.25\ \rm R_{\rm vir}$ following equation~(\ref{eq:out_def}). 

The left panel shows the results for D9, the least massive galaxy of our sample, whose evolution can be separated into two different regimes. At early times ($t<750\rm \ Myr$), when both the halo and stellar masses are relatively small ($M_{\rm vir}\lesssim 10^{11}M_{\odot}$, $M_{\star}\lesssim 10^{10}M_{\odot}$), stellar feedback is efficient in removing gas from the galaxy and therefore regulating the SFR. The connections among these three quantities are clearly visible: gas inflows (blue) are followed by bursts of star formation (black), which are readily shut down by powerful outflows ($\sim 1$ order of magnitude larger than the SFR, red). Interestingly, these outflows propagate into the CGM, suppressing the inflow rate. The same episodic (or bursty) behaviour has been found in simulations with similar halo masses at lower redshift (e.g., \citealt{2015MURATOV}). However, at $t\sim 750\rm \ Myrs$, indicated by the vertical dashed black line, as a consequence of the rapid mass growth due to a major merger, this behavior changes significantly. In this mass regime ($M_{\star}\gtrsim 10^{10} M_{\odot}$), the outflow rates do not follow the rapid increase in inflows and SFR, and stellar feedback becomes rapidly inefficient in both preventing gas inflows and regulating the SFR within the galaxy. The same behaviour is seen for D3, the most massive galaxy in our sample (right panel of Fig.~\ref{fig:outflow_profiles}). In this case there also is a clear transition between a feedback regulated phase before the merger at $t\sim 500$ Myr (black dashed vertical line), to a phase of rapid and uncontrolled star formation at later times. D7 (central panel) experiences a phases of rapid accretion between $500 \rm \ Myr<t<600 \rm \ Myr$ during which the stellar mass increases by three orders of magnitudes (see Fig.~\ref{fig:general_properties}). Once $t\gtrsim600$ Myr the outflows rates are always an order of magnitude lower than both inflow rates and SFR.

Moreover, it is important to note that the few outflow events that are visible in the high-mass regime of all three simulations (e.g., at $\sim 900\ \rm Myrs$ for D9, at $\sim 650\rm \ Myrs$ for D7, and at $\sim 850\ \rm Myrs$ for D3), are not powered by stellar feedback, but are created by close interaction with smaller structures or by strong inflow events. Indeed, for all these cases we visually checked that the increase in SFR and inflow rates corresponds to an object infalling toward the main halo (or to a large amount of gas inflowing toward the centre of the main galaxy), and the later outflow event corresponds to the same structure (or gas) now moving with a positive radial velocity with respect to the main halo.  

\begin{figure}
	\includegraphics[width=\columnwidth]{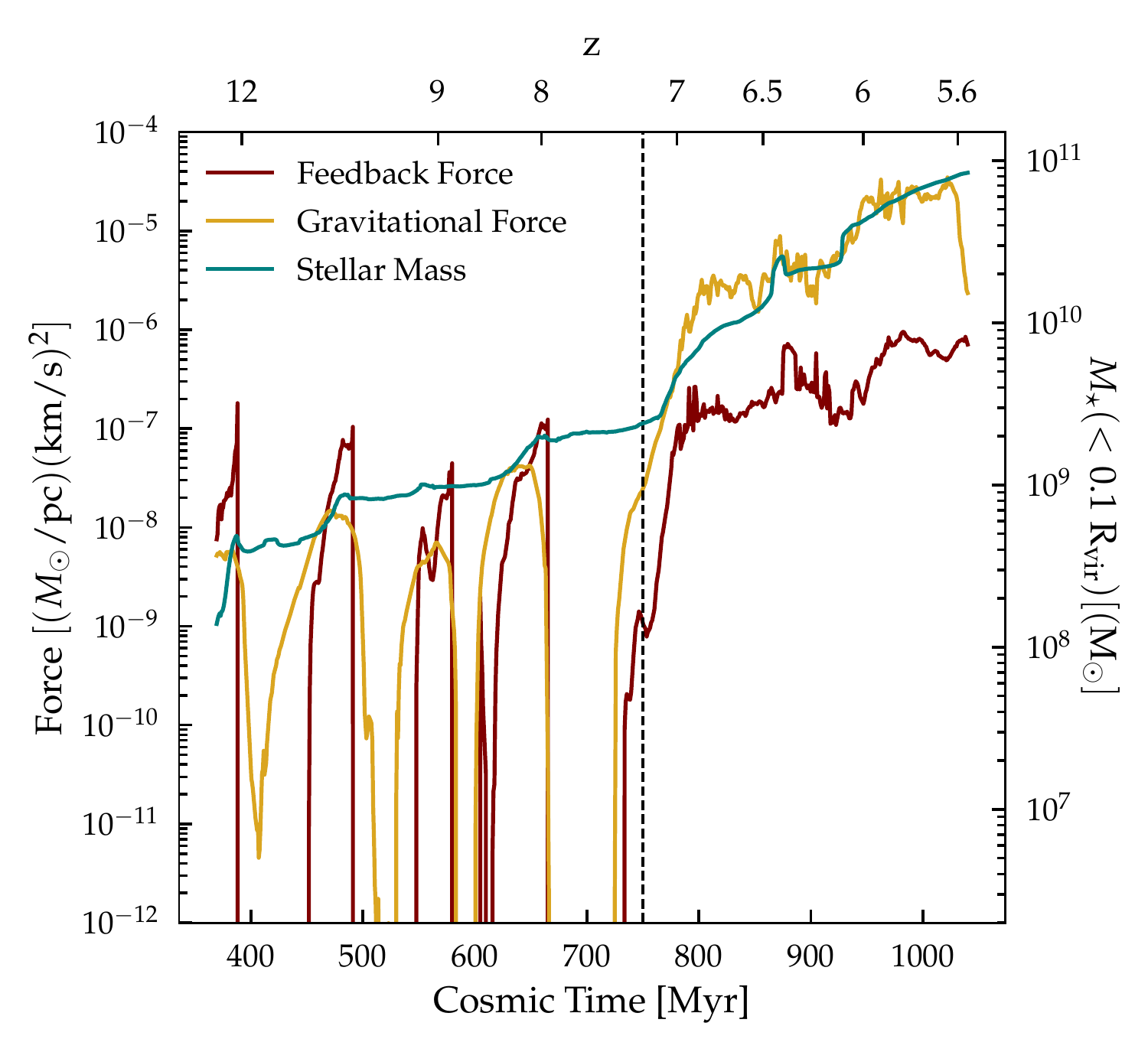}
    \caption{Feedback (maroon) and gravitational (gold) forces as a function of cosmic time for D9. The teal line shows the evolution of the stellar mass within $0.1 R_{\rm vir}$. The gravitational force is the total gravitational force exercised on the gas within $0.1\rm R_{\rm vir}$ (Eq.~\ref{eq:gravitational_force}), while the feedback force is the momentum rate injected into the gas from core-collapse SNe within the same region (Eq.~\ref{eq:feedback_force}). Two regimes, divided by the black vertical line, are clearly distinguishable: at $\lesssim 750$ Myr feedback generates enough momentum to expel the gas from the halo centre. At $t\gtrsim 750$ Myr, the stellar mass grows by an order of magnitude, and stellar feedback is no longer strong enough to drive significant gas outflows.}
    \label{fig:grav_vs_feedback}
\end{figure}

\subsection{Gravitational and feedback forces}\label{sec:grav_feed_forces}

The different regimes that we see in Fig.~\ref{fig:outflow_profiles} (efficient outflows at $M_{\star}\lesssim10^{10}M_{\odot}$ and inefficient outflows at higher masses) can be understood in terms of galaxy mass and size. Indeed, the more massive and/or compact the galaxy is, the deeper will be the potential well, thus requiring more energy to drive outflows. To show this analytically, following \cite{2005MURRAY} we first approximate the gas within a galaxy as a thin shell at distance R from a point source injecting momentum into the gas at a rate $\dot{P}_{\rm fb}$. The equation of motion for the gas shell will be:
\begin{equation}
    \frac{dP}{dt} = -G\frac{M(<R)M_{\rm gas}}{R^2}+\dot{P}_{\rm fb},
\end{equation}
where $M(<R)$ is the total mass enclosed within $R$, and $M_{\rm gas}$ is the total mass within the shell. In a galactic environment, most of the momentum will come from core-collapse SNe, with $\dot{P}_{\rm fb}= \left<\dot{p}/m_{\star}\right>M_{\star, \rm young}$, where $\left<\dot{p}/m_{\star}\right>$ is the average momentum injected per unit mass formed, and $M_{\star, \rm young}$ is the total mass in stars eligible to explode as SNeII (i.e., $5\ \rm Myr \lesssim t_{\rm age} \lesssim 40 \ \rm Myr$). Therefore, the condition to meet in order to have an outward moving gas shell is 
\begin{equation}\label{eq:condition_murray}
    \frac{\Sigma_{\rm crit}}{\Sigma_{\rm tot}(R)}\frac{M_{\star, \rm young}}{M_{\rm gas}} = \frac{\Sigma_{\rm crit}}{\Sigma_{\rm tot}(R)} \frac{\left< \rm SFR \right> t_{\star}}{M_{\rm gas}} =  \frac{\Sigma_{\rm crit}}{\Sigma_{\rm tot}(R)} \rm \frac{t_{\star}}{t_{\rm dep}}> 1.
\end{equation}
In Eq.~\ref{eq:condition_murray}, $\Sigma_{\rm tot} = M(<R)/\pi R^2$, $t_{\rm dep}=\rm M_{\rm gas} / \rm SFR$ is the depletion time, and we approximated $M_{\star, \rm young}$ with $\left< \rm SFR \right> t_{\star}$, where $t_{\star}$ is the interval of time where SNeII rates are non-zero and $\left< \rm SFR \right >$ is the average SFR during $t_{\star}$. Finally, $\Sigma_{\rm crit}=\left<\dot{p}/m_{\star}\right>/\pi G$ is a quantity with the units of surface density which describes the strength of stellar feedback as opposed to gravity (e.g., \citealt{2018GRUDIC}, \citealt{2020GRUDIC}, \citealt{2022HOPKINS}). From Eq.~\ref{eq:condition_murray} it is possible to see that at constant $t_{\rm dep}$, the capability of stellar feedback in generating an outward momentum (and therefore powering galactic outflows) roughly scales as $\propto \Sigma_{\rm crit}/\Sigma_{\rm tot}(R)$. 

We directly compare the values of the gravitational and stellar feedback forces in our simulations. In order to compute both quantities, we employ a more accurate procedure than the simplified assumption used for Eq.~\ref{eq:condition_murray}, by taking into consideration that gas is typically not uniformly distributed on a thin spherical shell. In particular, the total weight of the gas within the ISM ($\rm R<0.1\ \rm R_{\rm vir}$) is computed as:
\begin{equation}\label{eq:gravitational_force}
    F_{\rm g} = -G\sum_{i} \frac{M(<r_i)M_{\rm gas}(r_{i-1}<r<r_i)}{r_i^2},
\end{equation}
where the radii $r_i$ are equidistant in log-space (with bin sizes of 0.02 dex) in the range $[0-0.1]\times R_{\rm vir}$. The force exercised by the feedback is computed as:
\begin{equation}\label{eq:feedback_force}
    F_{\rm fb}(T) = \dot{P}_{\rm fb}(T) = p_t \int_0^T R_{\rm II}(T-\tau) \rm SFR(\tau)d\tau,
\end{equation}
where $p_t$ is the terminal momentum of the material ejected by one SNeII explosion, $R_{\rm II}$ are the SNeII rate and the SFR is the galaxy instantaneous SFR (SFR$_{1\rm Myr}$) computed within $0.1\rm R_{\rm vir}$. In this calculation, the supernovae rates are taken directly from Appendix A in \cite{2018HOPKINS}, while the terminal momentum $p_t$ is taken from Eq.~D23 of \cite{2018HOPKINS} which is derived from high-resolution simulations of individual SNe explosions. We note that at the current numerical resolution, the cooling radius is not resolved, and most of the energy is deposited in the form of momentum. Since $p_t$ depends weakly on the local gas density and metallicity ($\propto \rho^{-1/7}(Z/Z_{\odot})^{-0.21}$), we compute both these quantities as the average within the stellar half-mass radius, which corresponds to the region where most of the energy is released.

In Fig.~\ref{fig:grav_vs_feedback} we show the results of this computation for D9. As previously noted, the evolution of D9 can be divided into two different regimes: at $t<750\ \rm Myr$, gas accretion is followed by SF events and the resultant stellar feedback events are strong enough to evacuate all the remaining gas from the central region of the halo ($0.1\rm R_{\rm vir}$ in this case). At later times however, the increase in stellar mass drastically increases the gravitational force, making it nearly two orders of magnitude larger than the force generated by stellar feedback. In this regime, stellar feedback is not strong enough to overcome the gravitational force, and thus too weak to power galactic outflows.

\subsection{Mass loading factor}

\begin{figure}
	\includegraphics[width=\columnwidth]{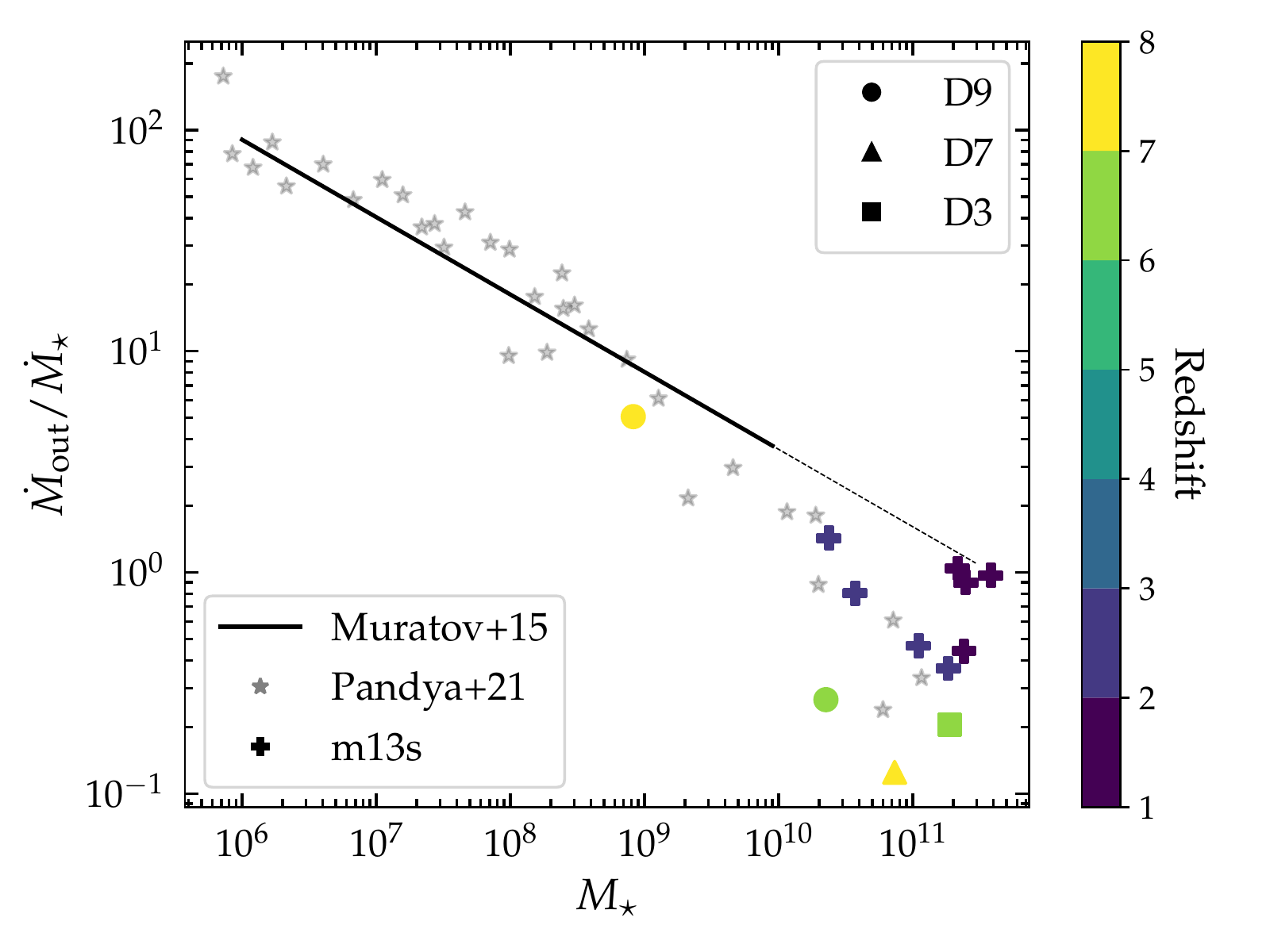}
    \caption{Mass loading factor as a function of stellar mass for different FIRE-1 (black solid line) and FIRE-2 (points) simulations. We report the mass range covered by FIRE-1 simulations as a solid line, while the extrapolation to higher stellar masses is shown as a dashed line. Grey stars show the results from \protect\cite{2021PANDYA}, measured following the \protect\cite{2015MURATOV} definition. Both FIRE-1 and FIRE-2 results refer to simulations in the redshift range $0<z<4$. Simulations analysed in this work are shown as circles, triangles, and squares. We show two points for D9 as we divide the time domain in two: before the merger ($t<750$ Myr) and after ($t>750$ Myr). Crosses highlight a subset of the FIRE-2 simulations analysed in \protect\cite{2021PANDYA}, with stellar masses and SFRs similar to the simulations analysed in this work. SFRs are computed within $0.1 \times R_{\rm vir}$. Outflow rates are computed following equation~(\ref{eq:out_def}) at $0.25\ \rm R_{\rm vir}$. At  high redshift, galaxies with $M_{\star}>10^{10}M_{\odot}$ are characterised by systematically lower mass loading factors.}
    \label{fig:muratov_outflow}
\end{figure}

A useful parameter that is commonly adopted to quantify the effect of feedback in galaxies is the mass loading factor, $\eta \equiv \dot{M}_{\rm out} / \dot{M}_{\star}$. This parameter has been extensively studied using both the FIRE-1 and FIRE-2 models (see \citealt{2015MURATOV}, \citealt{2017ALCAZAR}, \citealt{2021PANDYA}), by means of zoom-in simulations encompassing a wide range of stellar masses and redshifts. All these studies show that $\eta$ is a declining function of halo properties, such as the halo mass and the circular velocity, with a higher normalisation at higher redshifts. These trends are in qualitative agreement with analytical results (e.g., \citealt{2005MURRAY}), other zoom-in cosmological simulations (e.g., \citealt{2016CHRISTENSEN}, \citealt{2019TOLLET}), and large cosmological boxes (e.g., \citealt{2019NELSON}, \citealt{2020MITCHELL}). Interestingly, this redshift dependence largely disappears once the galaxy stellar mass is considered, with typical values of $\eta$ of $\sim 100$ for dwarfs and $\eta \lesssim 1$ for massive galaxies ($M_{\star}\sim 10^{11}\ M_{\odot}$, e.g., Fig.~5 of \citealt{2021PANDYA}).

In this section we aim to compare the mass loading factor as measured in our simulations with previous results from FIRE-1 (\citealt{2015MURATOV}, \citealt{2017ALCAZAR}) and FIRE-2 (\citealt{2021PANDYA}) simulations. In the previously cited works, the mass loading factors are computed as the ratio between the integrals of the SFR and mass outflow rate over three redshift bins: $4>z>2$, $2>z>0.5$, and $0.5>z>0$. The use of the integrated quantities is needed given the ``bursty'' nature of FIRE SFHs, with a relatively large scatter of the mass loading factor among different bursts of star formation. For all of the simulations, the outflow rates are computed following equation~(\ref{eq:out_def}) at $0.25\ \rm R_{\rm vir}$. In \cite{2021PANDYA}, the SFR is computed by summing over the instantaneous SFR of all gas particles within $0.1\times R_{\rm vir}$. For our simulations, we compute the instantaneous values of SFR using star particles and averaging over $1\ \rm Myr$. Finally, in \cite{2015MURATOV} the SFR is the average over the time between two contiguous snapshots ($\sim 50\rm \ Myrs$).

Our simulations are only evolved to $z \sim 5.5$ at most (D7 only to $7.2$), thus we cannot use the same redshift bins as \cite{2021PANDYA} and \cite{2015MURATOV}. For D9, since we can clearly distinguish two different regimes (before and after the merger), we divide the time domain in two: $t<750\ \rm Myrs$ and $t>750\ \rm Myrs$. For D7 and D3, we only consider the time interval at which $M_{\star} > 10^{10}\ \rm M_{\odot}$. This choice is motivated as at earlier times the stellar masses of these galaxies grow by 2 orders of magnitude in less than 100 Myr. In all cases the average SFR and outflow rates are computed by integrating the instantaneous outflow rate and SFR$_{1\rm Myr}$ over the time intervals described. 

Among the simulations analysed by \cite{2021PANDYA}, we highlight four FIRE-2 massive haloes: A1, A2, A4, and A8 (which are resimulations of original FIRE-1 runs included in the MassiveFIRE suite, \citealt{2016FELDMANN}, \citealt{2017FELDMANN}). These simulations, to which we will refer as m13s throughout the paper, are presented by \cite{2017ALCAZAR_BH} and further analysed in \cite{2019COCHRANE}, \cite{2020WELLONS}, \cite{2021STERN}, and \cite{2021PARSOTAN}, contain very massive haloes ($M_{\rm vir}\sim 10^{12.5}-10^{13} M_{\odot}$ at $z=1$) and galaxies with relatively high SFRs ($\sim 100\ M_{\odot}\rm yr^{-1}$, see \citealt{2019COCHRANE}). 

The values of the mass loading factor as a function of stellar mass are shown in Fig.~\ref{fig:muratov_outflow}, with the simulations presented in this work shown as coloured circles, triangles and squares. The mass loading factor of D9 computed before the merger agrees remarkably well with results obtained for simulations at lower redshift (\citealt{2015MURATOV}, \citealt{2021PANDYA}), suggesting that this relation is redshift independent at least at the low mass end. However, at stellar masses $M_{\star}\gtrsim 10^{10}M_{\odot}$ high redshift galaxies are characterised by consistently lower mass loading factors compared to their lower redshift counterparts. The values measured for $\eta$ are in the range $0.1-0.3$, about a factor of 5 lower than what is found in the m13s. These low values are similar to what has been measured in SPT-selected high redshift DSFGs (\citealt{2020SPILKER2}, $\eta \sim 0.2-0.5$). However, we caution that observed outflows are only measured for the molecular phase, and forward modelling of simulations results to mimic the observational procedure would be needed in order to directly compare to observations.

The plot also shows that results from FIRE-1 and FIRE-2 simulations agree remarkably well in the mass range covered by both sets of simulations (the highest stellar mass used by \citealt{2015MURATOV} in their fit is $M_{\star} \sim 10^{10}\ M_{\odot}$). However, at stellar masses higher than $\sim 10^{9} M_{\odot}$, the sample analysed by \cite{2021PANDYA} shows a bending in the relation. Moreover, at stellar masses $\gtrsim 10^{10} M_{\odot}$ we see that the scatter of the values of $\eta$ at fixed stellar mass increases considerably, with the lowest values found for D9, D7, and D3, and the highest values found in the m13s. In the
next subsections we investigate which physical property is the main driver of the large scatter of $\eta$ in the high stellar mass regime.

\subsection{Dependence of the mass loading factor on local physical properties} 

In order to identify the reason for the different values of $\eta$ among the m13s and D9, D7, and D3, it is important to understand the physical properties that most affect the mass loading factor. In (non-cosmological) simulations of isolated galaxies, \cite{2012HOPKINS} showed that the mass loading factor mainly depends on the circular velocity and gas surface density, computing all relevant quantities at the half-SFR radius, with $\eta = v_c^{-1.1 \pm 0.25}\Sigma_{\rm gas}^{-0.5\pm 0.15}$. 

The dependence on the circular velocity can be understood theoretically, and is driven by two concomitant physical processes. First of all, at larger circular velocities, more energy is required to escape the deeper potential wells, with a dependence that ranges from $v_c^{-1}$, for a momentum conserving outflow, to $v_c^{-2}$ for an energy conserving outflow (\citealt{2005MURRAY}). Secondly, at sufficiently large halo masses ($\sim 10^{12}\ \rm M_{\odot}$, \citealt{2021STERN}), a hot, uniform, and virialized halo forms. The pressure exercised from a virialized CGM can suppress the expansion of superbubbles created from SNe explosions, thus efficiently suppressing the launching of galactic outflows. The level of virialization of the CGM can be analytically expressed as a function of $v_c$, with a predicted scaling $v_c^{-3.4}$ (\citealt{2021STERN}). However, both confinement effects are comparable in magnitude and difficult to disentangle (Byrne et al. in prep.). Therefore, in what follows, we will study the dependence of the mass loading factor on $v_c$, without trying to infer which of these two physical processes is the main driver of this dependence. 

\begin{figure*}
\centering
\begin{minipage}{.5\textwidth}
  \centering
  \includegraphics[width=\linewidth]{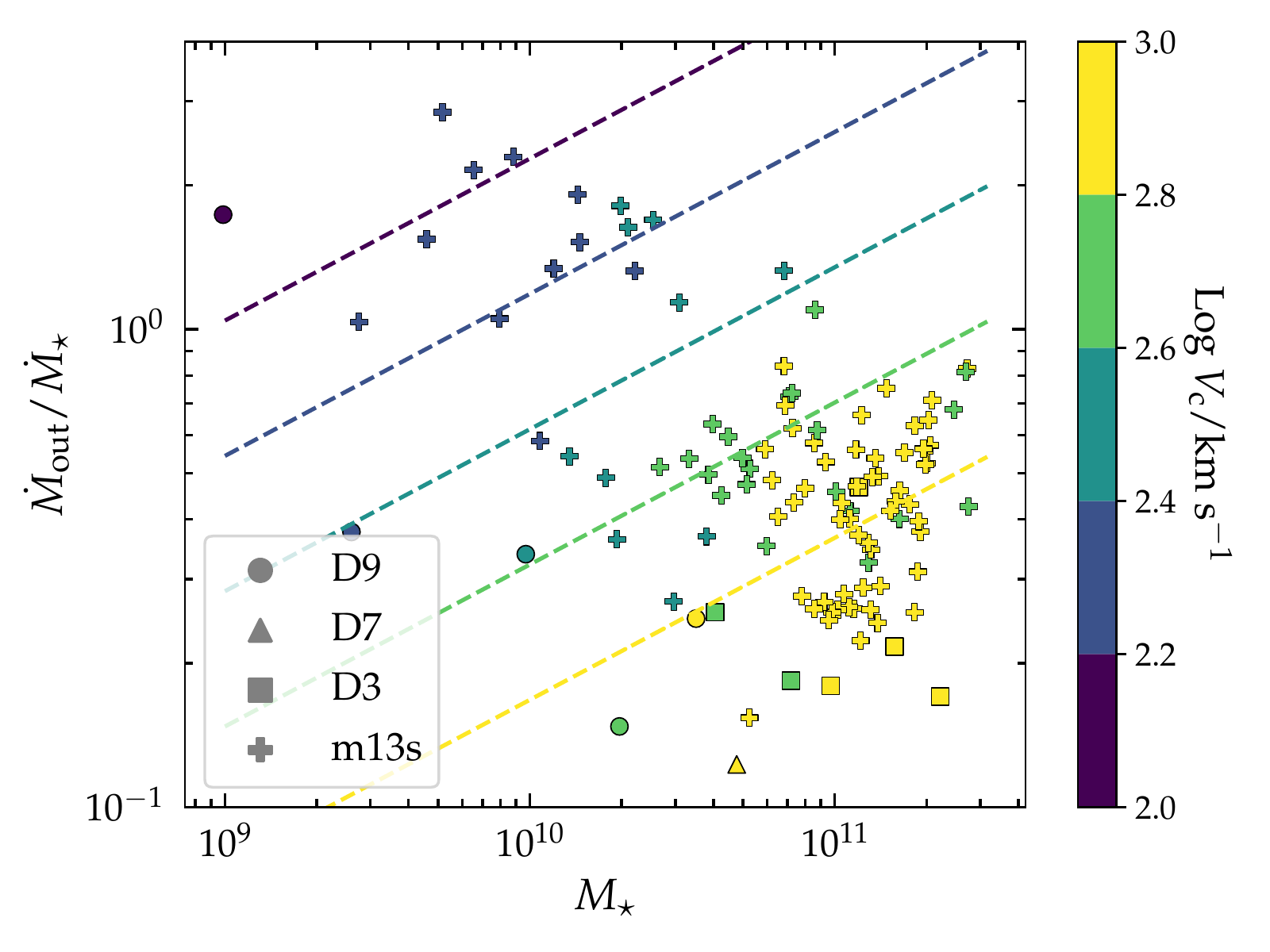}
\end{minipage}%
\begin{minipage}{.5\textwidth}
  \centering
  \includegraphics[width=\linewidth]{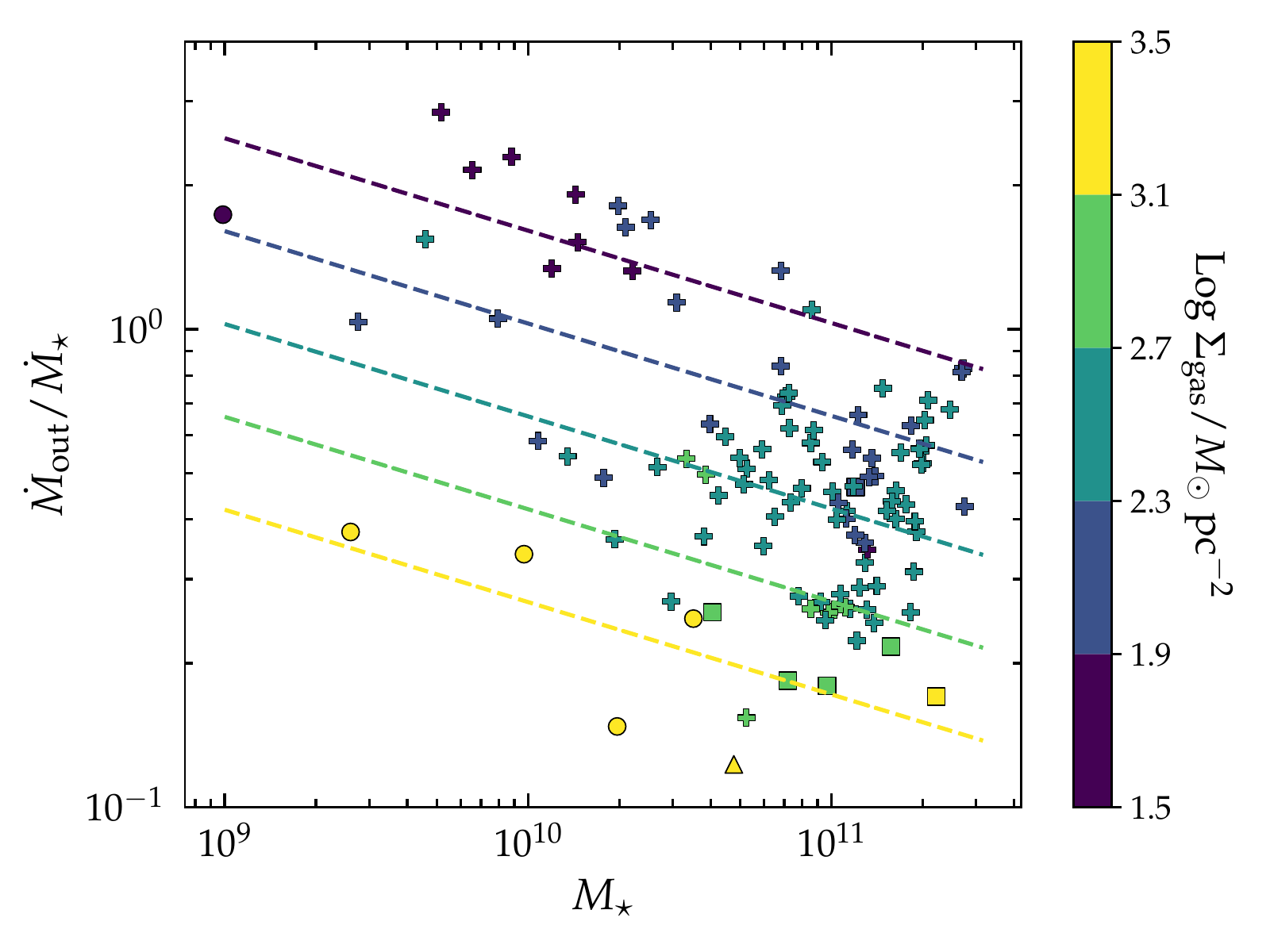}
\end{minipage}
\caption{Mass loading factor as a function of stellar mass for D9, D7, D3, and the m13s. The values of the mass loading factor are measured for single bursts of star formation, as explained in the text. Both mass loading factors and the stellar masses are measured within $0.1 R_{\rm vir}$. $v_c$ and $\Sigma_{\rm gas}$ are computed within the half stellar mass radius and are the average during single bursts. Points are colour-coded with respect to the value of the circular velocity at the half-stellar mass radius (left panel), and with respect to gas surface density (right panel). Dashed coloured lines are obtained using a multi-linear fit between $\eta$, $M_{\star}$, and $v_{\rm c}$ (or $\Sigma_{\rm gas}$). This figure shows that most of the scatter in the values of $\eta$ at fixed stellar mass is driven by different values of $\Sigma_{\rm gas}$.}
\label{fig:vcirc_sigma}
\end{figure*}

The scaling with the gas surface density is less straightforward to derive from first principles, and is due to the fact that different feedback processes driving galactic outflows are directly influenced by the local gas density. Specifically:
\begin{itemize}
    \item SNe explosions will create expanding (super)bubbles within the ISM. These will lead to strong galactic outflows only if the expanding shells will break out of the galactic disk before stalling (e.g., \citealt{2018FIELDING}, \citealt{2022ORR}). At fixed SFR, and therefore fixed amount of momentum injected into the ISM, the higher the gas density, the larger the gas mass swept up by the bubble. Since the momentum is conserved during the expansion of the superbubbles, this will lead to a stalling at smaller radii, reducing the probability of breaks out (e.g., \citealt{2022ORR}\footnote{Note that in \cite{2022ORR} it is assumed that $\Sigma_{\rm gas} \ll \Sigma_{\rm crit}\sim 3000\ \rm M_{\odot}\ pc^{-2}$. However, in our high redshift galaxies, this approximation does not hold. Therefore, in Eqs.~12, 13, 15, $\Sigma_{\rm crit}$ should be substituted with $\Sigma_{\rm crit} + \Sigma_{\rm gas}$, thus introducing an additional dependence on the gas surface density.}).
    
    \item At higher gas densities SN remnants lose more energy via radiative cooling during the energy-conserving phase (e.g., \citealt{2015MARTIZZI}). On scales larger that the cooling radius, this implies a lower amount of momentum injected in the ambient gas per unit stellar mass formed, therefore lowering the energy available to drive galactic outflows. Even though the dependence of the terminal momentum depends weakly on the gas density ($\propto \rho^{-1/7}$), in our simulations the gas density at stellar masses $\sim 10^{11}\rm \ M_{\odot}$ can vary by 2 orders of magnitudes (see right panel of Fig.~\ref{fig:vcirc_sigma}), thus implying a factor of $\sim 2$ difference in the injected momentum.
    
    \item Finally, the second possible feedback channel that can drive galactic outflows is the long-range radiation pressure, whose effect is to accelerate low-density gas particles outside the galaxy. While this channel is typically less relevant than SNeII feedback in driving galactic outflows, it might play an important role in local starburst and high redshift galaxies (e.g., \citealt{2012HOPKINS}). The long-range radiation pressure term is proportional to the emergent luminosity emitted from stars (i.e., radiation that is not locally absorbed by gas particles surrounding the emitting star particles, see Appendix D of \citealt{2018HOPKINS}):
\begin{equation}\label{eq:long_range}
    L_{\rm emergent} = \exp(-\tau)L,
\end{equation}
where $\tau$ is the optical depth, and $L$ is the intrinsic luminosity of star particles. Since the optical depth $\tau$ is directly proportional to the gas column density, Eq.~\ref{eq:long_range} implies that at high densities the energy available to drive galactic outflows exponentially declines, as most of the radiation energy is already absorbed within the nuclear star-forming region. 
\end{itemize} 

From a theoretical point of view, both the circular velocity and the gas surface density are expected to evolve with redshift, thus potentially explaining the large spread in the values of $\eta$ shown in Fig.~\ref{fig:muratov_outflow}. As discussed in Sect.~\ref{sec:grav_feed_forces}, the efficiency of stellar feedback in driving galactic outflows depends on the ratio between the force exercised by stellar feedback and the gravitational pull. Equivalently, the strength of galactic outflows depends on the ratio between the energy released by SNe feedback and the gravitational energy. This ratio can be expressed as (\citealt{1991WHITE}):

\begin{equation}\label{eq:analytical_eta}
    \eta = \frac{\epsilon_{\rm SN}\ \eta_{\rm SN}\ E_{\rm SN}}{E_{\rm bind}},
\end{equation}

where $\epsilon_{\rm SN}$ is the energy fraction effectively coupled to the ISM, $\eta_{\rm SN}$ is the occurrence of SN explosion per unit solar mass formed in stars, $E_{\rm SN}$ is the energy per single supernova explosion, and $E_{\rm bind}$ is the specific binding energy of the gas. In equation~(\ref{eq:analytical_eta}), the only term that explicitly depends on redshift is $E_{\rm bind}$, which can be written as (e.g, \citealt{1998MO}, \citealt{2003ZHAO}, \citealt{2020LAPI}):

\begin{equation}\label{eq:binding_energy}
    E_{\rm bind} \approx 10^{2}M_{\rm vir}^{2/3}E_{z}^{1/3} \rm \ cm^2 s^{-2},
\end{equation}

where $E_{z}=\sqrt{\Omega_M(1+z)^3 + \Omega_{\Lambda}}$. Therefore, in a matter dominated universe, at fixed halo mass $\eta$ is expected to scale as $\sim 1/\sqrt{1+z}$. This behaviour is driven by the larger matter density that characterises the high redshift Universe, which in turn implies that at fixed halo mass $R_{\rm vir}$ is smaller, and therefore the circular velocity is higher. Similarly, since at high$-z$ matter is more concentrated, higher redshift galaxies are expected to have higher gas surface densities. 

To test which physical properties primarily drive the differences in the mass loading factors, we use the python tools $\it{scipy.signal.find\_peaks}$\footnote{For this function we impose a {\it prominence} of 10, and a minimum time interval among different peaks of $\sim 50$ Myr.} and ${\it scipy.signal.correlate}$ to identify single bursts of star formation, compute their duration, and the time shift between the SF burst and the following outflow events. For this analysis, we compute both the instantaneous SFR ($\rm SFR_{1\rm Myr}$) and the outflow rate within $0.1 R_{\rm vir}$. Next, we compute the average of different physical properties during the duration of the burst, including: the circular velocity at the stellar half-mass radius, $v_c$, the gas surface density within the stellar half-mass radius, $\Sigma_{\rm gas}$, the stellar half-mass radius, $r_{\star, 1/2}$, and the SFR within $0.1R_{\rm vir}$. The results of a power-law fit in the form $\eta = 10^{\alpha}X^{\beta}$ between these quantities and the instantaneous values of the mass loading factors performed with the python package $\it{linmix}$\footnote{\url{https://github.com/jmeyers314/linmix}} are summarised in Table~\ref{tab:fit_results}. Unsurprisingly, the two quantities that show a stronger anti-correlation with $\eta$ are $v_c$ and $\Sigma_{\rm gas}$. 

\begin{table}
	\centering
	\caption{Result of power-law fit to the correlation between instantaneous values of the mass loading factors and different physical properties. The functional form is $\eta = 10^{\alpha} X^{\beta}$, and the fit is performed with the public python package linmix. In the table we also report the intrinsic scatter, $\sigma$, and the linear correlation coefficient, $\rho$.}
	\label{tab:fit_results}
	\begin{tabular}{lccccc} % four columns, alignment for each
		\hline
		$X$ & $\alpha$ & $\beta$ & $\sigma$ & $\rho$ \\
		\hline
		$\Sigma_{\rm gas}[M_{\odot}\ \rm pc^{-2}]$ & $1.15\pm0.13$ & $-0.59\pm0.05$ & $0.19$ & $-0.74\pm0.05$ \\
		$v_c[\rm km\ s^{-1}]$ & $1.8\pm0.2$ & $-0.76\pm 0.08$ & $0.20$ & $-0.69\pm0.05$ \\
		$\dot{M}_{\star}[M_{\odot}\ \rm yr^{-1}]$ & $0.5\pm0.1$ & $-0.36\pm0.06$ & $0.24$ & $-0.54\pm0.08$ \\
		$r_{\star, 1/2}[\rm kpc]$ & $-0.28\pm0.02$ & $0.5\pm0.1$ & $0.25$ & $0.49\pm0.08$ \\
		\hline
	\end{tabular}
\end{table}

In Fig.~\ref{fig:vcirc_sigma} we show how the position on the $\eta-M_{\star}$ plane depends on both the circular velocity (left panel) and on the gas surface density (right panel). In order to show the trends more clearly, we also show dashed coloured lines representing the results of a multi-linear fitting between $\eta$, $M_{\star}$, and $v_c$ (or $\Sigma_{\rm gas}$). From this plot it is clear that at fixed stellar mass, the values of circular velocities are close to constant. We emphasise that some scatter is still present, especially at stellar masses $M_{\star}\gtrsim 10^{10} M_{\odot}$, driven by the difference in galaxy sizes (since $v_c \propto \sqrt{M_{\star}/R}$). More importantly, while our high redshift simulations show lower values of the mass loading factor with respect to m13s, they are all characterised by similar values of $v_c$ at fixed $M_{\star}$. The right panel of Fig.~\ref{fig:vcirc_sigma} shows that at fixed stellar mass the value of the mass loading factor strongly depends on the gas surface density, with systems characterised by larger values of $\Sigma_{\rm gas}$ having lower values of $\eta$. Strikingly, D9, D7, and D3 are all characterised by the highest values of $\Sigma_{\rm gas}$, and, therefore, by the lowest values of $\eta$.

\begin{figure}
    \centering
    \includegraphics[width=\columnwidth]{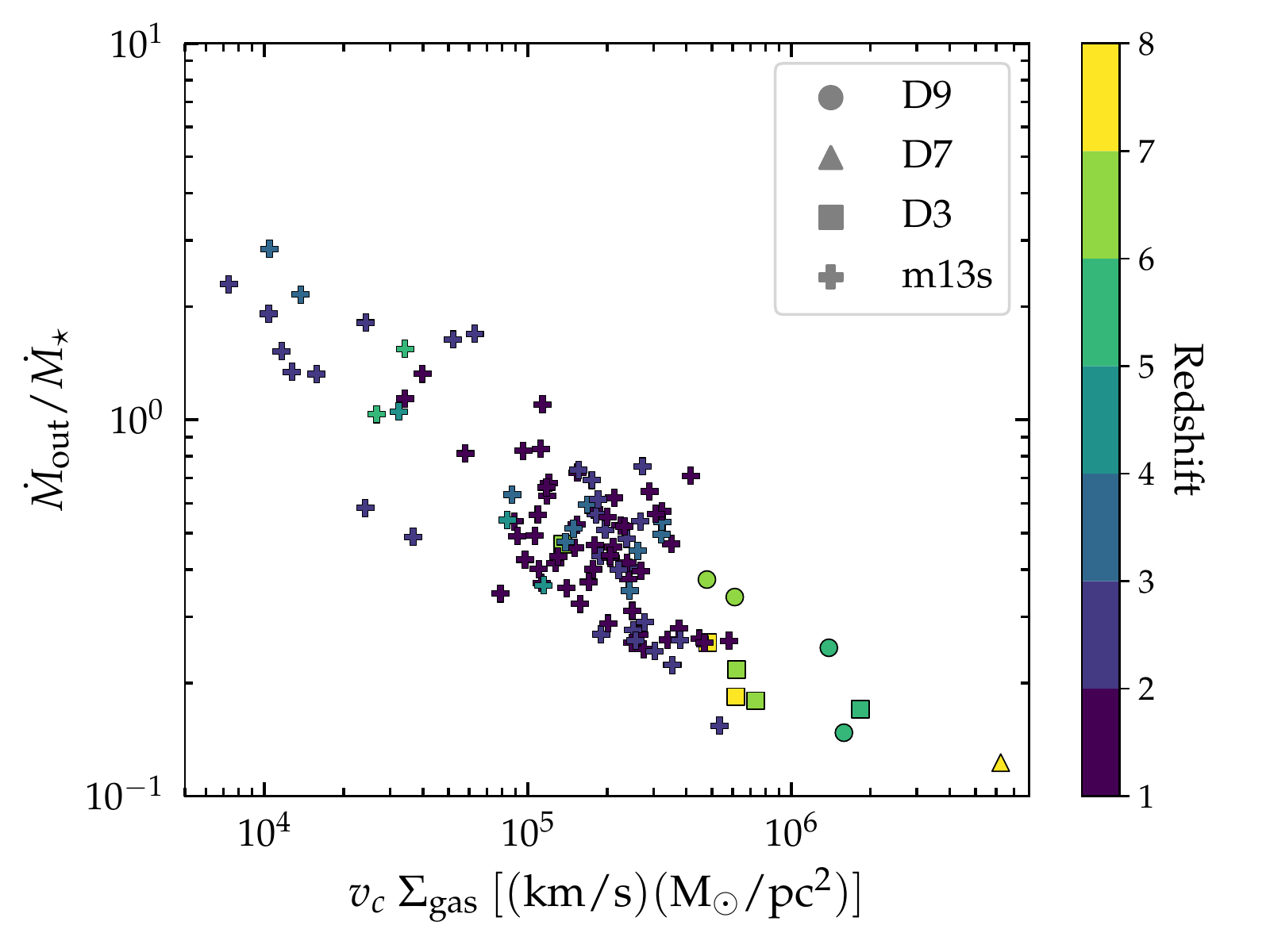}
    \caption{Instantaneous mass loading factor as a function of circular velocity and gas surface density for D9, D7, D3, and the m13s. The values of the mass loading factor are measured for single bursts of star formation, as explained in the text. Both mass loading factors and the stellar masses are measured within $0.1 R_{\rm vir}$. $v_c$ and $\Sigma_{\rm gas}$ are computed within the stellar half-mass radius and are averaged during single bursts. Points are colour-coded by the redshift, measured as the median value in the time interval considered. Instantaneous values of $\eta$ can be derived from $v_c$ and $\Sigma_{\rm gas}$, with the best fit values given in Eq.~\ref{eq:eta_best_fit}. }
    \label{fig:vcirc_sigma_fit}
\end{figure}

Finally, in Fig.~\ref{fig:vcirc_sigma_fit} we show the anti-correlation with the mass loading factor and the product between the circular velocity and the gas surface density. Similarly to \cite{2012HOPKINS} results, we see that with respect to previous figures, the scatter is greatly reduced, and all systems, independently on the redshift, follow the same relation. A multi-linear fit performed with the public python code ${\it scipy.optimize.curve\_fit}$\footnote{\url{https://docs.scipy.org/doc/scipy/reference/generated/scipy.optimize.curve\_fit.html}} results in:

\begin{multline}\label{eq:eta_best_fit}
    {\log} (\eta) = (2.14\pm0.15) - (0.43\pm0.05){\rm \log}(\Sigma_{\rm gas}) \\ 
    - (0.50\pm0.06){\rm \log}(v_c).
\end{multline}

Even though this result is in qualitative agreement with \cite{2012HOPKINS}, the dependence on the circular velocity is weaker. However, it is important to note that the dynamical range covered by the circular velocity in our sample is small ($v_c \gtrsim 300$ km/s for most of our points, see Fig.~\ref{fig:vcirc_sigma}), as we are focusing here on massive galaxies. Therefore, while we constrain the mass loading factors measured at high stellar masses (see Fig.~\ref{fig:muratov_outflow}), our results do not necessarily apply to lower stellar masses. 

\subsection{Effect of different feedback channels in driving galactic outflows}\label{sec:different_FB}

\begin{figure*}
\centering
\begin{minipage}{.5\textwidth}
  \centering
  \includegraphics[width=\linewidth]{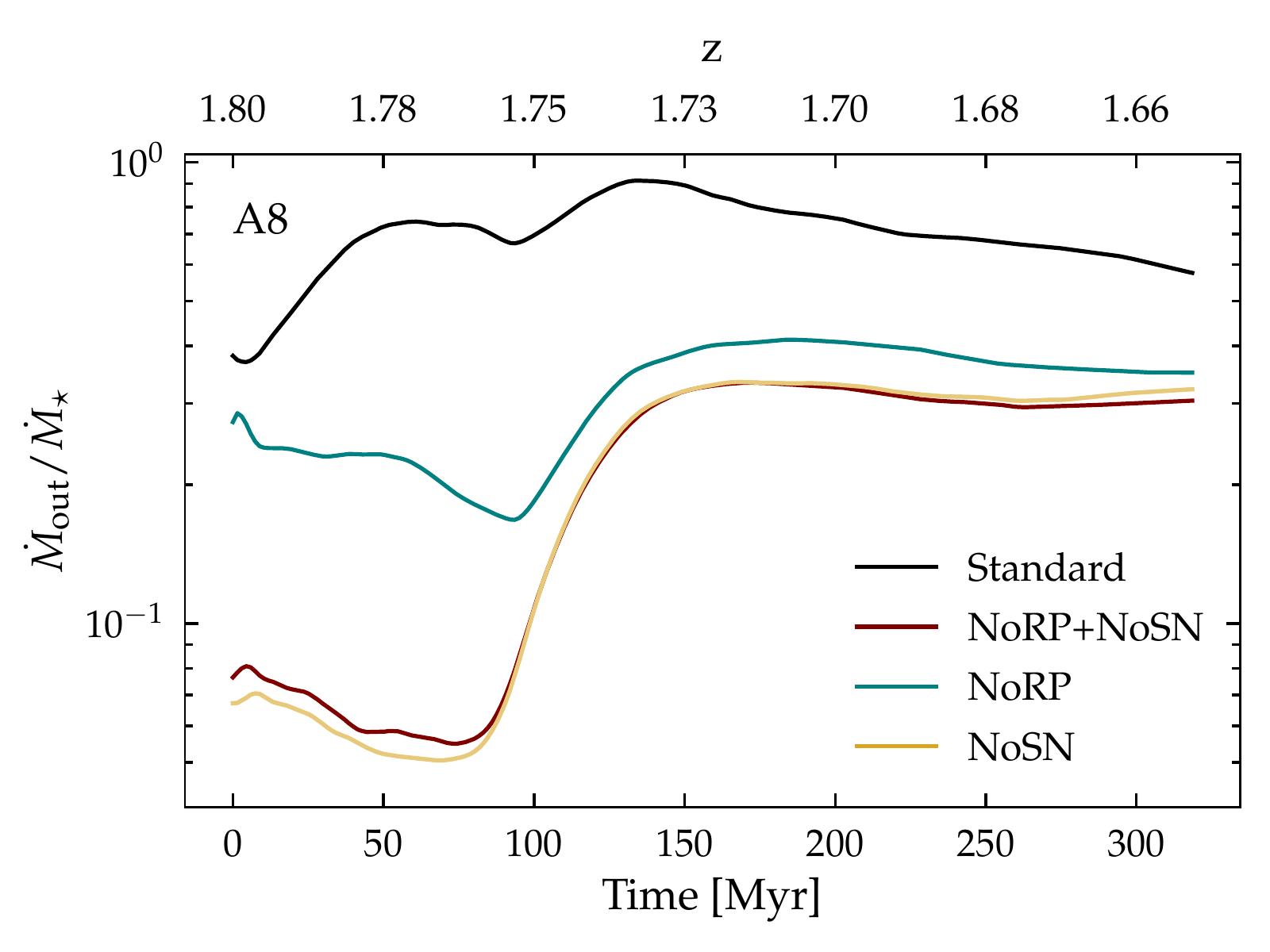}
\end{minipage}%
\begin{minipage}{.5\textwidth}
  \centering
  \includegraphics[width=\linewidth]{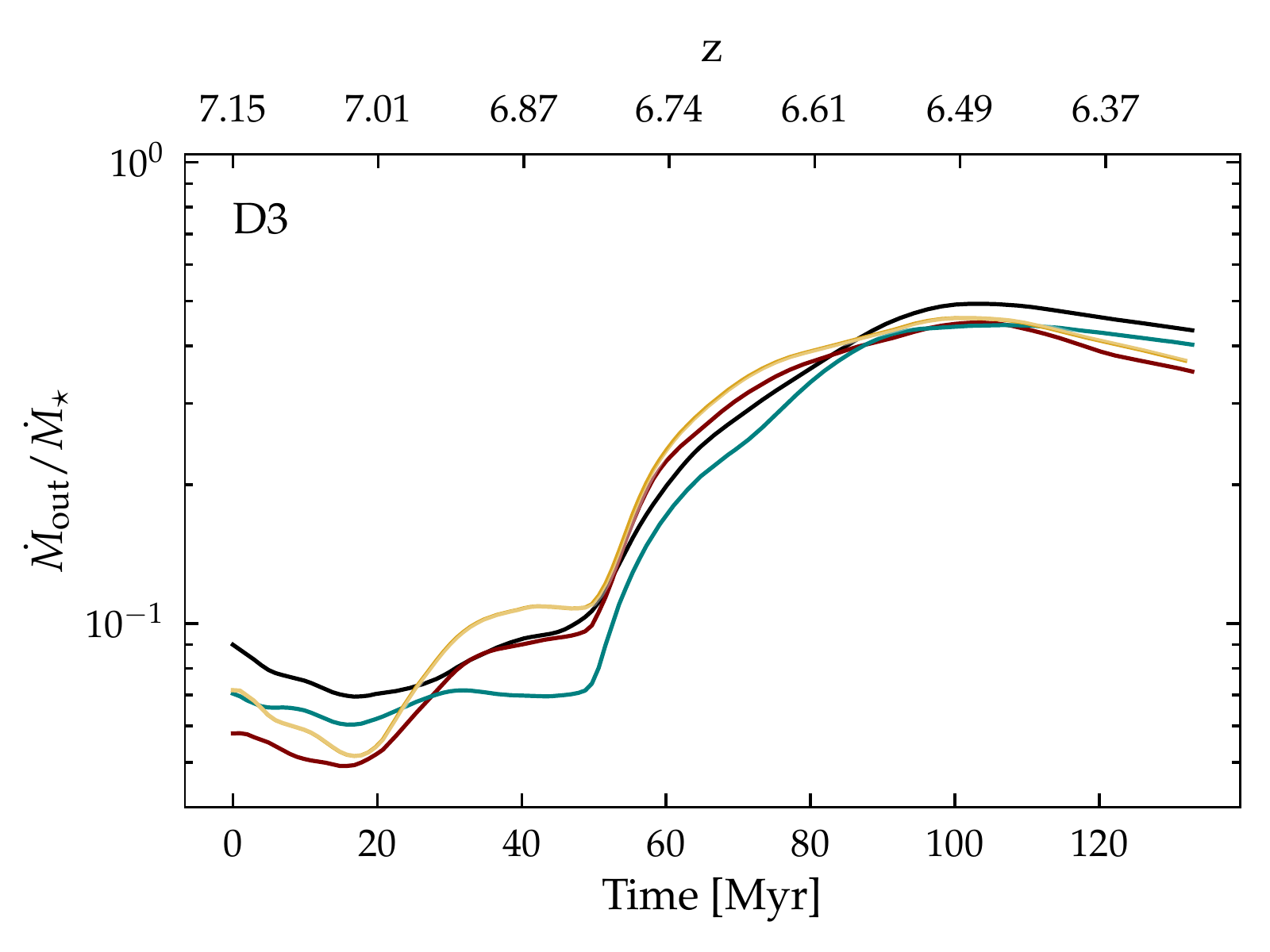}
\end{minipage}
\caption{Mass loading factor as a function of time for one of the m13s zoom-ins (A8, left), and for D3 (right). Both simulations are re-run with 4 different stellar feedback schemes. At each time T, the mass loading factor is computed by integrating both the outflow and SFR rates in the time interval $0<t<T$. Both simulations have been re-started at the snapshot corresponding to a galactic stellar mass $M_{\star}=10^{11} M_{\odot}$.  We compare re-runs of our standard feedback implementation (Standard) to those where different feedback channels have been switched off: no supernovae (NoSn), no long range radiation pressure (NoRP), and no SNe and no long-range radiation pressure (NoRP+NoSN).}
\label{fig:h2_eta}
\end{figure*}

To test the efficiency of different stellar feedback channels in driving galactic outflows in massive systems we re-ran two simulations and switched off different feedback mechanisms. For these tests we chose one system at high redshift (D3) and one system among the m13s simulations (A8). The simulations have not been re-run from initial conditions, but we re-started them when they already had a similar stellar mass of $10^{11} M_{\odot}$, as we are interested in the effect of the feedback in the high stellar mass regime. 

Regarding the particular stellar feedback channels, we have given priority to the ones that are expected to drive galactic outflows (SNeII and long-range radiation pressure). Therefore, the four re-runs for each simulation have been performed as: $(i)$ with no SNe feedback (NoSN), $(ii)$ with no long range radiation pressure (NoRP), $(iii)$ without both SNe and long range radiation pressure (NoRP+NoSN). The results of these tests are summarised in Fig.~\ref{fig:h2_eta}, where we show the ratio $\int_0^T\dot{M}_{\rm out}dt / \int_0^T \dot{M}_{\star}dt$ as a function of time T. These results clearly demonstrate that while for the m13 simulation switching off any of the feedback channels, and in particular SNe feedback, cause a decrease in the mass loading factor (left panel), the mass loading factor of D3 is independent of the stellar feedback (right panel). As can be seen from Fig.~\ref{fig:vcirc_sigma}, this galaxy is characterised by an extremely high gas surface density ($\gtrsim 10^3 M_{\odot}\rm \ pc^{-2}$), which greatly limits the energy which is available to launch galactic outflows. Instead, A8 at the time of the re-run ($z\sim 1.75$, see Fig.~\ref{fig:h2_eta}) is characterised by a significantly lower gas surface density ($\sim 10^2 M_{\odot}\rm \ pc^{-2}$), thus explaining the different behaviour.

\begin{figure}
    \centering
    \includegraphics[width=\columnwidth]{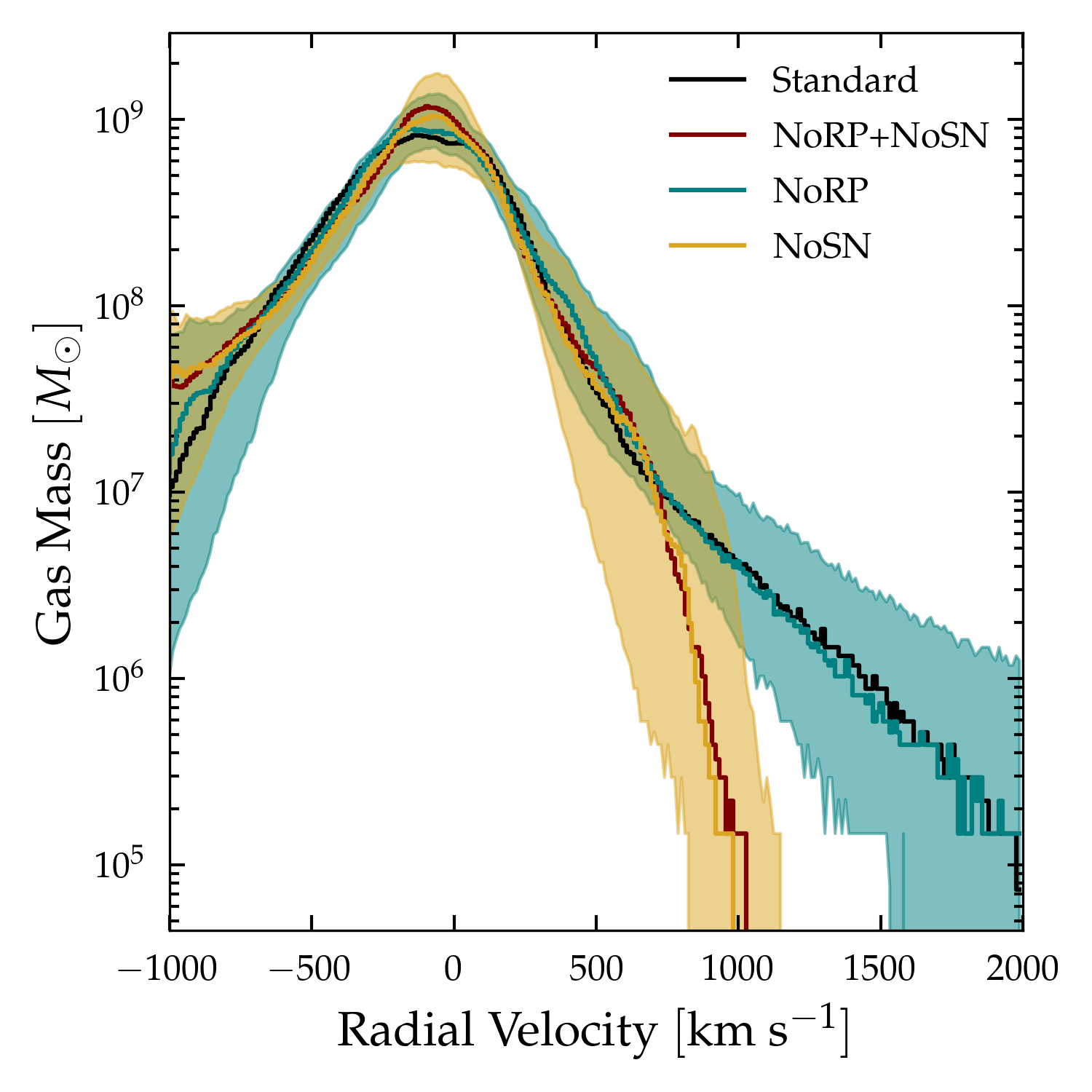}
    \caption{Radial velocity distribution of gas particles from the D3 simulation re-runs. Each line represents a different feedback implementation, as described in Fig.~\ref{fig:h2_eta}. For this analysis, the last $100$ Myr of the re-runs have been considered. Solid lines shows the median values among the snapshots analysed, while shaded regions represent the dispersion and encompass the $16^{\rm th}-84^{\rm th}$ percentiles region. For clarity, we show the dispersion only for two characteristic re-runs, as the other two show a similar level of variation. While turning off SN feedback does not affect the overall outflow rate, this feedback channel can accelerate a small fraction of gas particles to extremely high velocities ($> 1000$ km/s).}
    \label{fig:velocity_distribution}
\end{figure}

Interestingly, although the overall mass loading factor of D3 does not depend on the particular feedback implementation, SNeII feedback is still important in producing extremely high-velocity winds. This is shown in Fig.~\ref{fig:velocity_distribution}, where it is clear that SNeII feedback is essential to produce outflowing gas with velocities higher than $\gtrsim 1000\ \rm km\ s^{-1}$. These high-velocity winds, which contribute up to $\sim 7\%$ of the total galactic outflows, could be observed with standard techniques based on the detection of CII broad emission lines. Since the presence of these winds depend on the implementation of SNe feedback in (the FIRE) simulations, a direct comparison with observational data can help in constrain and validate the model implemented. Indeed, recent observational studies successfully detected molecular outflows in $z>4$ DSFGs (\citealt{2020SPILKER2}), with energetics that are consistent with expectations for SNe momentum-driven winds. Even though from a theoretical perspective we do not expect stellar feedback to impact integral properties (such as stellar and gas masses) of these high redshift massive systems, we will investigate the possibility of observationally probing high-velocity winds and their relation with the overall outflowing material in a future work.

In conclusion, we have shown that in our simulated high redshift massive galaxies stellar feedback is highly inefficient in driving galactic outflows. Moreover, it is even less efficient than in lower redshift haloes of similar mass. Indeed, on top of very high circular velocities ($\sim 1000\rm \ km\ s^{-1}$), high redshift galaxies are characterised by extremely high central gas surface densities, which limit the energy available to launch galactic outflows.

\section{Effect of low values of mass loading factor in regulating the SFR}\label{sec:SFE}

Within the FIRE model stellar feedback directly affects the instantaneous SFR within the galaxy. In particular, energy and momentum injected by young stars and SNe explosions balance the energy lost from radiative gas cooling. This prevents the gas from becoming Toomre-unstable, fragmenting, becoming self-shielding, and eventually forming stars (\citealt{2018ORR}). Indeed, it has been shown in several works that feedback and star formation self-regulate each other, enabling to correctly reproduce both the integrated and the resolved Kennicutt-Schmidt (KS) relation (e.g., \citealt{2011OSTRIKER}, \citealt{2013FAUCHER-GIGUERE}, \citealt{2013AGERTZ}, \citealt{2014HOPKINS}, \citealt{2018HOPKINS}, \citealt{2018ORR}, \citealt{2020GURVICH}) and the Elmegreen-Silk (ES) relation (e.g., \citealt{2018ORR}).

\iffalse
\begin{figure}
    \centering
    \includegraphics[width=\columnwidth]{Figures/new_KS_relation.pdf}
    \caption{Integrated Kennicutt-Schmidt relation of the simulations. Blue crosses represent snapshots from m13s simulations, black symbols show results for D9, D7, and D3. Maroon circles represent results at $z=0$ for three Milky-Way like zoom-in simulations run with FIRE-2 (\citealt{2016WETZEL}, \citealt{2017GARRISON_KIMMEL}, \citealt{2018HOPKINS}). All relevant quantities are computed within 3 times the stellar half-mass radius. The solid and dotted grey lines show observations from \protect\cite{2010DADDI} (starbursts and normal star forming galaxies respectively). Black lines show constant values of the depletion time. $\Sigma_{\rm gas}$ used in the plots is computed considering the total gas mass.}
    \label{fig:new_KS_relation}
\end{figure}
\fi

\begin{figure}
    \centering
    \includegraphics[width=\columnwidth]{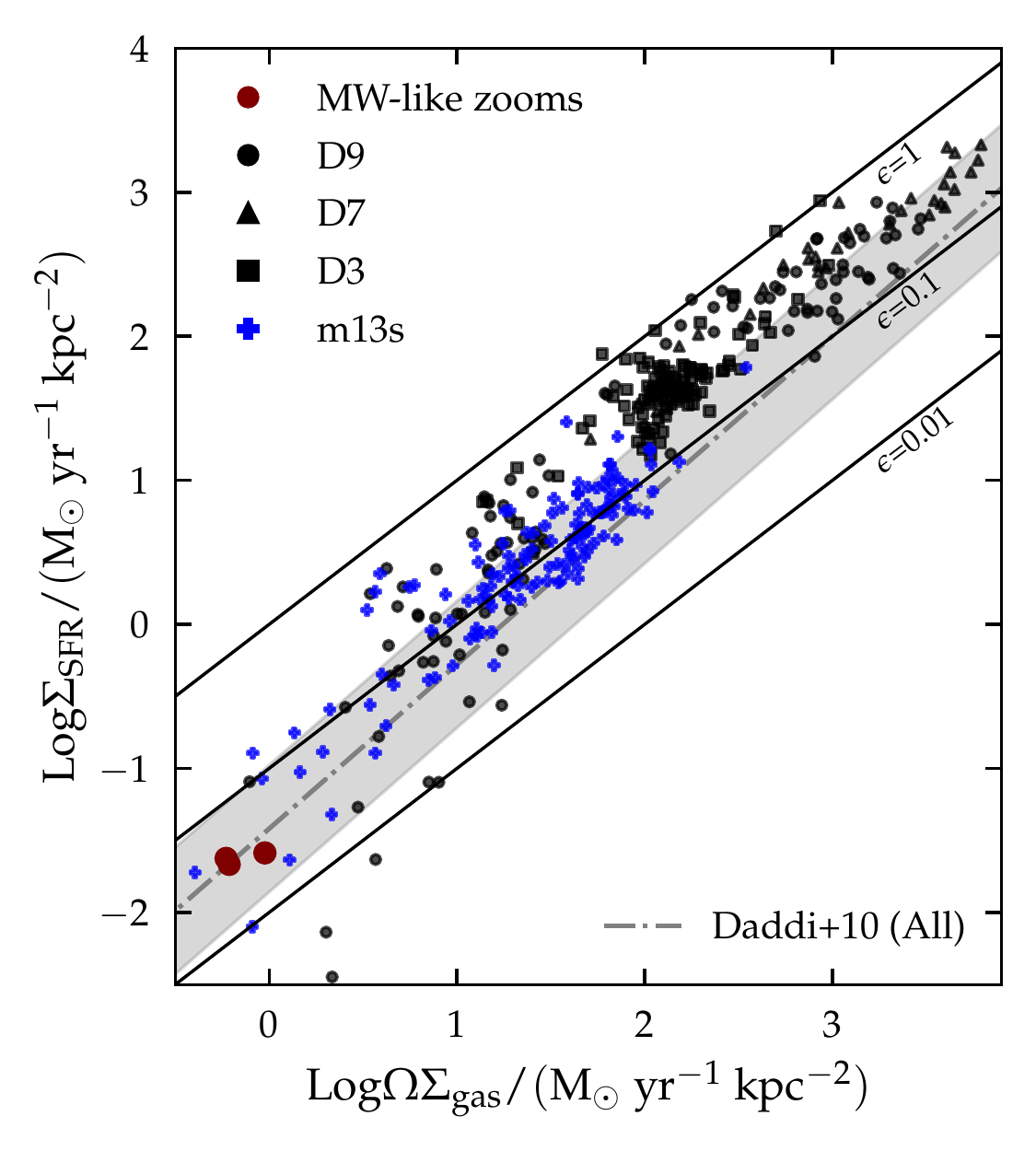}
    \caption{Integrated Elmegreen-Silk relation of the simulations. $\Omega$ is the orbital frequency (inverse of the orbital time, see Eq.~\ref{eq:Omega}). Blue crosses represent snapshots from m13s simulations, black symbols show results for D9, D7, and D3. Maroon points represents results at $z=0$ for three Milky-Way like zoom-in simulations run with FIRE-2. All relevant quantities are computed within 3 times the stellar half-mass radius. The grey dash-dotted line shows the relation from \protect\cite{2010DADDI} Black lines show constant values of the star formation efficiency per dynamical time. $\Sigma_{\rm gas}$ used in the plots is computed considering the total gas mass. While Milky-Way like simulations follow the observational relation, simulated massive galaxies, especially at high redshift, are characterised by higher SFE.}
    \label{fig:new_ES_relation}
\end{figure}

Observationally, it has been found that systems that lie above the KS relation (i.e. starbursts) tend to be characterised by shorter dynamical times ($t_{\rm dyn}$, \citealt{1997SILK}, \citealt{2002ELMEGREEN}, \citealt{2010DADDI}, see Eq.~\ref{eq:Omega}), such that $\rm SFE_{\rm dyn} \equiv t_{\rm dyn}/t_{\rm dep}$ is roughly constant. In other words, it has been empirically established that on galactic scales the SFE per dynamical time, $\rm SFE_{\rm dyn}$,  is roughly constant, and of the order of a few per cent (e.g., \citealt{1998KENNICUTT}, \citealt{2002WONG}, \citealt{2010DADDI}). Most notably, this relation does not depend on redshift and on galaxy type (i.e., main sequence galaxies and starbursts, e.g. \citealt{2010DADDI}). To test whether this is also the case for our simulations, in Fig.~\ref{fig:new_ES_relation} we show the integrated Elmegreen-Silk relation (\citealt{1997SILK}, \citealt{2002ELMEGREEN}) for FIRE-2 simulations in comparison with observations. In the plot, $\Omega$ is the dynamical angular frequency defined as:
\begin{equation}\label{eq:Omega}
    \Omega = \frac{2\pi}{t_{\rm dyn}} = \frac{v_c}{R} = \frac{(GM(<R))^{1/2}}{R^{3/2}},
\end{equation}
where $v_c$ is the circular velocity, $M(<R)$ is the total mass within $R$, and we set $R=3\times R_{\star, 1/2}$. In addition to the m13s and the D9, D7, and D3 simulations discussed so far, we also plot three Milky Way-mass (MW) simulations results at $z=0$: m12i, m12f, and m12m. The latter simulations (\citealt{2016WETZEL}, \citealt{2017GARRISON_KIMMEL}, and \citealt{2018HOPKINS}) have a mass resolution for gas particles of $7100 M_{\odot}$, and are run with the FIRE-2 sub-resolution physics described in Sect.~\ref{sec:subgrid_model}. The MW-mass simulations are in good agreement with the observational relation, with a star formation efficiency per dynamical time of $\sim 2\%$. However, results from both m13 and D9, D7, and D3 simulations show higher efficiencies, in the range $10\%<\epsilon<100\%$, with the highest values found in D3, D7, and D9. Specifically, we find a median value of $\epsilon \sim 10\%$ for m13s, while the median values of $\epsilon$ for D3, D7, and D9 is $\sim 30\%$. In order to interpret this result, it is important to stress that within the framework of the FIRE-2 model the low SFE$_{\rm dyn}$ time found for MW-like galaxies is a direct consequence of stellar feedback. Therefore, the results that are shown in Fig.~\ref{fig:new_ES_relation} directly imply that stellar feedback is less efficient in self-regulating star formation within massive galaxies, especially at high redshift.

The low efficiency of stellar feedback can be explained by the high mass surface density in these galaxies. In galactic nuclei, the fraction of gas that can be retained without being expelled as a consequence of stellar feedback is (e.g., \citealt{2022HOPKINS}):
\begin{equation}
    f_{\rm retained} \sim \frac{\Sigma_{\rm tot}}{\Sigma_{\rm crit} + \Sigma_{\rm tot}},
\end{equation}
where $\Sigma_{\rm crit} \sim 3000\ \rm M_{\odot}\rm \ pc^{-2}$. In our simulations, galaxies reach a stellar mass $M_{\star}\gtrsim 10^{11}\ \rm M_{\odot}$, with a stellar half-mass radius $\lesssim 1\ \rm kpc$. Therefore, $\Sigma_{\rm tot} \gtrsim 3 \times 10^4\ \rm  M_{\odot} \rm \ pc^{-2} \gg \Sigma_{\rm crit}$. In this regime, stellar feedback is much weaker than the gravitational pull. Thus, gas is retained at the galactic centre, and collapses and form stars on time-scales which are comparable to $t_{\rm dyn}$. This analysis shows that within the FIRE-2 model at $z \gtrsim 6$ haloes at masses $M_{\rm halo}\sim 10^{12}M_{\odot}$ can reach matter surface density large enough such that stellar feedback saturates and becomes inefficient. 

\begin{figure*}
\centering
\begin{minipage}{.5\textwidth}
  \centering
  \includegraphics[width=\linewidth]{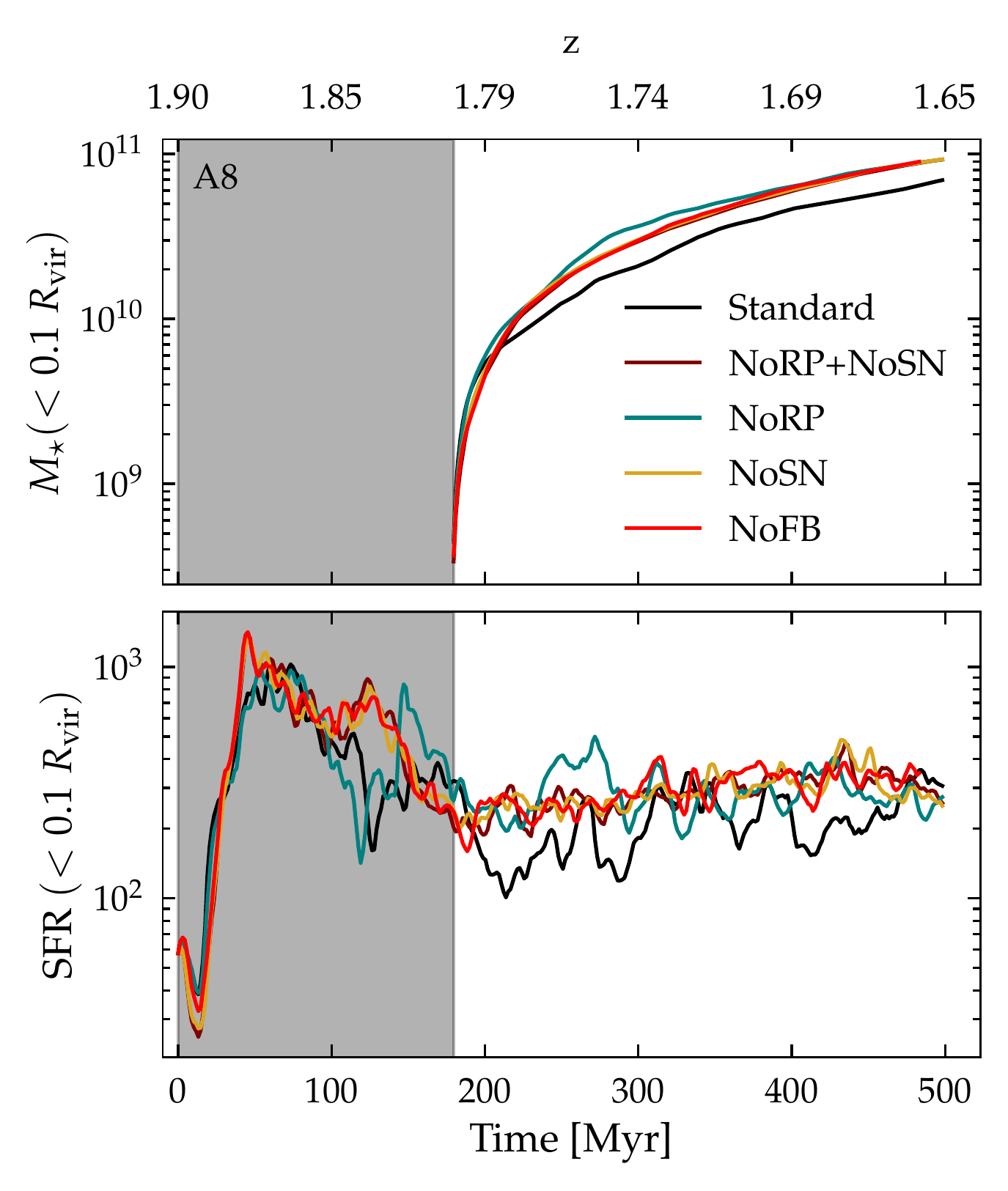}
\end{minipage}%
\begin{minipage}{.5\textwidth}
  \centering
  \includegraphics[width=\linewidth]{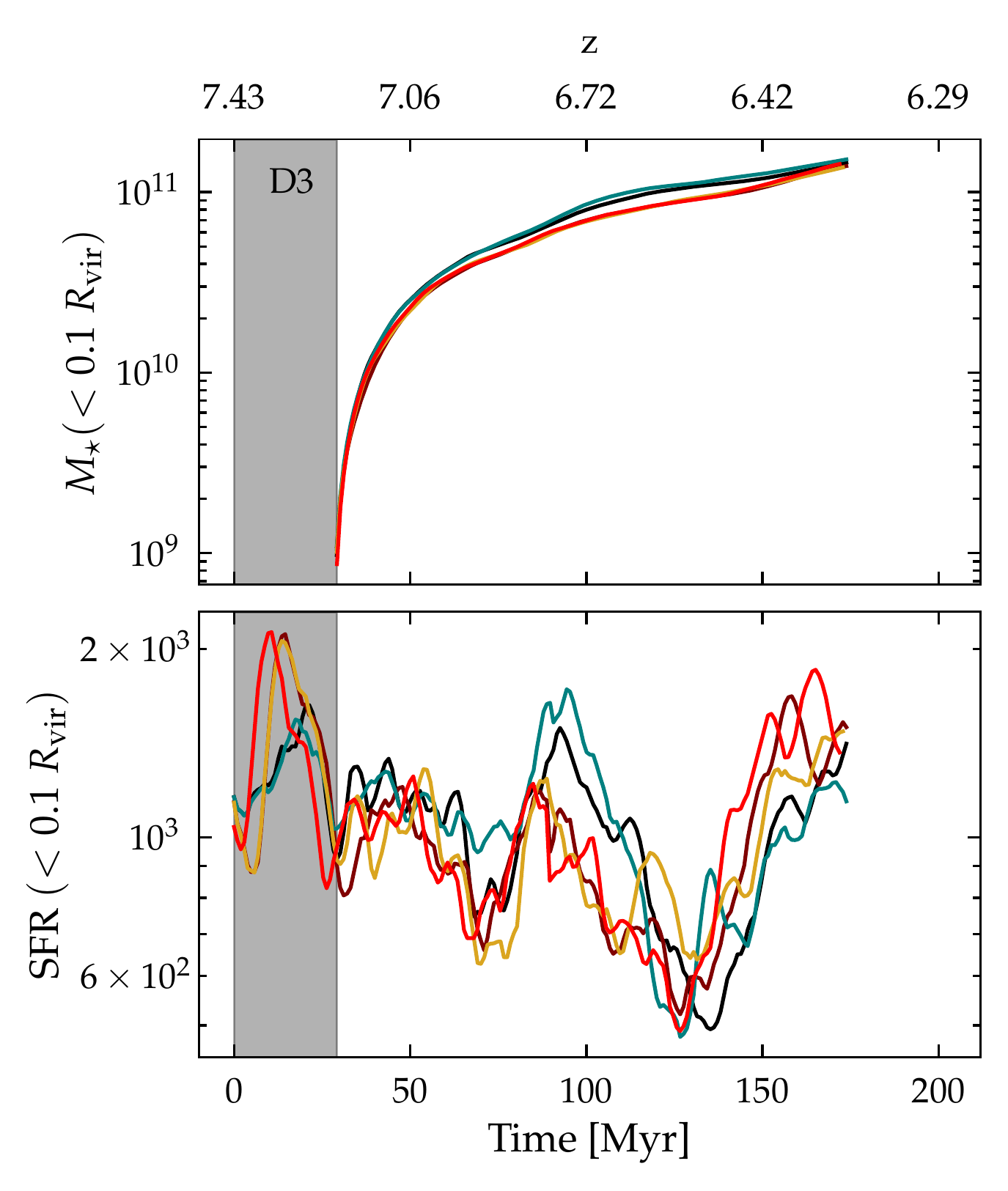}
\end{minipage}
\caption{Results of the m13 halo (A8, left) and D3 (right) re-runs. In both figures the lower panels show the evolution of the instantaneous SFR within $0.1\times R_{\rm vir}$ as a function of time, while the upper panels show the cumulative mass in new-born stars. For this computation, the first burst of star formation (grey shaded region) has been excluded, as it represents a numerical artefact related to the re-starting of the simulations from snapshot.}
\label{fig:smas_reruns_prova}
\end{figure*}

Similarly to the analysis of the mass loading factor, we re-run a m13 (A8) and the D3 simulation with different feedback prescriptions, to directly test the effect of stellar feedback on the instantaneous SFR. These tests are used to demonstrate what can be inferred from Fig.~\ref{fig:new_ES_relation}: weakening (or removing) stellar feedback will have a smaller impact on high redshift massive galaxies with respect to lower redshift and similar masses haloes, as the former have higher SFE per dynamical time. For this analysis, we make use of the simulations already presented in Sect.~\ref{sec:different_FB}, that are run with different stellar feedback implementations. To those simulations we also add a fifth re-run, where on top of switching off SNe and long-range radiation pressure feedback, we also switch off early feedback. While this feedback channel is not the main driver of galactic outflows, it is important for the regulation of star formation, as it defines the time-scale at which GMCs are disrupted. Therefore, the final sample comprises 5 different setups: $(i)$ our fiducial model (Standard), $(ii)$ with no SNe feedback (NoSN), $(iii)$ with no long-range radiation pressure (NoRP), $(iv)$ without both SNe and long-range radiation pressure (NoRP+NoSN), and $(v)$ without any stellar feedback (NoFB, which also removes early feedback). The results of these tests are shown in Fig.~\ref{fig:smas_reruns_prova}. For A8 (left panel), we find that removing any feedback channel results in SFRs that are, on average, $30\%$ higher, which can also be inferred from the cumulative stellar mass. Although still relevant, this difference is smaller than the effects that changes in the stellar feedback parameters can have in low redshift milky-way like galaxies (which can be up to a factor of two, see, e.g., Fig.~35 of \citealt{2018HOPKINS}). Therefore, even at intermediate redshift ($1 \lesssim z \lesssim 4$), stellar feedback is quite inefficient in regulating star formation in massive galaxies, although this effect is less pronounced than at high redshift. Indeed, for D3 (right panel), the instantaneous SFR is unchanged regardless of which feedback sub-resolution model is active. This confirms that stellar feedback is very inefficient in these high redshift massive objects.

\section{Summary and discussion}\label{summary}
%%%%%%%%%%%%%%%%%%%%%%%%%%%%%%%%%%%%%%%%%%%%%%%%%%%%%%%%%%%%

In this work, we studied the role of stellar feedback in driving galactic outflows and regulating the SFR in a sample of 3 simulated high redshift massive galaxies extending previous FIRE results (\citealt{2015MURATOV}, \citealt{2017ALCAZAR}, \citealt{2021PANDYA}) to higher redshift. The three systems analysed have been selected from the most massive haloes in a cosmological box of $400/h$ cMpc a side at $z=6$. Therefore, they represent rare systems and are suitable for a comparison with the most massive and most star-forming galaxies observed at high redshift.

The main conclusions of this work can be summarised as follow:
\begin{itemize}
    \item The three simulated systems with a halo mass $M_{h}>10^{12}M_{\odot}$ at $z>6$ resemble gas-related properties of observed high redshift DSFGs. They are characterised by extremely high central SFRs ($>1000\ \rm M_{\odot}\ yr^{-1}$, Fig.~\ref{fig:general_properties}), and short depletion times ($\sim 10$ Myrs, Fig.~\ref{fig:depletion_time}). However, we also find that the circular velocity of our systems, traced by the 1D velocity dispersion of the molecular gas, is a few times larger than what observations suggest (Fig.~\ref{fig:fwhm}). This difference hints that simulated galaxies are either too massive, too compact, or both. Their cold gas reservoirs, which are rapidly consumed by star formation, are promptly re-filled by strong inflows. The magnitude of the inflow determines the overall SFR within our simulated galaxies (Fig.~\ref{fig:outflow_profiles}). 
    
    \item In our high redshift simulations, we find that the mass loading factors, $\eta=\dot{M}_{\rm out}/ \dot{M}_{\star}$, are of the order of $0.1$ for massive galaxies  ($M_{\star}\gtrsim10^{10}\ M_{\odot}$, Fig.~\ref{fig:muratov_outflow}). Given these low values of $\eta$, the metals produced during stellar evolution are trapped in the star-forming region and rapidly recycled in newly born stars. Therefore, simulated galaxies reach solar (or super solar) metallicity by ($z\sim 6$), a direct prediction for upcoming JWST observations.

    \item We find that the mass loading factor shows the strongest anti-correlation with the circular velocity, $v_c$, and total gas mass surface density, $\Sigma_{\rm gas}$. While the high ($z\gtrsim 6$) and intermediate ($1 \lesssim z \lesssim 4$) redshift galaxies are characterised by similar circular velocities, the gas surface densities in high redshift galaxies are much larger, thus explaining the lower mass loading factors found for the high compared to the intermediate redshift galaxies from previous FIRE studies.

     \item The high gas surface densities prevent the efficient launching of outflows, resulting in low mass loading factors in high redshift massive galaxies. Even though stellar feedback is inefficient in driving galactic outflows, we show that SNeII are needed in order to reproduce high wind velocities ($v > 1000$ km/s), which are found in extremely luminous  FIR systems ($\log L_{\rm IR}/L_{\odot}>13$, e.g. \citealt{2020SPILKER2}).
    
    \item With respect to the SFE of $\sim2$ per cent measured in main sequence star forming galaxies, we find that stellar feedback is highly inefficient in self-regulating star formation in high redshift massive galaxies. We find that the SFE per dynamical (orbital) time is on average $30\%$, in some cases reaching values as high as $100\%$ (Fig.~\ref{fig:new_ES_relation}). These values are a factor of three higher than what we measure for massive galaxies at intermediate redshift, and are driven by the large total mass surface densities that characterise these objects. 
    
 \end{itemize}
  
We show that massive galaxies $(\sim 10^{11}M_{\odot})$ can form very rapidly, and resemble observed high redshift DSFGs. However, we also find evidence that these galaxies are either too compact or too massive with respect to observations. Moreover, while DSFGs are likely the natural progenitors of the massive quiescent galaxies that quench prior to redshift $\geq 3$ (e.g., \citealt{2017GLAZEBROOK}, \citealt{2018SCHREIBER}, \citealt{2020FORRESTb}, \citealt{2020VALENTINO}, \citealt{2021DEUGENIO}), it is highly unlikely that our simulated galaxies would reach quiescence on a similar timescale, as stellar feedback is too inefficient (in both driving outflows and in regulating the SFR). Indeed, we show that at $z\gtrsim 6$, within the FIRE-2 model, stellar feedback already saturates at haloes masses $\sim 10^{12}\rm \ M_{\odot}$ (i.e., MW-like haloes masses).

While the results shown in this work are valid in the physical conditions reached by our simulations (high mass surface densities), we caution that these conditions might not be realistic (see Fig.~\ref{fig:fwhm}). This can be related to two different factors: either the numerical implementation of the FIRE-2 feedback model might not be sophisticated enough, or an additional energy input beyond the currently implemented stellar feedback is required to quench massive galaxies at high redshifts. Regarding the former, \cite{2022HOPKINS_FIRE3} presented the FIRE-3 model, where the major improvements with respect to FIRE-2 regard an update in the stellar evolution inputs. This resulted in a shallower $M_{\star}-M_{\rm halo}$ relation (see their Fig.~9), which could result in lower stellar masses and larger sizes in the massive haloes presented in this paper. We will quantify the impact on the results presented in this work in a following paper. Regarding the inclusion of an additional energy input, AGN feedback represents a very plausible source, due to its ability to prevent gas accretion and/or heat the gas in the galaxy (which will in turn make stellar feedback more efficient), and their inclusion in cosmological simulations has been key in accurately reproducing the galaxy luminosity and stellar mass function at the massive end at $z=0$ (e.g., \citealt{2015SCHAYE}, \citealt{2018PILLEPICH}, \citealt{2019DAVE}, \citealt{2022WELLONS}). Therefore, investigating the role of AGN feedback in the evolution of high redshift massive galaxies is the crucial next step in exploring how massive galaxies quench at high redshift, which we plan to do in a future work.

\section*{Acknowledgements}

LB thanks L. Boco for helpful discussions. LB, RF, EC, JG acknowledge financial support from the Swiss National Science Foundation (grant no PP00P2\_194814). RF, EC, MB acknowledge financial support from the Swiss National Science Foundation (grant no 200021\_188552). JG gratefully acknowledges financial support from the Swiss National Science Foundation (grant no CRSII5\_193826). CAFG was supported by NSF through grants AST-1715216, AST-2108230,  and CAREER award AST-1652522; by NASA through grants 17-ATP17-006 7 and 21-ATP21-0036; by STScI through grants HST-AR-16124.001-A and HST-GO-16730.016-A; by CXO through grant TM2-23005X; and by the Research Corporation for Science Advancement through a Cottrell Scholar Award. The Flatiron Institute is supported by the Simons Foundation. LL acknowledges financial support from the Swiss National
Science Foundation (SNSF) (grant no P2ZHP2\_199729). This work was supported in part by a grant from the Swiss National Supercomputing Centre (CSCS) under project IDs s1098. We acknowledge access to Eiger and Piz Daint at the Swiss National Supercomputing Centre, Switzerland under the University of Zurich’s share with
the project ID uzh18. This work made use of infrastructure services provided by S3IT (\url{www.s3it.uzh.ch}), the Service and Support for
Science IT team at the University of Zurich. All plots were created with the Matplotlib library for visualization with Python (\citealt{2007HUNTER}).

%%%%%%%%%%%%%%%%%%%%%%%%%%%%%%%%%%%%%%%%%%%%%%%%%%
\section*{Data Availability}

The data supporting the plots within this article are available on reasonable request to the corresponding author. A public version of the
GIZMO code is available at \url{http://www.tapir.caltech.edu/~phopkins/Site/GIZMO.html}. FIRE-2 simulations are publicly available (\citealt{2022WETZEL}) at \url{http://flathub.flatironinstitute.org/fire}. Additional data, including initial conditions and derived data products, are available at \url{https://fire.northwestern.edu/data/}.

%%%%%%%%%%%%%%%%%%%% REFERENCES %%%%%%%%%%%%%%%%%%

% The best way to enter references is to use BibTeX:

\bibliographystyle{mnras}
\bibliography{bibliography} % if your bibtex file is called example.bib

\begin{thebibliography}{}
\makeatletter
\relax
\def\mn@urlcharsother{\let\do\@makeother \do\$\do\&\do\#\do\^\do\_\do\%\do\~}
\def\mn@doi{\begingroup\mn@urlcharsother \@ifnextchar [ {\mn@doi@}
  {\mn@doi@[]}}
\def\mn@doi@[#1]#2{\def\@tempa{#1}\ifx\@tempa\@empty \href
  {http://dx.doi.org/#2} {doi:#2}\else \href {http://dx.doi.org/#2} {#1}\fi
  \endgroup}
\def\mn@eprint#1#2{\mn@eprint@#1:#2::\@nil}
\def\mn@eprint@arXiv#1{\href {http://arxiv.org/abs/#1} {{\tt arXiv:#1}}}
\def\mn@eprint@dblp#1{\href {http://dblp.uni-trier.de/rec/bibtex/#1.xml}
  {dblp:#1}}
\def\mn@eprint@#1:#2:#3:#4\@nil{\def\@tempa {#1}\def\@tempb {#2}\def\@tempc
  {#3}\ifx \@tempc \@empty \let \@tempc \@tempb \let \@tempb \@tempa \fi \ifx
  \@tempb \@empty \def\@tempb {arXiv}\fi \@ifundefined
  {mn@eprint@\@tempb}{\@tempb:\@tempc}{\expandafter \expandafter \csname
  mn@eprint@\@tempb\endcsname \expandafter{\@tempc}}}

\bibitem[\protect\citeauthoryear{{Agertz} \& {Kravtsov}}{{Agertz} \&
  {Kravtsov}}{2015}]{2015AGERTZ}
{Agertz} O.,  {Kravtsov} A.~V.,  2015, \mn@doi [\apj]
  {10.1088/0004-637X/804/1/18}, \href
  {https://ui.adsabs.harvard.edu/abs/2015ApJ...804...18A} {804, 18}

\bibitem[\protect\citeauthoryear{{Agertz}, {Kravtsov}, {Leitner}  \&
  {Gnedin}}{{Agertz} et~al.}{2013}]{2013AGERTZ}
{Agertz} O.,  {Kravtsov} A.~V.,  {Leitner} S.~N.,   {Gnedin} N.~Y.,  2013,
  \mn@doi [\apj] {10.1088/0004-637X/770/1/25}, \href
  {https://ui.adsabs.harvard.edu/abs/2013ApJ...770...25A} {770, 25}

\bibitem[\protect\citeauthoryear{{Angl{\'e}s-Alc{\'a}zar},
  {Faucher-Gigu{\`e}re}, {Kere{\v{s}}}, {Hopkins}, {Quataert}  \&
  {Murray}}{{Angl{\'e}s-Alc{\'a}zar} et~al.}{2017a}]{2017ALCAZAR}
{Angl{\'e}s-Alc{\'a}zar} D.,  {Faucher-Gigu{\`e}re} C.-A.,  {Kere{\v{s}}} D.,
  {Hopkins} P.~F.,  {Quataert} E.,   {Murray} N.,  2017a, \mn@doi [\mnras]
  {10.1093/mnras/stx1517}, \href
  {https://ui.adsabs.harvard.edu/abs/2017MNRAS.470.4698A} {470, 4698}

\bibitem[\protect\citeauthoryear{{Angl{\'e}s-Alc{\'a}zar},
  {Faucher-Gigu{\`e}re}, {Quataert}, {Hopkins}, {Feldmann}, {Torrey}, {Wetzel}
  \& {Kere{\v{s}}}}{{Angl{\'e}s-Alc{\'a}zar} et~al.}{2017b}]{2017ALCAZAR_BH}
{Angl{\'e}s-Alc{\'a}zar} D.,  {Faucher-Gigu{\`e}re} C.-A.,  {Quataert} E.,
  {Hopkins} P.~F.,  {Feldmann} R.,  {Torrey} P.,  {Wetzel} A.,   {Kere{\v{s}}}
  D.,  2017b, \mn@doi [\mnras] {10.1093/mnrasl/slx161}, \href
  {https://ui.adsabs.harvard.edu/abs/2017MNRAS.472L.109A} {472, L109}

\bibitem[\protect\citeauthoryear{{Applebaum}, {Brooks}, {Christensen},
  {Munshi}, {Quinn}, {Shen}  \& {Tremmel}}{{Applebaum}
  et~al.}{2021}]{2021APPLEBAUM}
{Applebaum} E.,  {Brooks} A.~M.,  {Christensen} C.~R.,  {Munshi} F.,  {Quinn}
  T.~R.,  {Shen} S.,   {Tremmel} M.,  2021, \mn@doi [\apj]
  {10.3847/1538-4357/abcafa}, \href
  {https://ui.adsabs.harvard.edu/abs/2021ApJ...906...96A} {906, 96}

\bibitem[\protect\citeauthoryear{{Aravena} et~al.,}{{Aravena}
  et~al.}{2016}]{2016ARAVENA}
{Aravena} M.,  et~al., 2016, \mn@doi [\mnras] {10.1093/mnras/stw275}, \href
  {https://ui.adsabs.harvard.edu/abs/2016MNRAS.457.4406A} {457, 4406}

\bibitem[\protect\citeauthoryear{{Arribas}, {Colina}, {Bellocchi}, {Maiolino}
  \& {Villar-Mart{\'\i}n}}{{Arribas} et~al.}{2014}]{2014ARRIBAS}
{Arribas} S.,  {Colina} L.,  {Bellocchi} E.,  {Maiolino} R.,
  {Villar-Mart{\'\i}n} M.,  2014, \mn@doi [\aap] {10.1051/0004-6361/201323324},
  \href {https://ui.adsabs.harvard.edu/abs/2014A&A...568A..14A} {568, A14}

\bibitem[\protect\citeauthoryear{{Aversa}, {Lapi}, {de Zotti}, {Shankar}  \&
  {Danese}}{{Aversa} et~al.}{2015}]{2015AVERSA}
{Aversa} R.,  {Lapi} A.,  {de Zotti} G.,  {Shankar} F.,   {Danese} L.,  2015,
  \mn@doi [\apj] {10.1088/0004-637X/810/1/74}, \href
  {https://ui.adsabs.harvard.edu/abs/2015ApJ...810...74A} {810, 74}

\bibitem[\protect\citeauthoryear{{Bischetti}, {Maiolino}, {Carniani}, {Fiore},
  {Piconcelli}  \& {Fluetsch}}{{Bischetti} et~al.}{2019}]{2019BISCHETTI}
{Bischetti} M.,  {Maiolino} R.,  {Carniani} S.,  {Fiore} F.,  {Piconcelli} E.,
   {Fluetsch} A.,  2019, \mn@doi [\aap] {10.1051/0004-6361/201833557}, \href
  {https://ui.adsabs.harvard.edu/abs/2019A&A...630A..59B} {630, A59}

\bibitem[\protect\citeauthoryear{{Bothwell} et~al.,}{{Bothwell}
  et~al.}{2013}]{2013BOTHWELL}
{Bothwell} M.~S.,  et~al., 2013, \mn@doi [\mnras] {10.1093/mnras/sts562}, \href
  {https://ui.adsabs.harvard.edu/abs/2013MNRAS.429.3047B} {429, 3047}

\bibitem[\protect\citeauthoryear{{Bryan} \& {Norman}}{{Bryan} \&
  {Norman}}{1998}]{1998BRYAN}
{Bryan} G.~L.,  {Norman} M.~L.,  1998, \mn@doi [\apj] {10.1086/305262}, \href
  {https://ui.adsabs.harvard.edu/abs/1998ApJ...495...80B} {495, 80}

\bibitem[\protect\citeauthoryear{{Caffau}, {Ludwig}, {Steffen}, {Freytag}  \&
  {Bonifacio}}{{Caffau} et~al.}{2011}]{2011CAFFAU}
{Caffau} E.,  {Ludwig} H.~G.,  {Steffen} M.,  {Freytag} B.,   {Bonifacio} P.,
  2011, \mn@doi [\solphys] {10.1007/s11207-010-9541-4}, \href
  {https://ui.adsabs.harvard.edu/abs/2011SoPh..268..255C} {268, 255}

\bibitem[\protect\citeauthoryear{{Carlstrom} et~al.,}{{Carlstrom}
  et~al.}{2011}]{2011CARLSTROM}
{Carlstrom} J.~E.,  et~al., 2011, \mn@doi [\pasp] {10.1086/659879}, \href
  {https://ui.adsabs.harvard.edu/abs/2011PASP..123..568C} {123, 568}

\bibitem[\protect\citeauthoryear{{Carniani}, {Maiolino}, {Smit}  \&
  {Amor{\'\i}n}}{{Carniani} et~al.}{2018}]{2018CARNIANI}
{Carniani} S.,  {Maiolino} R.,  {Smit} R.,   {Amor{\'\i}n} R.,  2018, \mn@doi
  [\apjl] {10.3847/2041-8213/aaab45}, \href
  {https://ui.adsabs.harvard.edu/abs/2018ApJ...854L...7C} {854, L7}

\bibitem[\protect\citeauthoryear{{Chapman}, {Blain}, {Smail}  \&
  {Ivison}}{{Chapman} et~al.}{2005}]{2005CHAPMAN}
{Chapman} S.~C.,  {Blain} A.~W.,  {Smail} I.,   {Ivison} R.~J.,  2005, \mn@doi
  [\apj] {10.1086/428082}, \href
  {https://ui.adsabs.harvard.edu/abs/2005ApJ...622..772C} {622, 772}

\bibitem[\protect\citeauthoryear{{Chisholm}, {Tremonti}, {Leitherer}, {Chen},
  {Wofford}  \& {Lundgren}}{{Chisholm} et~al.}{2015}]{2015CHISHOLM}
{Chisholm} J.,  {Tremonti} C.~A.,  {Leitherer} C.,  {Chen} Y.,  {Wofford} A.,
  {Lundgren} B.,  2015, \mn@doi [\apj] {10.1088/0004-637X/811/2/149}, \href
  {https://ui.adsabs.harvard.edu/abs/2015ApJ...811..149C} {811, 149}

\bibitem[\protect\citeauthoryear{{Christensen}, {Dav{\'e}}, {Governato},
  {Pontzen}, {Brooks}, {Munshi}, {Quinn}  \& {Wadsley}}{{Christensen}
  et~al.}{2016}]{2016CHRISTENSEN}
{Christensen} C.~R.,  {Dav{\'e}} R.,  {Governato} F.,  {Pontzen} A.,  {Brooks}
  A.,  {Munshi} F.,  {Quinn} T.,   {Wadsley} J.,  2016, \mn@doi [\apj]
  {10.3847/0004-637X/824/1/57}, \href
  {https://ui.adsabs.harvard.edu/abs/2016ApJ...824...57C} {824, 57}

\bibitem[\protect\citeauthoryear{{Cicone} et~al.,}{{Cicone}
  et~al.}{2015}]{2015CICONE}
{Cicone} C.,  et~al., 2015, \mn@doi [\aap] {10.1051/0004-6361/201424980}, \href
  {https://ui.adsabs.harvard.edu/abs/2015A&A...574A..14C} {574, A14}

\bibitem[\protect\citeauthoryear{{Cicone}, {Maiolino}  \& {Marconi}}{{Cicone}
  et~al.}{2016}]{2016CICONE}
{Cicone} C.,  {Maiolino} R.,   {Marconi} A.,  2016, \mn@doi [\aap]
  {10.1051/0004-6361/201424514}, \href
  {https://ui.adsabs.harvard.edu/abs/2016A&A...588A..41C} {588, A41}

\bibitem[\protect\citeauthoryear{{Cochrane} et~al.,}{{Cochrane}
  et~al.}{2019}]{2019COCHRANE}
{Cochrane} R.~K.,  et~al., 2019, \mn@doi [\mnras] {10.1093/mnras/stz1736},
  \href {https://ui.adsabs.harvard.edu/abs/2019MNRAS.488.1779C} {488, 1779}

\bibitem[\protect\citeauthoryear{{D'Eugenio} et~al.,}{{D'Eugenio}
  et~al.}{2021}]{2021DEUGENIO}
{D'Eugenio} C.,  et~al., 2021, \mn@doi [\aap] {10.1051/0004-6361/202040067},
  \href {https://ui.adsabs.harvard.edu/abs/2021A&A...653A..32D} {653, A32}

\bibitem[\protect\citeauthoryear{{Daddi} et~al.,}{{Daddi}
  et~al.}{2010}]{2010DADDI}
{Daddi} E.,  et~al., 2010, \mn@doi [\apjl] {10.1088/2041-8205/714/1/L118},
  \href {https://ui.adsabs.harvard.edu/abs/2010ApJ...714L.118D} {714, L118}

\bibitem[\protect\citeauthoryear{{Dav{\'e}}, {Thompson}  \&
  {Hopkins}}{{Dav{\'e}} et~al.}{2016}]{2016DAVE}
{Dav{\'e}} R.,  {Thompson} R.,   {Hopkins} P.~F.,  2016, \mn@doi [\mnras]
  {10.1093/mnras/stw1862}, \href
  {https://ui.adsabs.harvard.edu/abs/2016MNRAS.462.3265D} {462, 3265}

\bibitem[\protect\citeauthoryear{{Dav{\'e}}, {Angl{\'e}s-Alc{\'a}zar},
  {Narayanan}, {Li}, {Rafieferantsoa}  \& {Appleby}}{{Dav{\'e}}
  et~al.}{2019}]{2019DAVE}
{Dav{\'e}} R.,  {Angl{\'e}s-Alc{\'a}zar} D.,  {Narayanan} D.,  {Li} Q.,
  {Rafieferantsoa} M.~H.,   {Appleby} S.,  2019, \mn@doi [\mnras]
  {10.1093/mnras/stz937}, \href
  {https://ui.adsabs.harvard.edu/abs/2019MNRAS.486.2827D} {486, 2827}

\bibitem[\protect\citeauthoryear{{Dekel} \& {Silk}}{{Dekel} \&
  {Silk}}{1986}]{1986DEKEL}
{Dekel} A.,  {Silk} J.,  1986, \mn@doi [\apj] {10.1086/164050}, \href
  {https://ui.adsabs.harvard.edu/abs/1986ApJ...303...39D} {303, 39}

\bibitem[\protect\citeauthoryear{{Dekel} et~al.,}{{Dekel}
  et~al.}{2009}]{2009DEKEL}
{Dekel} A.,  et~al., 2009, \mn@doi [\nat] {10.1038/nature07648}, \href
  {https://ui.adsabs.harvard.edu/abs/2009Natur.457..451D} {457, 451}

\bibitem[\protect\citeauthoryear{{Downes} \& {Solomon}}{{Downes} \&
  {Solomon}}{1998}]{1998DOWNES}
{Downes} D.,  {Solomon} P.~M.,  1998, \mn@doi [\apj] {10.1086/306339}, \href
  {https://ui.adsabs.harvard.edu/abs/1998ApJ...507..615D} {507, 615}

\bibitem[\protect\citeauthoryear{{Dubois} et~al.,}{{Dubois}
  et~al.}{2021}]{2021DUBOIS}
{Dubois} Y.,  et~al., 2021, \mn@doi [\aap] {10.1051/0004-6361/202039429}, \href
  {https://ui.adsabs.harvard.edu/abs/2021A&A...651A.109D} {651, A109}

\bibitem[\protect\citeauthoryear{{Elmegreen}}{{Elmegreen}}{2002}]{2002ELMEGREEN}
{Elmegreen} B.~G.,  2002, \mn@doi [\apj] {10.1086/342177}, \href
  {https://ui.adsabs.harvard.edu/abs/2002ApJ...577..206E} {577, 206}

\bibitem[\protect\citeauthoryear{{Erb}}{{Erb}}{2015}]{2015ERB}
{Erb} D.~K.,  2015, \mn@doi [\nat] {10.1038/nature14454}, \href
  {https://ui.adsabs.harvard.edu/abs/2015Natur.523..169E} {523, 169}

\bibitem[\protect\citeauthoryear{{Faucher-Gigu{\`e}re}, {Lidz}, {Zaldarriaga}
  \& {Hernquist}}{{Faucher-Gigu{\`e}re} et~al.}{2009}]{2009FAUCHER}
{Faucher-Gigu{\`e}re} C.-A.,  {Lidz} A.,  {Zaldarriaga} M.,   {Hernquist} L.,
  2009, \mn@doi [\apj] {10.1088/0004-637X/703/2/1416}, \href
  {https://ui.adsabs.harvard.edu/abs/2009ApJ...703.1416F} {703, 1416}

\bibitem[\protect\citeauthoryear{{Faucher-Gigu{\`e}re}, {Kere{\v{s}}}  \&
  {Ma}}{{Faucher-Gigu{\`e}re} et~al.}{2011}]{2011CLAUDEANDRE}
{Faucher-Gigu{\`e}re} C.-A.,  {Kere{\v{s}}} D.,   {Ma} C.-P.,  2011, \mn@doi
  [\mnras] {10.1111/j.1365-2966.2011.19457.x}, \href
  {https://ui.adsabs.harvard.edu/abs/2011MNRAS.417.2982F} {417, 2982}

\bibitem[\protect\citeauthoryear{{Faucher-Gigu{\`e}re}, {Quataert}  \&
  {Hopkins}}{{Faucher-Gigu{\`e}re} et~al.}{2013}]{2013FAUCHER-GIGUERE}
{Faucher-Gigu{\`e}re} C.-A.,  {Quataert} E.,   {Hopkins} P.~F.,  2013, \mn@doi
  [\mnras] {10.1093/mnras/stt866}, \href
  {https://ui.adsabs.harvard.edu/abs/2013MNRAS.433.1970F} {433, 1970}

\bibitem[\protect\citeauthoryear{{Feldmann}, {Hopkins}, {Quataert},
  {Faucher-Gigu{\`e}re}  \& {Kere{\v{s}}}}{{Feldmann}
  et~al.}{2016}]{2016FELDMANN}
{Feldmann} R.,  {Hopkins} P.~F.,  {Quataert} E.,  {Faucher-Gigu{\`e}re} C.-A.,
   {Kere{\v{s}}} D.,  2016, \mn@doi [\mnras] {10.1093/mnrasl/slw014}, \href
  {https://ui.adsabs.harvard.edu/abs/2016MNRAS.458L..14F} {458, L14}

\bibitem[\protect\citeauthoryear{{Feldmann}, {Quataert}, {Hopkins},
  {Faucher-Gigu{\`e}re}  \& {Kere{\v{s}}}}{{Feldmann}
  et~al.}{2017}]{2017FELDMANN}
{Feldmann} R.,  {Quataert} E.,  {Hopkins} P.~F.,  {Faucher-Gigu{\`e}re} C.-A.,
   {Kere{\v{s}}} D.,  2017, \mn@doi [\mnras] {10.1093/mnras/stx1120}, \href
  {https://ui.adsabs.harvard.edu/abs/2017MNRAS.470.1050F} {470, 1050}

\bibitem[\protect\citeauthoryear{{Feldmann} et~al.,}{{Feldmann}
  et~al.}{2022}]{2022FELDMANN}
{Feldmann} R.,  et~al., 2022, arXiv e-prints, \href
  {https://ui.adsabs.harvard.edu/abs/2022arXiv220515325F} {p. arXiv:2205.15325}

\bibitem[\protect\citeauthoryear{{Fielding}, {Quataert}  \&
  {Martizzi}}{{Fielding} et~al.}{2018}]{2018FIELDING}
{Fielding} D.,  {Quataert} E.,   {Martizzi} D.,  2018, \mn@doi [\mnras]
  {10.1093/mnras/sty2466}, \href
  {https://ui.adsabs.harvard.edu/abs/2018MNRAS.481.3325F} {481, 3325}

\bibitem[\protect\citeauthoryear{{Forrest} et~al.,}{{Forrest}
  et~al.}{2020}]{2020FORRESTb}
{Forrest} B.,  et~al., 2020, \mn@doi [\apjl] {10.3847/2041-8213/ab5b9f}, \href
  {https://ui.adsabs.harvard.edu/abs/2020ApJ...890L...1F} {890, L1}

\bibitem[\protect\citeauthoryear{{Fujimoto} et~al.,}{{Fujimoto}
  et~al.}{2019}]{2019FUJIMOTO}
{Fujimoto} S.,  et~al., 2019, \mn@doi [\apj] {10.3847/1538-4357/ab480f}, \href
  {https://ui.adsabs.harvard.edu/abs/2019ApJ...887..107F} {887, 107}

\bibitem[\protect\citeauthoryear{{Gallerani}, {Pallottini}, {Feruglio},
  {Ferrara}, {Maiolino}, {Vallini}, {Riechers}  \& {Pavesi}}{{Gallerani}
  et~al.}{2018}]{2018GALLERANI}
{Gallerani} S.,  {Pallottini} A.,  {Feruglio} C.,  {Ferrara} A.,  {Maiolino}
  R.,  {Vallini} L.,  {Riechers} D.~A.,   {Pavesi} R.,  2018, \mn@doi [\mnras]
  {10.1093/mnras/stx2458}, \href
  {https://ui.adsabs.harvard.edu/abs/2018MNRAS.473.1909G} {473, 1909}

\bibitem[\protect\citeauthoryear{{Garrison-Kimmel} et~al.,}{{Garrison-Kimmel}
  et~al.}{2017}]{2017GARRISON_KIMMEL}
{Garrison-Kimmel} S.,  et~al., 2017, \mn@doi [\mnras] {10.1093/mnras/stx1710},
  \href {https://ui.adsabs.harvard.edu/abs/2017MNRAS.471.1709G} {471, 1709}

\bibitem[\protect\citeauthoryear{{Gill}, {Knebe}  \& {Gibson}}{{Gill}
  et~al.}{2004}]{2004GILL}
{Gill} S. P.~D.,  {Knebe} A.,   {Gibson} B.~K.,  2004, \mn@doi [\mnras]
  {10.1111/j.1365-2966.2004.07786.x}, \href
  {https://ui.adsabs.harvard.edu/abs/2004MNRAS.351..399G} {351, 399}

\bibitem[\protect\citeauthoryear{{Ginolfi} et~al.,}{{Ginolfi}
  et~al.}{2020}]{2020GINOLFI}
{Ginolfi} M.,  et~al., 2020, \mn@doi [\aap] {10.1051/0004-6361/201936872},
  \href {https://ui.adsabs.harvard.edu/abs/2020A&A...633A..90G} {633, A90}

\bibitem[\protect\citeauthoryear{{Glazebrook} et~al.,}{{Glazebrook}
  et~al.}{2017}]{2017GLAZEBROOK}
{Glazebrook} K.,  et~al., 2017, \mn@doi [\nat] {10.1038/nature21680}, \href
  {https://ui.adsabs.harvard.edu/abs/2017Natur.544...71G} {544, 71}

\bibitem[\protect\citeauthoryear{{Grudi{\'c}}, {Hopkins},
  {Faucher-Gigu{\`e}re}, {Quataert}, {Murray}  \& {Kere{\v{s}}}}{{Grudi{\'c}}
  et~al.}{2018}]{2018GRUDIC}
{Grudi{\'c}} M.~Y.,  {Hopkins} P.~F.,  {Faucher-Gigu{\`e}re} C.-A.,  {Quataert}
  E.,  {Murray} N.,   {Kere{\v{s}}} D.,  2018, \mn@doi [\mnras]
  {10.1093/mnras/sty035}, \href
  {https://ui.adsabs.harvard.edu/abs/2018MNRAS.475.3511G} {475, 3511}

\bibitem[\protect\citeauthoryear{{Grudi{\'c}}, {Boylan-Kolchin},
  {Faucher-Gigu{\`e}re}  \& {Hopkins}}{{Grudi{\'c}} et~al.}{2020}]{2020GRUDIC}
{Grudi{\'c}} M.~Y.,  {Boylan-Kolchin} M.,  {Faucher-Gigu{\`e}re} C.-A.,
  {Hopkins} P.~F.,  2020, \mn@doi [\mnras] {10.1093/mnrasl/slaa103}, \href
  {https://ui.adsabs.harvard.edu/abs/2020MNRAS.496L.127G} {496, L127}

\bibitem[\protect\citeauthoryear{{Guedes}, {Callegari}, {Madau}  \&
  {Mayer}}{{Guedes} et~al.}{2011}]{2011GUEDES}
{Guedes} J.,  {Callegari} S.,  {Madau} P.,   {Mayer} L.,  2011, \mn@doi [\apj]
  {10.1088/0004-637X/742/2/76}, \href
  {https://ui.adsabs.harvard.edu/abs/2011ApJ...742...76G} {742, 76}

\bibitem[\protect\citeauthoryear{{Gurvich} et~al.,}{{Gurvich}
  et~al.}{2020}]{2020GURVICH}
{Gurvich} A.~B.,  et~al., 2020, \mn@doi [\mnras] {10.1093/mnras/staa2578},
  \href {https://ui.adsabs.harvard.edu/abs/2020MNRAS.498.3664G} {498, 3664}

\bibitem[\protect\citeauthoryear{{Gurvich} et~al.,}{{Gurvich}
  et~al.}{2022}]{2022GURVICH}
{Gurvich} A.~B.,  et~al., 2022, arXiv e-prints, \href
  {https://ui.adsabs.harvard.edu/abs/2022arXiv220304321G} {p. arXiv:2203.04321}

\bibitem[\protect\citeauthoryear{{Hahn} \& {Abel}}{{Hahn} \&
  {Abel}}{2011}]{2011HAHN}
{Hahn} O.,  {Abel} T.,  2011, \mn@doi [\mnras]
  {10.1111/j.1365-2966.2011.18820.x}, \href
  {https://ui.adsabs.harvard.edu/abs/2011MNRAS.415.2101H} {415, 2101}

\bibitem[\protect\citeauthoryear{{Hashimoto} et~al.,}{{Hashimoto}
  et~al.}{2018}]{2018HASHIMOTO}
{Hashimoto} T.,  et~al., 2018, \mn@doi [\nat] {10.1038/s41586-018-0117-z},
  \href {https://ui.adsabs.harvard.edu/abs/2018Natur.557..392H} {557, 392}

\bibitem[\protect\citeauthoryear{{Hayward} \& {Hopkins}}{{Hayward} \&
  {Hopkins}}{2017}]{2017HAYWARD}
{Hayward} C.~C.,  {Hopkins} P.~F.,  2017, \mn@doi [\mnras]
  {10.1093/mnras/stw2888}, \href
  {https://ui.adsabs.harvard.edu/abs/2017MNRAS.465.1682H} {465, 1682}

\bibitem[\protect\citeauthoryear{{Hayward} et~al.,}{{Hayward}
  et~al.}{2014}]{2014HAYWARD}
{Hayward} C.~C.,  et~al., 2014, \mn@doi [\mnras] {10.1093/mnras/stu1843}, \href
  {https://ui.adsabs.harvard.edu/abs/2014MNRAS.445.1598H} {445, 1598}

\bibitem[\protect\citeauthoryear{{Hayward} et~al.,}{{Hayward}
  et~al.}{2021}]{2021HAYWARD}
{Hayward} C.~C.,  et~al., 2021, \mn@doi [\mnras] {10.1093/mnras/stab246}, \href
  {https://ui.adsabs.harvard.edu/abs/2021MNRAS.502.2922H} {502, 2922}

\bibitem[\protect\citeauthoryear{{Heckman}, {Alexandroff}, {Borthakur},
  {Overzier}  \& {Leitherer}}{{Heckman} et~al.}{2015}]{2015HECKMAN}
{Heckman} T.~M.,  {Alexandroff} R.~M.,  {Borthakur} S.,  {Overzier} R.,
  {Leitherer} C.,  2015, \mn@doi [\apj] {10.1088/0004-637X/809/2/147}, \href
  {https://ui.adsabs.harvard.edu/abs/2015ApJ...809..147H} {809, 147}

\bibitem[\protect\citeauthoryear{{Henden}, {Puchwein}, {Shen}  \&
  {Sijacki}}{{Henden} et~al.}{2018}]{2018HENDEN}
{Henden} N.~A.,  {Puchwein} E.,  {Shen} S.,   {Sijacki} D.,  2018, \mn@doi
  [\mnras] {10.1093/mnras/sty1780}, \href
  {https://ui.adsabs.harvard.edu/abs/2018MNRAS.479.5385H} {479, 5385}

\bibitem[\protect\citeauthoryear{{Herrera-Camus} et~al.,}{{Herrera-Camus}
  et~al.}{2021}]{2021HERRERA}
{Herrera-Camus} R.,  et~al., 2021, \mn@doi [\aap]
  {10.1051/0004-6361/202039704}, \href
  {https://ui.adsabs.harvard.edu/abs/2021A&A...649A..31H} {649, A31}

\bibitem[\protect\citeauthoryear{{Hopkins}}{{Hopkins}}{2015}]{2015HOPKINS}
{Hopkins} P.~F.,  2015, \mn@doi [\mnras] {10.1093/mnras/stv195}, \href
  {https://ui.adsabs.harvard.edu/abs/2015MNRAS.450...53H} {450, 53}

\bibitem[\protect\citeauthoryear{{Hopkins}, {Quataert}  \& {Murray}}{{Hopkins}
  et~al.}{2012}]{2012HOPKINS}
{Hopkins} P.~F.,  {Quataert} E.,   {Murray} N.,  2012, \mn@doi [\mnras]
  {10.1111/j.1365-2966.2012.20593.x}, \href
  {https://ui.adsabs.harvard.edu/abs/2012MNRAS.421.3522H} {421, 3522}

\bibitem[\protect\citeauthoryear{{Hopkins}, {Kere{\v{s}}}, {O{\~n}orbe},
  {Faucher-Gigu{\`e}re}, {Quataert}, {Murray}  \& {Bullock}}{{Hopkins}
  et~al.}{2014}]{2014HOPKINS}
{Hopkins} P.~F.,  {Kere{\v{s}}} D.,  {O{\~n}orbe} J.,  {Faucher-Gigu{\`e}re}
  C.-A.,  {Quataert} E.,  {Murray} N.,   {Bullock} J.~S.,  2014, \mn@doi
  [\mnras] {10.1093/mnras/stu1738}, \href
  {https://ui.adsabs.harvard.edu/abs/2014MNRAS.445..581H} {445, 581}

\bibitem[\protect\citeauthoryear{{Hopkins} et~al.,}{{Hopkins}
  et~al.}{2018}]{2018HOPKINS}
{Hopkins} P.~F.,  et~al., 2018, \mn@doi [\mnras] {10.1093/mnras/sty1690}, \href
  {https://ui.adsabs.harvard.edu/abs/2018MNRAS.480..800H} {480, 800}

\bibitem[\protect\citeauthoryear{{Hopkins} et~al.,}{{Hopkins}
  et~al.}{2022a}]{2022HOPKINS_FIRE3}
{Hopkins} P.~F.,  et~al., 2022a, arXiv e-prints, \href
  {https://ui.adsabs.harvard.edu/abs/2022arXiv220300040H} {p. arXiv:2203.00040}

\bibitem[\protect\citeauthoryear{{Hopkins}, {Wellons},
  {Angl{\'e}s-Alc{\'a}zar}, {Faucher-Gigu{\`e}re}  \& {Grudi{\'c}}}{{Hopkins}
  et~al.}{2022b}]{2022HOPKINS}
{Hopkins} P.~F.,  {Wellons} S.,  {Angl{\'e}s-Alc{\'a}zar} D.,
  {Faucher-Gigu{\`e}re} C.-A.,   {Grudi{\'c}} M.~Y.,  2022b, \mn@doi [\mnras]
  {10.1093/mnras/stab3458}, \href
  {https://ui.adsabs.harvard.edu/abs/2022MNRAS.510..630H} {510, 630}

\bibitem[\protect\citeauthoryear{{Hunter}}{{Hunter}}{2007}]{2007HUNTER}
{Hunter} J.~D.,  2007, \mn@doi [Computing in Science and Engineering]
  {10.1109/MCSE.2007.55}, \href
  {https://ui.adsabs.harvard.edu/abs/2007CSE.....9...90H} {9, 90}

\bibitem[\protect\citeauthoryear{{Kennicutt}}{{Kennicutt}}{1998}]{1998KENNICUTT}
{Kennicutt} Robert~C. J.,  1998, \mn@doi [\apj] {10.1086/305588}, \href
  {https://ui.adsabs.harvard.edu/abs/1998ApJ...498..541K} {498, 541}

\bibitem[\protect\citeauthoryear{{Kennicutt} \& {De Los Reyes}}{{Kennicutt} \&
  {De Los Reyes}}{2021}]{2021KENNICUTT}
{Kennicutt} Robert~C. J.,  {De Los Reyes} M. A.~C.,  2021, \mn@doi [\apj]
  {10.3847/1538-4357/abd3a2}, \href
  {https://ui.adsabs.harvard.edu/abs/2021ApJ...908...61K} {908, 61}

\bibitem[\protect\citeauthoryear{{Kennicutt} \& {Evans}}{{Kennicutt} \&
  {Evans}}{2012}]{2012KENNICUTT}
{Kennicutt} R.~C.,  {Evans} N.~J.,  2012, \mn@doi [\araa]
  {10.1146/annurev-astro-081811-125610}, \href
  {https://ui.adsabs.harvard.edu/abs/2012ARA&A..50..531K} {50, 531}

\bibitem[\protect\citeauthoryear{{Kim} \& {Ostriker}}{{Kim} \&
  {Ostriker}}{2015}]{2015KIM}
{Kim} C.-G.,  {Ostriker} E.~C.,  2015, \mn@doi [\apj]
  {10.1088/0004-637X/802/2/99}, \href
  {https://ui.adsabs.harvard.edu/abs/2015ApJ...802...99K} {802, 99}

\bibitem[\protect\citeauthoryear{{Knollmann} \& {Knebe}}{{Knollmann} \&
  {Knebe}}{2009}]{2009AKNOLLMANN}
{Knollmann} S.~R.,  {Knebe} A.,  2009, \mn@doi [\apjs]
  {10.1088/0067-0049/182/2/608}, \href
  {https://ui.adsabs.harvard.edu/abs/2009ApJS..182..608K} {182, 608}

\bibitem[\protect\citeauthoryear{{Kroupa}}{{Kroupa}}{2001}]{2001KROUPA}
{Kroupa} P.,  2001, \mn@doi [\mnras] {10.1046/j.1365-8711.2001.04022.x}, \href
  {https://ui.adsabs.harvard.edu/abs/2001MNRAS.322..231K} {322, 231}

\bibitem[\protect\citeauthoryear{{Krumholz} \& {Gnedin}}{{Krumholz} \&
  {Gnedin}}{2011}]{2011KRUMHOLZ}
{Krumholz} M.~R.,  {Gnedin} N.~Y.,  2011, \mn@doi [\apj]
  {10.1088/0004-637X/729/1/36}, \href
  {https://ui.adsabs.harvard.edu/abs/2011ApJ...729...36K} {729, 36}

\bibitem[\protect\citeauthoryear{{Lapi}, {Pantoni}, {Boco}  \& {Danese}}{{Lapi}
  et~al.}{2020}]{2020LAPI}
{Lapi} A.,  {Pantoni} L.,  {Boco} L.,   {Danese} L.,  2020, \mn@doi [\apj]
  {10.3847/1538-4357/ab9812}, \href
  {https://ui.adsabs.harvard.edu/abs/2020ApJ...897...81L} {897, 81}

\bibitem[\protect\citeauthoryear{{Larson}}{{Larson}}{1974}]{1974LARSON}
{Larson} R.~B.,  1974, \mn@doi [\mnras] {10.1093/mnras/169.2.229}, \href
  {https://ui.adsabs.harvard.edu/abs/1974MNRAS.169..229L} {169, 229}

\bibitem[\protect\citeauthoryear{{Leitherer} et~al.,}{{Leitherer}
  et~al.}{1999}]{1999LEITHERE}
{Leitherer} C.,  et~al., 1999, \mn@doi [\apjs] {10.1086/313233}, \href
  {https://ui.adsabs.harvard.edu/abs/1999ApJS..123....3L} {123, 3}

\bibitem[\protect\citeauthoryear{{Ma}, {Hopkins}, {Faucher-Gigu{\`e}re},
  {Zolman}, {Muratov}, {Kere{\v{s}}}  \& {Quataert}}{{Ma}
  et~al.}{2016}]{2016MA}
{Ma} X.,  {Hopkins} P.~F.,  {Faucher-Gigu{\`e}re} C.-A.,  {Zolman} N.,
  {Muratov} A.~L.,  {Kere{\v{s}}} D.,   {Quataert} E.,  2016, \mn@doi [\mnras]
  {10.1093/mnras/stv2659}, \href
  {https://ui.adsabs.harvard.edu/abs/2016MNRAS.456.2140M} {456, 2140}

\bibitem[\protect\citeauthoryear{{Maiolino} et~al.,}{{Maiolino}
  et~al.}{2012}]{2012MAIOLINO}
{Maiolino} R.,  et~al., 2012, \mn@doi [\mnras]
  {10.1111/j.1745-3933.2012.01303.x}, \href
  {https://ui.adsabs.harvard.edu/abs/2012MNRAS.425L..66M} {425, L66}

\bibitem[\protect\citeauthoryear{{Man} \& {Belli}}{{Man} \&
  {Belli}}{2018}]{2018MAN}
{Man} A.,  {Belli} S.,  2018, \mn@doi [Nature Astronomy]
  {10.1038/s41550-018-0558-1}, \href
  {https://ui.adsabs.harvard.edu/abs/2018NatAs...2..695M} {2, 695}

\bibitem[\protect\citeauthoryear{{Mancuso}, {Lapi}, {Shi}, {Gonzalez-Nuevo},
  {Aversa}  \& {Danese}}{{Mancuso} et~al.}{2016}]{2016MANCUSO}
{Mancuso} C.,  {Lapi} A.,  {Shi} J.,  {Gonzalez-Nuevo} J.,  {Aversa} R.,
  {Danese} L.,  2016, \mn@doi [\apj] {10.3847/0004-637X/823/2/128}, \href
  {https://ui.adsabs.harvard.edu/abs/2016ApJ...823..128M} {823, 128}

\bibitem[\protect\citeauthoryear{{Marrone} et~al.,}{{Marrone}
  et~al.}{2018}]{2018MARRONE}
{Marrone} D.~P.,  et~al., 2018, \mn@doi [\nat] {10.1038/nature24629}, \href
  {https://ui.adsabs.harvard.edu/abs/2018Natur.553...51M} {553, 51}

\bibitem[\protect\citeauthoryear{{Martizzi}, {Faucher-Gigu{\`e}re}  \&
  {Quataert}}{{Martizzi} et~al.}{2015}]{2015MARTIZZI}
{Martizzi} D.,  {Faucher-Gigu{\`e}re} C.-A.,   {Quataert} E.,  2015, \mn@doi
  [\mnras] {10.1093/mnras/stv562}, \href
  {https://ui.adsabs.harvard.edu/abs/2015MNRAS.450..504M} {450, 504}

\bibitem[\protect\citeauthoryear{{Matthee} et~al.,}{{Matthee}
  et~al.}{2019}]{2019MATTHEE}
{Matthee} J.,  et~al., 2019, \mn@doi [\apj] {10.3847/1538-4357/ab2f81}, \href
  {https://ui.adsabs.harvard.edu/abs/2019ApJ...881..124M} {881, 124}

\bibitem[\protect\citeauthoryear{{McKee} \& {Ostriker}}{{McKee} \&
  {Ostriker}}{1977}]{1977MCKEE}
{McKee} C.~F.,  {Ostriker} J.~P.,  1977, \mn@doi [\apj] {10.1086/155667}, \href
  {https://ui.adsabs.harvard.edu/abs/1977ApJ...218..148M} {218, 148}

\bibitem[\protect\citeauthoryear{{Micha{\l}owski}, {Hjorth}  \&
  {Watson}}{{Micha{\l}owski} et~al.}{2010}]{2010MICHALOWSKI}
{Micha{\l}owski} M.,  {Hjorth} J.,   {Watson} D.,  2010, \mn@doi [\aap]
  {10.1051/0004-6361/200913634}, \href
  {https://ui.adsabs.harvard.edu/abs/2010A&A...514A..67M} {514, A67}

\bibitem[\protect\citeauthoryear{{Mitchell}, {Schaye}, {Bower}  \&
  {Crain}}{{Mitchell} et~al.}{2020}]{2020MITCHELL}
{Mitchell} P.~D.,  {Schaye} J.,  {Bower} R.~G.,   {Crain} R.~A.,  2020, \mn@doi
  [\mnras] {10.1093/mnras/staa938}, \href
  {https://ui.adsabs.harvard.edu/abs/2020MNRAS.494.3971M} {494, 3971}

\bibitem[\protect\citeauthoryear{{Mo}, {Mao}  \& {White}}{{Mo}
  et~al.}{1998}]{1998MO}
{Mo} H.~J.,  {Mao} S.,   {White} S. D.~M.,  1998, \mn@doi [\mnras]
  {10.1046/j.1365-8711.1998.01227.x}, \href
  {https://ui.adsabs.harvard.edu/abs/1998MNRAS.295..319M} {295, 319}

\bibitem[\protect\citeauthoryear{{Muratov}, {Kere{\v{s}}},
  {Faucher-Gigu{\`e}re}, {Hopkins}, {Quataert}  \& {Murray}}{{Muratov}
  et~al.}{2015}]{2015MURATOV}
{Muratov} A.~L.,  {Kere{\v{s}}} D.,  {Faucher-Gigu{\`e}re} C.-A.,  {Hopkins}
  P.~F.,  {Quataert} E.,   {Murray} N.,  2015, \mn@doi [\mnras]
  {10.1093/mnras/stv2126}, \href
  {https://ui.adsabs.harvard.edu/abs/2015MNRAS.454.2691M} {454, 2691}

\bibitem[\protect\citeauthoryear{{Murray}, {Quataert}  \& {Thompson}}{{Murray}
  et~al.}{2005}]{2005MURRAY}
{Murray} N.,  {Quataert} E.,   {Thompson} T.~A.,  2005, \mn@doi [\apj]
  {10.1086/426067}, \href
  {https://ui.adsabs.harvard.edu/abs/2005ApJ...618..569M} {618, 569}

\bibitem[\protect\citeauthoryear{{Naab} \& {Ostriker}}{{Naab} \&
  {Ostriker}}{2017}]{2017NAAB}
{Naab} T.,  {Ostriker} J.~P.,  2017, \mn@doi [\araa]
  {10.1146/annurev-astro-081913-040019}, \href
  {https://ui.adsabs.harvard.edu/abs/2017ARA&A..55...59N} {55, 59}

\bibitem[\protect\citeauthoryear{{Nelson} et~al.,}{{Nelson}
  et~al.}{2019}]{2019NELSON}
{Nelson} D.,  et~al., 2019, \mn@doi [\mnras] {10.1093/mnras/stz2306}, \href
  {https://ui.adsabs.harvard.edu/abs/2019MNRAS.490.3234N} {490, 3234}

\bibitem[\protect\citeauthoryear{{Orr} et~al.,}{{Orr} et~al.}{2018}]{2018ORR}
{Orr} M.~E.,  et~al., 2018, \mn@doi [\mnras] {10.1093/mnras/sty1241}, \href
  {https://ui.adsabs.harvard.edu/abs/2018MNRAS.478.3653O} {478, 3653}

\bibitem[\protect\citeauthoryear{{Orr}, {Fielding}, {Hayward}  \&
  {Burkhart}}{{Orr} et~al.}{2022}]{2022ORR}
{Orr} M.~E.,  {Fielding} D.~B.,  {Hayward} C.~C.,   {Burkhart} B.,  2022,
  \mn@doi [\apj] {10.3847/1538-4357/ac6c26}, \href
  {https://ui.adsabs.harvard.edu/abs/2022ApJ...932...88O} {932, 88}

\bibitem[\protect\citeauthoryear{{Ostriker} \& {Shetty}}{{Ostriker} \&
  {Shetty}}{2011}]{2011OSTRIKER}
{Ostriker} E.~C.,  {Shetty} R.,  2011, \mn@doi [\apj]
  {10.1088/0004-637X/731/1/41}, \href
  {https://ui.adsabs.harvard.edu/abs/2011ApJ...731...41O} {731, 41}

\bibitem[\protect\citeauthoryear{{Pandya} et~al.,}{{Pandya}
  et~al.}{2021}]{2021PANDYA}
{Pandya} V.,  et~al., 2021, \mn@doi [\mnras] {10.1093/mnras/stab2714}, \href
  {https://ui.adsabs.harvard.edu/abs/2021MNRAS.508.2979P} {508, 2979}

\bibitem[\protect\citeauthoryear{{Pantoni} et~al.,}{{Pantoni}
  et~al.}{2021}]{2021PANTONI}
{Pantoni} L.,  et~al., 2021, \mn@doi [\mnras] {10.1093/mnras/stab2346}, \href
  {https://ui.adsabs.harvard.edu/abs/2021MNRAS.507.3998P} {507, 3998}

\bibitem[\protect\citeauthoryear{{Parsotan}, {Cochrane}, {Hayward},
  {Angl{\'e}s-Alc{\'a}zar}, {Feldmann}, {Faucher-Gigu{\`e}re}, {Wellons}  \&
  {Hopkins}}{{Parsotan} et~al.}{2021}]{2021PARSOTAN}
{Parsotan} T.,  {Cochrane} R.~K.,  {Hayward} C.~C.,  {Angl{\'e}s-Alc{\'a}zar}
  D.,  {Feldmann} R.,  {Faucher-Gigu{\`e}re} C.~A.,  {Wellons} S.,   {Hopkins}
  P.~F.,  2021, \mn@doi [\mnras] {10.1093/mnras/staa3765}, \href
  {https://ui.adsabs.harvard.edu/abs/2021MNRAS.501.1591P} {501, 1591}

\bibitem[\protect\citeauthoryear{{Perrotta} et~al.,}{{Perrotta}
  et~al.}{2021}]{2021PERROTTA}
{Perrotta} S.,  et~al., 2021, \mn@doi [\apj] {10.3847/1538-4357/ac2fa4}, \href
  {https://ui.adsabs.harvard.edu/abs/2021ApJ...923..275P} {923, 275}

\bibitem[\protect\citeauthoryear{{Pillepich} et~al.,}{{Pillepich}
  et~al.}{2018}]{2018PILLEPICH}
{Pillepich} A.,  et~al., 2018, \mn@doi [\mnras] {10.1093/mnras/stx2656}, \href
  {https://ui.adsabs.harvard.edu/abs/2018MNRAS.473.4077P} {473, 4077}

\bibitem[\protect\citeauthoryear{{Planck Collaboration} et~al.,}{{Planck
  Collaboration} et~al.}{2020}]{2020PLANCK}
{Planck Collaboration} et~al., 2020, \mn@doi [\aap]
  {10.1051/0004-6361/201833910}, \href
  {https://ui.adsabs.harvard.edu/abs/2020A&A...641A...6P} {641, A6}

\bibitem[\protect\citeauthoryear{{Riechers} et~al.,}{{Riechers}
  et~al.}{2013}]{2013RIECHERS}
{Riechers} D.~A.,  et~al., 2013, \mn@doi [\nat] {10.1038/nature12050}, \href
  {https://ui.adsabs.harvard.edu/abs/2013Natur.496..329R} {496, 329}

\bibitem[\protect\citeauthoryear{{Riechers} et~al.,}{{Riechers}
  et~al.}{2020}]{2020RIECHERS}
{Riechers} D.~A.,  et~al., 2020, \mn@doi [\apj] {10.3847/1538-4357/ab8c48},
  \href {https://ui.adsabs.harvard.edu/abs/2020ApJ...895...81R} {895, 81}

\bibitem[\protect\citeauthoryear{{Rubin}, {Prochaska}, {Koo}, {Phillips},
  {Martin}  \& {Winstrom}}{{Rubin} et~al.}{2014}]{2014RUBIN}
{Rubin} K. H.~R.,  {Prochaska} J.~X.,  {Koo} D.~C.,  {Phillips} A.~C.,
  {Martin} C.~L.,   {Winstrom} L.~O.,  2014, \mn@doi [\apj]
  {10.1088/0004-637X/794/2/156}, \href
  {https://ui.adsabs.harvard.edu/abs/2014ApJ...794..156R} {794, 156}

\bibitem[\protect\citeauthoryear{{Schaye} et~al.,}{{Schaye}
  et~al.}{2015}]{2015SCHAYE}
{Schaye} J.,  et~al., 2015, \mn@doi [\mnras] {10.1093/mnras/stu2058}, \href
  {https://ui.adsabs.harvard.edu/abs/2015MNRAS.446..521S} {446, 521}

\bibitem[\protect\citeauthoryear{{Schreiber} et~al.,}{{Schreiber}
  et~al.}{2018}]{2018SCHREIBER}
{Schreiber} C.,  et~al., 2018, \mn@doi [\aap] {10.1051/0004-6361/201833070},
  \href {https://ui.adsabs.harvard.edu/abs/2018A&A...618A..85S} {618, A85}

\bibitem[\protect\citeauthoryear{{Silk}}{{Silk}}{1997}]{1997SILK}
{Silk} J.,  1997, \mn@doi [\apj] {10.1086/304073}, \href
  {https://ui.adsabs.harvard.edu/abs/1997ApJ...481..703S} {481, 703}

\bibitem[\protect\citeauthoryear{{Silk} \& {Rees}}{{Silk} \&
  {Rees}}{1998}]{1998SILK}
{Silk} J.,  {Rees} M.~J.,  1998, \aap, \href
  {https://ui.adsabs.harvard.edu/abs/1998A&A...331L...1S} {331, L1}

\bibitem[\protect\citeauthoryear{{Smith}, {Hayward}, {Jarvis}  \&
  {Simpson}}{{Smith} et~al.}{2017}]{2017SMITH}
{Smith} D.~J.~B.,  {Hayward} C.~C.,  {Jarvis} M.~J.,   {Simpson} C.,  2017,
  \mn@doi [\mnras] {10.1093/mnras/stx1689}, \href
  {https://ui.adsabs.harvard.edu/abs/2017MNRAS.471.2453S} {471, 2453}

\bibitem[\protect\citeauthoryear{{Somerville} \& {Dav{\'e}}}{{Somerville} \&
  {Dav{\'e}}}{2015}]{2015SOMERVILLE}
{Somerville} R.~S.,  {Dav{\'e}} R.,  2015, \mn@doi [\araa]
  {10.1146/annurev-astro-082812-140951}, \href
  {https://ui.adsabs.harvard.edu/abs/2015ARA&A..53...51S} {53, 51}

\bibitem[\protect\citeauthoryear{{Sparre}, {Hayward}, {Feldmann},
  {Faucher-Gigu{\`e}re}, {Muratov}, {Kere{\v{s}}}  \& {Hopkins}}{{Sparre}
  et~al.}{2017}]{2017SPARRE}
{Sparre} M.,  {Hayward} C.~C.,  {Feldmann} R.,  {Faucher-Gigu{\`e}re} C.-A.,
  {Muratov} A.~L.,  {Kere{\v{s}}} D.,   {Hopkins} P.~F.,  2017, \mn@doi
  [\mnras] {10.1093/mnras/stw3011}, \href
  {https://ui.adsabs.harvard.edu/abs/2017MNRAS.466...88S} {466, 88}

\bibitem[\protect\citeauthoryear{{Spilker} et~al.,}{{Spilker}
  et~al.}{2020a}]{2020SPILKER1}
{Spilker} J.~S.,  et~al., 2020a, \mn@doi [\apj] {10.3847/1538-4357/abc47f},
  \href {https://ui.adsabs.harvard.edu/abs/2020ApJ...905...85S} {905, 85}

\bibitem[\protect\citeauthoryear{{Spilker} et~al.,}{{Spilker}
  et~al.}{2020b}]{2020SPILKER2}
{Spilker} J.~S.,  et~al., 2020b, \mn@doi [\apj] {10.3847/1538-4357/abc4e6},
  \href {https://ui.adsabs.harvard.edu/abs/2020ApJ...905...86S} {905, 86}

\bibitem[\protect\citeauthoryear{{Springel} \& {Hernquist}}{{Springel} \&
  {Hernquist}}{2003}]{2003SPRINGEL}
{Springel} V.,  {Hernquist} L.,  2003, \mn@doi [\mnras]
  {10.1046/j.1365-8711.2003.06206.x}, \href
  {https://ui.adsabs.harvard.edu/abs/2003MNRAS.339..289S} {339, 289}

\bibitem[\protect\citeauthoryear{{Steidel}, {Erb}, {Shapley}, {Pettini},
  {Reddy}, {Bogosavljevi{\'c}}, {Rudie}  \& {Rakic}}{{Steidel}
  et~al.}{2010}]{2010STEIDEL}
{Steidel} C.~C.,  {Erb} D.~K.,  {Shapley} A.~E.,  {Pettini} M.,  {Reddy} N.,
  {Bogosavljevi{\'c}} M.,  {Rudie} G.~C.,   {Rakic} O.,  2010, \mn@doi [\apj]
  {10.1088/0004-637X/717/1/289}, \href
  {https://ui.adsabs.harvard.edu/abs/2010ApJ...717..289S} {717, 289}

\bibitem[\protect\citeauthoryear{{Stern} et~al.,}{{Stern}
  et~al.}{2021}]{2021STERN}
{Stern} J.,  et~al., 2021, \mn@doi [\apj] {10.3847/1538-4357/abd776}, \href
  {https://ui.adsabs.harvard.edu/abs/2021ApJ...911...88S} {911, 88}

\bibitem[\protect\citeauthoryear{{Stinson}, {Brook}, {Macci{\`o}}, {Wadsley},
  {Quinn}  \& {Couchman}}{{Stinson} et~al.}{2013}]{2013STINSON}
{Stinson} G.~S.,  {Brook} C.,  {Macci{\`o}} A.~V.,  {Wadsley} J.,  {Quinn}
  T.~R.,   {Couchman} H.~M.~P.,  2013, \mn@doi [\mnras] {10.1093/mnras/sts028},
  \href {https://ui.adsabs.harvard.edu/abs/2013MNRAS.428..129S} {428, 129}

\bibitem[\protect\citeauthoryear{{Su} et~al.,}{{Su} et~al.}{2021}]{2021SU}
{Su} K.-Y.,  et~al., 2021, \mn@doi [\mnras] {10.1093/mnras/stab2021}, \href
  {https://ui.adsabs.harvard.edu/abs/2021MNRAS.507..175S} {507, 175}

\bibitem[\protect\citeauthoryear{{Talia} et~al.,}{{Talia}
  et~al.}{2017}]{2017TALIA}
{Talia} M.,  et~al., 2017, \mn@doi [\mnras] {10.1093/mnras/stx1788}, \href
  {https://ui.adsabs.harvard.edu/abs/2017MNRAS.471.4527T} {471, 4527}

\bibitem[\protect\citeauthoryear{{Tollet}, {Cattaneo}, {Macci{\`o}}, {Dutton}
  \& {Kang}}{{Tollet} et~al.}{2019}]{2019TOLLET}
{Tollet} {\'E}.,  {Cattaneo} A.,  {Macci{\`o}} A.~V.,  {Dutton} A.~A.,   {Kang}
  X.,  2019, \mn@doi [\mnras] {10.1093/mnras/stz545}, \href
  {https://ui.adsabs.harvard.edu/abs/2019MNRAS.485.2511T} {485, 2511}

\bibitem[\protect\citeauthoryear{{Valentino} et~al.,}{{Valentino}
  et~al.}{2020}]{2020VALENTINO}
{Valentino} F.,  et~al., 2020, \mn@doi [\apj] {10.3847/1538-4357/ab64dc}, \href
  {https://ui.adsabs.harvard.edu/abs/2020ApJ...889...93V} {889, 93}

\bibitem[\protect\citeauthoryear{{Veilleux}, {Cecil}  \&
  {Bland-Hawthorn}}{{Veilleux} et~al.}{2005}]{2005VEILLEUX}
{Veilleux} S.,  {Cecil} G.,   {Bland-Hawthorn} J.,  2005, \mn@doi [\araa]
  {10.1146/annurev.astro.43.072103.150610}, \href
  {https://ui.adsabs.harvard.edu/abs/2005ARA&A..43..769V} {43, 769}

\bibitem[\protect\citeauthoryear{{Veilleux}, {Maiolino}, {Bolatto}  \&
  {Aalto}}{{Veilleux} et~al.}{2020}]{2020VEILLEUX}
{Veilleux} S.,  {Maiolino} R.,  {Bolatto} A.~D.,   {Aalto} S.,  2020, \mn@doi
  [\aapr] {10.1007/s00159-019-0121-9}, \href
  {https://ui.adsabs.harvard.edu/abs/2020A&ARv..28....2V} {28, 2}

\bibitem[\protect\citeauthoryear{{Vogelsberger}, {Marinacci}, {Torrey}  \&
  {Puchwein}}{{Vogelsberger} et~al.}{2020}]{2020VOGELSBERGER}
{Vogelsberger} M.,  {Marinacci} F.,  {Torrey} P.,   {Puchwein} E.,  2020,
  \mn@doi [Nature Reviews Physics] {10.1038/s42254-019-0127-2}, \href
  {https://ui.adsabs.harvard.edu/abs/2020NatRP...2...42V} {2, 42}

\bibitem[\protect\citeauthoryear{{Wellons}, {Faucher-Gigu{\`e}re},
  {Angl{\'e}s-Alc{\'a}zar}, {Hayward}, {Feldmann}, {Hopkins}  \&
  {Kere{\v{s}}}}{{Wellons} et~al.}{2020}]{2020WELLONS}
{Wellons} S.,  {Faucher-Gigu{\`e}re} C.-A.,  {Angl{\'e}s-Alc{\'a}zar} D.,
  {Hayward} C.~C.,  {Feldmann} R.,  {Hopkins} P.~F.,   {Kere{\v{s}}} D.,  2020,
  \mn@doi [\mnras] {10.1093/mnras/staa2229}, \href
  {https://ui.adsabs.harvard.edu/abs/2020MNRAS.497.4051W} {497, 4051}

\bibitem[\protect\citeauthoryear{{Wellons} et~al.,}{{Wellons}
  et~al.}{2022}]{2022WELLONS}
{Wellons} S.,  et~al., 2022, arXiv e-prints, \href
  {https://ui.adsabs.harvard.edu/abs/2022arXiv220306201W} {p. arXiv:2203.06201}

\bibitem[\protect\citeauthoryear{{Wetzel}, {Hopkins}, {Kim},
  {Faucher-Gigu{\`e}re}, {Kere{\v{s}}}  \& {Quataert}}{{Wetzel}
  et~al.}{2016}]{2016WETZEL}
{Wetzel} A.~R.,  {Hopkins} P.~F.,  {Kim} J.-h.,  {Faucher-Gigu{\`e}re} C.-A.,
  {Kere{\v{s}}} D.,   {Quataert} E.,  2016, \mn@doi [\apjl]
  {10.3847/2041-8205/827/2/L23}, \href
  {https://ui.adsabs.harvard.edu/abs/2016ApJ...827L..23W} {827, L23}

\bibitem[\protect\citeauthoryear{{Wetzel} et~al.,}{{Wetzel}
  et~al.}{2022}]{2022WETZEL}
{Wetzel} A.,  et~al., 2022, arXiv e-prints, \href
  {https://ui.adsabs.harvard.edu/abs/2022arXiv220206969W} {p. arXiv:2202.06969}

\bibitem[\protect\citeauthoryear{{White} \& {Frenk}}{{White} \&
  {Frenk}}{1991}]{1991WHITE}
{White} S. D.~M.,  {Frenk} C.~S.,  1991, \mn@doi [\apj] {10.1086/170483}, \href
  {https://ui.adsabs.harvard.edu/abs/1991ApJ...379...52W} {379, 52}

\bibitem[\protect\citeauthoryear{{Wong} \& {Blitz}}{{Wong} \&
  {Blitz}}{2002}]{2002WONG}
{Wong} T.,  {Blitz} L.,  2002, \mn@doi [\apj] {10.1086/339287}, \href
  {https://ui.adsabs.harvard.edu/abs/2002ApJ...569..157W} {569, 157}

\bibitem[\protect\citeauthoryear{{Yesuf}, {Faber}, {Trump}, {Koo}, {Fang},
  {Liu}, {Wild}  \& {Hayward}}{{Yesuf} et~al.}{2014}]{2014YESUF}
{Yesuf} H.~M.,  {Faber} S.~M.,  {Trump} J.~R.,  {Koo} D.~C.,  {Fang} J.~J.,
  {Liu} F.~S.,  {Wild} V.,   {Hayward} C.~C.,  2014, \mn@doi [\apj]
  {10.1088/0004-637X/792/2/84}, \href
  {https://ui.adsabs.harvard.edu/abs/2014ApJ...792...84Y} {792, 84}

\bibitem[\protect\citeauthoryear{{Zavala} et~al.,}{{Zavala}
  et~al.}{2018}]{2018ZAVALA}
{Zavala} J.~A.,  et~al., 2018, \mn@doi [Nature Astronomy]
  {10.1038/s41550-017-0297-8}, \href
  {https://ui.adsabs.harvard.edu/abs/2018NatAs...2...56Z} {2, 56}

\bibitem[\protect\citeauthoryear{{Zhao}, {Mo}, {Jing}  \& {B{\"o}rner}}{{Zhao}
  et~al.}{2003}]{2003ZHAO}
{Zhao} D.~H.,  {Mo} H.~J.,  {Jing} Y.~P.,   {B{\"o}rner} G.,  2003, \mn@doi
  [\mnras] {10.1046/j.1365-8711.2003.06135.x}, \href
  {https://ui.adsabs.harvard.edu/abs/2003MNRAS.339...12Z} {339, 12}

\bibitem[\protect\citeauthoryear{{{\c{C}}atmabacak}, {Feldmann},
  {Angl{\'e}s-Alc{\'a}zar}, {Faucher-Gigu{\`e}re}, {Hopkins}  \&
  {Kere{\v{s}}}}{{{\c{C}}atmabacak} et~al.}{2022}]{2022CATMABACAK}
{{\c{C}}atmabacak} O.,  {Feldmann} R.,  {Angl{\'e}s-Alc{\'a}zar} D.,
  {Faucher-Gigu{\`e}re} C.-A.,  {Hopkins} P.~F.,   {Kere{\v{s}}} D.,  2022,
  \mn@doi [\mnras] {10.1093/mnras/stac040}, \href
  {https://ui.adsabs.harvard.edu/abs/2022MNRAS.511..506C} {511, 506}

\makeatother
\end{thebibliography}

% Alternatively you could enter them by hand, like this:
% This method is tedious and prone to error if you have lots of references
%\begin{thebibliography}{99}
%\bibitem[\protect\citeauthoryear{Author}{2012}]{Author2012}
%Author A.~N., 2013, Journal of Improbable Astronomy, 1, 1
%\bibitem[\protect\citeauthoryear{Others}{2013}]{Others2013}
%Others S., 2012, Journal of Interesting Stuff, 17, 198
%\end{thebibliography}

%%%%%%%%%%%%%%%%%%%%%%%%%%%%%%%%%%%%%%%%%%%%%%%%%%

%%%%%%%%%%%%%%%%% APPENDICES %%%%%%%%%%%%%%%%%%%%%

%%%%%%%%%%%%%%%%%%%%%%%%%%%%%%%%%%%%%%%%%%%%%%%%%%

% Don't change these lines
\bsp	% typesetting comment
\label{lastpage}
\end{document}